\documentclass[twocolumn, superscriptaddress]{revtex4-2}

\usepackage[utf8]{inputenc}
\usepackage[ngerman, english]{babel}
\usepackage[T1]{fontenc}

\usepackage{amsthm}
\usepackage{graphicx, color}
\usepackage{amsmath}
\usepackage{amssymb}
\usepackage{mathtools}
\usepackage{caption}

\usepackage{tikz}
\usetikzlibrary{quantikz2}

\usepackage{adjustbox}

\usepackage{url}
\usepackage[colorlinks=false, linkbordercolor=blue]{hyperref}

\addto\captionsenglish{}

\begin{document}

\title{Resource Implications of Different Encodings for Quantum Computational Fluid Dynamics}

\author{Hans A. K\"osel}

\email[corresponding author -- e-mail address: ]{hans.koesel@dlr.de}
\affiliation{Institute of Aerodynamics and Flow Technology, German Aerospace Center (DLR), Lilienthalplatz 7, 38108 Braunschweig, Germany}

\author{Roland Ewert}
\affiliation{Institute of Aerodynamics and Flow Technology, German Aerospace Center (DLR), Lilienthalplatz 7, 38108 Braunschweig, Germany}

\author{Jan W. Delfs}
\affiliation{Institute of Aerodynamics and Flow Technology, German Aerospace Center (DLR), Lilienthalplatz 7, 38108 Braunschweig, Germany}

\date{\today}

\begin{abstract}

For quantum algorithms for problems in which the task is to compute an entire field of values, like e.g. computational fluid dynamics (CFD), it is often proposed amplitude encoding w.r.t. multiple qubits; however, the efforts implied by it for initialization and read-out are not addressed. 
This work is devoted specifically to this issue: 
It reviews different encoding schemes in quantum computing, discussing their computational costs for initialization and read-out as well as resulting aspects for their usage via minimal examples. 
The considerations in previous literature on the required computational resources for amplitude encoding w.r.t. multiple qubits are extended in the presented quantification by explicitly deducing the circuit depth that results for the decomposed initialization procedure of V. V. Shende et al. \cite{Shende_et_al_Synthesis_of_q_circuits_long, Shende_et_al_Synthesis_of_q_circuits_short} and deriving an upper bound for the necessary number of executions of a quantum algorithm to extract the encoded values with a specific accuracy. 
For these two results, an empirical verification via the means provided by IBM's quantum computing simulation framework {\emph{Qiskit}} \cite{qiskit} is given. 
In the framework of the study on the required number of runs to achieve a desired accuracy, it is however found that the derived upper bound, scaling like $ {{{\tilde{n}}}^2} ~ {\ln( {\tilde{n}} )} $ with the number of encoded values $ {\tilde{n}} $, is too conservative to be used for precise estimations. 
Therefore, a corresponding study of the required runs for the reference distribution of equal probabilities for all basis states is done in particular, which suggests $ {\tilde{n}} ~ { \ln( {\tilde{n}} ) } $ as an empirical scaling law. 
Since the view regarding CFD applications is taken here, it is presented in particular that the insights from this work lead to a new encoding approach, which is proposed specifically for a quantum algorithm for the lattice Boltzmann method.

\end{abstract}

\maketitle

\section{Introduction}

As quantum mechanics (QM) is a generalized theory w.r.t. classical mechanics, i.e. a theory that is superordinate to it and comprises it as a limiting case, quantum computing (QC) generalizes classical computing by augmenting the classical bits, being either in a state $ \left| 0  \right\rangle $ or $ \left| 1 \right\rangle $, with the wave phenomena of superposition and in particular its special form of entanglement that can occur for many-particle systems. 
Thus, the question arises whether these additional phenomena can be exploited to reduce computation times w.r.t. classical algorithms, where we take the perspective regarding the applicability of QC for computational fluid dynamics (CFD).

The fundamental computation task in CFD is the calculation of data fields of values for quantities that describe a fluid, e.g. the pressure, the fluid velocity or a potential, where the dynamics are given by differential equations. 
Based on the necessary computation of the entire flow field, parameters of interest that characterize the system, like e.g. lift coefficients, pressure fluctuations at specific positions or correlations, can then be derived in a subsequent computation. 
Therefore, many of the works on quantum algorithms for CFD \cite{Succi_et_al_Review, Griffin_et_al_Review} focus on the elemental first computation problem, i.e., they present quantum circuits that prepare a final quantum state that encodes the entire flow field.

Concerning these, we have the impression that the majority resorts to the common amplitude encoding w.r.t. multiple qubits, often proposed for QC since it allows to encode $ 2^n $ complex numbers in $ n \in \mathbb{N} $ qubits \cite{Nielsen_Chuang_Book}. 
Among the various strategies for simulating the dynamics of flows by means of quantum algorithms that process the data for the flow field in this encoding format, there are in particular proposals for time-marching schemes, where some are set up specifically for the computation of one time step (e.g. \cite{Budinski_Q_alg_by_streamfunction_vorticity_LBM, Wawrzyniak_et_al_Q_alg_for_LBM} and \cite{Wawrzyniak_et_al_Unitary_q_alg_for_LBM} for the non-linear problem), since they explicitly demand the read-out of the encoded flow field and its re-initialization as the input for the next step, while others even allow to compute multiple time steps without the need of the classical data (e.g. \cite{Over_et_al_Q_alg_for_AD_eq, Bengoechea_et_al_Q_algs_BCs_arxiv_v1, Nagel_Loewe_Q_LBM} and \cite{Wawrzyniak_et_al_Unitary_q_alg_for_LBM} for the linear problem).

However, for amplitude encoding w.r.t. multiple qubits, it is also known that in the general case, i.e. if no structure of the given field can be exploited, the required resources in terms of subsequent operations for the preparation of a state in this encoding format scale exponentially with the number of qubits $ n $ \cite{Shende_et_al_Synthesis_of_q_circuits_long, Shende_et_al_Synthesis_of_q_circuits_short, Nielsen_Chuang_Book, Succi_et_al_Review, Griffin_et_al_Review} and also the read-out of a state in this encoding format, i.e. the extraction of the accessible information about it, requires an immense number of executions of the quantum algorithm relative to the $ 2^n $ slots of the state space used for encoding \cite{Wawrzyniak_et_al_Q_alg_for_LBM, Nagel_Loewe_Q_LBM, Kocherla_et_al_Q_alg_for_mesoscale_simulations}. 
Therefore, it is to conclude that quantum algorithms that use this encoding should be able to leverage the runtime improvements found for one run of them if they are used as a building block in a larger quantum algorithm, e.g. for the purpose of calculating selected observables of the flow system, but for their use as a stand-alone, this seems questionable, and in particular regarding situations where an in general turbulent flow field has to be in fact read out and re-initialized at every time step like in computational aeroacoustics (CAA). 
Nevertheless, at least for testing the functionality of such procedures that involve some amount of amplitude encoding for an unstructured initial state and the final state, the required time for the initialization and number of runs of the algorithm should remain relevant, respectively.

This paper provides a framework for the explicit quantitative estimation of such computational resources for different encoding schemes. 
It has the following structure: 
In section {\hyperref[sec:Theoretical_Considerations]{II -- \emph{Theoretical Considerations}}}, the considered kinds of encoding are reviewed, where the efforts for initialization and read-out that accompany them, respectively, are derived. 
In it, the QC terms used already in this introduction are also more precisely introduced via definitions and it is commented in particular on aspects regarding the usability of the classical data storing format for QC, where the detailed considerations are out-sourced to the {\hyperref[App_sec:Remarks_on_Bitstring_Encoding]{appendix A}}. 
Specifically for amplitude encoding w.r.t. multiple qubits, an initialization procedure is explained in the work of V. V. Shende et al. \cite{Shende_et_al_Synthesis_of_q_circuits_long} (and see also \cite{Shende_et_al_Synthesis_of_q_circuits_short}), in which a formula for the number of required controlled NOT-gates (abbreviated here as CX-gates) is derived. 
The present work from us reviews this procedure and explicitly deduces the circuit depth of it in the section {\hyperref[sec:Theoretical_Considerations]{II}}. 
For this encoding scheme, this section presents furthermore a derivation of an upper bound for the required number of runs, i.e. a specific number of executions of the quantum algorithm that is sufficient to achieve a certain accuracy for the values extracted by that many runs. 
We are not aware that such a quantitative bound would be already stated elsewhere in the literature. 
Moreover, these two claims for the costs of amplitude encoding are confirmed by empirical studies, which were done using the {\emph{Qiskit}}-framework from IBM \cite{qiskit} and are presented in section {\hyperref[sec:Empirical_Studies_for_GAE]{III -- \emph{Empirical Studies for General Amplitude Encoding}}}. 
Since the obtained upper bound for the number of runs that is needed to extract a state in amplitude encoding with a certain precision is relatively pessimistic, section {\hyperref[sec:Empirical_Studies_for_GAE]{III}} also gives a protocol of empirically found numbers for the reference situation that all basis state probabilities are equal. 
For this situation, an empirical scaling law is hence provided in this section and an additional analytical consideration can be found in the appendix  {\hyperref[App_sec:Probability_Fulfilling_error_bound_for_reference_situation]{\ref*{App_sec:Probability_Fulfilling_error_bound_for_reference_situation}}}. 
Section {\hyperref[sec:New_Encoding_for_LBM]{IV -- \emph{A Novel Encoding Scheme for Treating Generic Problems according to the Lattice Boltzmann Method}}} explains then a design proposal for a quantum algorithm according to the {\emph{lattice Boltzmann method}} (LBM) \cite{Succi_Book_long, Succi_Book_short, Krueger_et_al_Book}, which results as a consequence of the considerations on encodings. 
The considerations in section {\hyperref[sec:New_Encoding_for_LBM]{IV}} and the algorithm presented there are found to represent an extension of the discussion and the proposed algorithm of M. Schalkers and M. Möller in \cite{Schalkers_Moeller_Encoding_BMs}, respectively. 
At last, the section {\hyperref[sec:Conclusion]{V -- \emph{Conclusion}}} gives a summary of our work and perspectives for further work.

\section{Theoretical Considerations}
\label{sec:Theoretical_Considerations}

Since computational resources like the necessary number of runs to obtain an encoded solution with a specific accuracy depend on the used encoding, the subsection {\hyperref[subsec:Encoding]{\emph{Encoding}}} gives at first a revision of different encoding techniques. 
Based on this, subsection {\hyperref[subsec:Initialization]{\emph{Initialization}}} presents estimations for the required number of native gates for preparing an initial state vector in these encoding formats and at last, in subsection {\hyperref[subsec:Read-out]{\emph{Read-out}}}, upper bounds for the required number of runs of a quantum algorithm to achieve a given accuracy for the extraction of a result that is encoded in a previously presented format are derived.

\subsection{Encoding}
\label{subsec:Encoding}

The principle idea of QC is to store the information about a computational problem via the values of the probability amplitudes $ a_{k} \in \mathbb{C} $ of the basis states $ \left| k \right\rangle $ of a general superposition state
\begin{align}
 | {{\psi}^{(n)}} \rangle
  &= {\sum_{k=0}^{{2^n}-1}} {a_k} { \left| k \right\rangle }  ,  \label{eq:general_state} \\
 \left| k \right\rangle &= | {\underbrace{ {q_{0,k}} ~ {q_{1,k}} \dots {q_{n-1,k}} }_{ \text{bitstring of }k }} \rangle \nonumber \\
 & = \left| {q_{0,k}} \right\rangle \otimes \left| {q_{1,k}} \right\rangle \otimes \dots \otimes \left| {q_{n-1,k}} \right\rangle , \label{eq:general_state_basis_state} \\[0.15cm] 
  & \quad ~ {q_{i, k}} \in \{ 0, 1 \} , \nonumber
\end{align}
where $ \otimes $ denotes the Kronecker tensor product and normalization
\begin{align}
{\sum_{k=0}^{{2^n}-1}} {a^{*}_k} {a_k} &= 1 
\label{eq:general_state_normalization}
\end{align}
holds, using $ {}^{*} $ to denote the complex conjugate.

'Encoding' refers to the specific rule how the numbers describing the problem are represented in the state of a qubit system ({\hyperref[eq:general_state]{\ref*{eq:general_state}}}). 
In the following, three encoding patterns are considered, which are briefly referred to here as {\emph{general amplitude encoding}}, {\emph{1-qubit amplitude encoding}} and {\emph{bitstring encoding}}, respectively. 
Starting from the {{general amplitude encoding}}, often proposed for QC, via its special form of {{1-qubit amplitude encoding}} to the further specialization of {{bitstring encoding}}, it is gone from the encoding format in which the information is stored most densely w.r.t. the needed qubits to the technique that corresponds to the classical data storing.

\subsubsection{General Amplitude Encoding}

'General amplitude encoding' means that in general all available amplitudes of a superposition state (\hyperref[eq:general_state]{\ref*{eq:general_state}}) are used to store numbers, allowing to encode up to $ 2^{n} $ numbers with $n$ qubits. 
It is to note that for $ 2^{n} - 1 $ known amplitudes, the value of the last amplitude is constrained in its absolute value via the normalization condition (\hyperref[eq:general_state_normalization]{\ref*{eq:general_state_normalization}}) but the missing piece of information to deduce the absolute values of all $ 2^{n} $ given encoded numbers from $ 2^{n} - 1 $ probabilities $ {{|{a_k}|}^2} = {a_k^{*}}{a_k} $ is given by the normalization of the vector of given numbers that has to be computed and kept in mind in order to finally multiply it again with the $ {{|{a_k}|}^2} $, representing $ 2^n $ parts of a whole.

Since a general superposition state results for this encoding, the amplitudes can be such that the system state cannot be written as a tensor product of 1-qubit states, i.e., {\bf{entanglement can be present}} in contrast to the two following encoding schemes.

\subsubsection{1-Qubit Amplitude Encoding}

'1-qubit amplitude encoding' means the possibility to use amplitude encoding only for individual qubits to store up to $ 2 $ numbers via a qubit according to the special case $ n = 1 $ of (\hyperref[eq:general_state]{\ref*{eq:general_state}})
\begin{align}
 \left| {\psi}_{\text{qubit}} \right\rangle = 
 | {{\psi}^{(1)}} \rangle 
  &= {a_0} \cdot \left| {0} \right\rangle + {a_1} \cdot \left| {1} \right\rangle , \label{eq:general_state_1_qubit} \\
   \quad {{|a_0|}^2} + {{|a_1|}^2} &= 1 \nonumber
\end{align}
and set then multiple independent qubits to store all given numbers. 
Thus, $ 2n $ numbers can be encoded in this way.

This way of encoding is characterized by the feature that no general superposition state but a product state is described, i.e., there is {\bf{no entanglement}}.

An example for a quantum algorithm that exploits {{1-qubit amplitude encoding}} is the work of S. Kocherla et al. \cite{Kocherla_et_al_Q_alg_for_mesoscale_simulations} for the lattice gas automaton representation of the LBM, where a qubit stores only the value of one distribution function in one of its two amplitudes. 
This encoding format fits naturally to the local collision operation of the LBM, which is also apparent from the fact that this lattice gas automata representation of the collision also allows at least for one time step to represent a change of the distribution functions according to a non-linear problem like the Burgers equation via a unitary operation.

\subsubsection{Bitstring Encoding}

'Bitstring encoding' means that only the basis states $ \left| 0 \right\rangle $ and $ \left| 1 \right\rangle $ of the qubits are taken to store information and accordingly corresponds to the storing pattern as in classical bits. 
Then, a specific amount of $ d \in \mathbb{N} $ qubits is needed to resolve a given number w.r.t. a given value range via a specific number of binary digits and multiple registers of independent qubits are set then to store all given numbers.

The fundamental feature of the state vector according to this encoding is that it is known that it is {\bf{no superposition state}} but the classical binary representation. 
Therefore, it is to remark in particular that at a stage of a quantum algorithm with this encoding, measuring the qubits does not perturb the state, so that the time window for performing QC operations that is given by the coherence time of the used device can be reset. 
Moreover, since it exists a finite set of known input states for this encoding, the implementation of non-linear functions is relatively easy (see Appendix {\hyperref[App_subsec:Representing_Non-lin_Functions]{\ref*{App_subsec:Representing_Non-lin_Functions}}}).\\

In general, it is to note that the encoding formats of a prepared initial state and the resulting final state of QC procedure do not have to be the same. 
E.g., in the probabilistic method of {\emph{linear combination of unitaries}} (LCU) \cite{Childs_Wiebe_LCU, Over_et_al_Q_alg_for_AD_eq, Bengoechea_et_al_Q_algs_BCs_arxiv_v1}, the initialization procedure for {{general amplitude encoding}} has to be applied to the ancilla qubits as the so-called {\emph{preparation gate}} at the beginning and end of this routine to encode a data set of summation coefficients but at the end, the ancilla qubit register does not store relevant information according to {{general amplitude encoding}}, but in terms of {{bitstring encoding}}, i.e., just the resulting bitstring of the ancilla register is of interest and read out.

\subsection{Initialization}
\label{subsec:Initialization}

'Initialization' refers to the procedure to prepare an arbitrary state according to a specific encoding starting from a given state. 
Usually, it is started with the state $ \left| 0 ~ 0 ~ 0 ~ \dots \right\rangle $, i.e. the state in which all qubits are in the $ \left| 0 \right\rangle $-state, which is also set here as the starting state.

For the three kinds of encoding introduced in the last subsection, the expected execution times for the respective initialization procedures are stated in this subsection. 
The time to do a QC procedure is basically given via a decomposition of the quantum circuit diagram into the {\emph{native}} gates, i.e. the gate operations that the hardware can actually do. 
More general, the procedure to express a quantum circuit via the physically implemented operations of a specific hardware and taking into account the connectivity of the real qubit systems of this device in it, is called {\emph{compiling}} or {\emph{transpiling}} of the quantum circuit. 
The specific execution time results then via summing up the execution times of the obtained layers of native gates, where the execution time of a layer is given by the largest execution time of a gate appearing in the considered layer since {\emph{gate layers}} are defined by performing all gates in a quantum circuit diagram as early as possible, i.e. moving them as far as possible to the start of the circuit diagram, and grouping then gates that can be done in parallel. 
The number of layers is further called  {\emph{depth}} of the circuit.

For a procedure to perform such a decomposition, it is referred here to the work of V. V. Shende et al. \cite{Shende_et_al_Synthesis_of_q_circuits_long}, in which it is explained how arbitrary gates can be expressed as {\emph{multiplexors}}, which are generalized multi-controlled gates, for which decomposition rules are derived, which have to be applied recursively. 
The native gates to which circuits are broken down in \cite{Shende_et_al_Synthesis_of_q_circuits_long} are the three 1-qubit rotation gates in the forms 
\begin{align}
{R_x}({\theta})&= 
\begin{pmatrix}
{\cos{\left( {\frac{\theta}{2}} \right)}} & {\text{i}} ~ {\sin{\left( {\frac{\theta}{2}} \right)}} \\
{\text{i}} ~ {\sin{\left( {\frac{\theta}{2}} \right)}} & {\cos{\left( {\frac{\theta}{2}} \right)}}
\end{pmatrix} , \label{eq:Rx_matrix_form} \\
{R_y}({\theta})&= 
\begin{pmatrix}
{\cos{\left( {\frac{\theta}{2}} \right)}} & {\sin{\left( {\frac{\theta}{2}} \right)}} 
 \\
-{\sin{\left( {\frac{\theta}{2}} \right)}} & {\cos{\left( {\frac{\theta}{2}} \right)}}
\end{pmatrix} , \label{eq:Ry_matrix_form} \\
{R_z}({\theta})&= 
\begin{pmatrix}
{\exp{\left( -{\text{i}} {\frac{\theta}{2}} \right)}} & 0 \\
0 & {\exp{\left( {\text{i}} {\frac{\theta}{2}} \right)}}
\end{pmatrix}
 \label{eq:Rz_matrix_form}
\end{align}
and the controlled NOT-gate, denoted here in the following as CX-gate. 
In \cite{Shende_et_al_Synthesis_of_q_circuits_long}, also an initialization procedure for {{general amplitude encoding}} is explained in particular:

\subsubsection{General Amplitude Encoding}

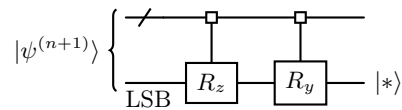
\begin{figure}[t]
\begin{center}
\tikzset{invisible/.style={fill=none,draw=none,line width=0pt,inner xsep=0pt,inner ysep=0pt}}
\begin{quantikz}
 \lstick[2]{$ {{ | {\psi}^{(n+1)} \rangle }} $} & \qwbundle{ } &[-0.2cm]
  |[operator]| & |[operator]| &  \\
 & \gategroup[1,steps=1, style={invisible}, label style={label position=below,anchor=mid,yshift=-0.05cm, xshift=-0.2cm}]{LSB} & \gate{{R_z}}\wire[u]{q} & \gate{{R_y}}\wire[u]{q} & \rstick{$ {{\left| {*} \right\rangle}} $}
\end{quantikz}
\end{center}
\captionsetup{justification=raggedright, singlelinecheck=false}
\caption[]{Quantum circuit schematics of a representation for the disentangling of the LSB of the general state $ {{ | {\psi}^{(n+1)} \rangle }} $. 
Figure adapted from \cite{Shende_et_al_Synthesis_of_q_circuits_long}.}
\label{Fig:Shende_et_al_adapted_Disentangling_LSB}
\end{figure}

Following \cite{Shende_et_al_Synthesis_of_q_circuits_long}, it is found that an arbitrary state vector formed by $ n $ qubits can be prepared by a quantum circuit that contains
\begin{align}
 {2^n} - 1 
\label{eq:GAEI_Number_Ry-gates_for_n}
\end{align}
$ {R_y} $-gates,
\begin{align}
 {2^n} - 1 
\label{eq:GAEI_Number_Rz-gates_for_n}
\end{align}
$ {R_z} $-gates 
and
\begin{align}
 {2^{n+1}} - 2{(n+1)}
\label{eq:GAEI_Number_CX-gates_for_n}
\end{align}
CX-gates, where the CX-gates can span over qubits between the control and the target qubit, i.e., they are not all coupling only adjacent qubits in the circuit diagram. 
In this circuit, at most $ 2 $ 1-qubit rotation gates are in a gate layer and all CX-gates form an own layer, resulting in a contribution to the circuit depth due to the 1-qubit rotation gates of $ {2^{n+1}} - (n + 1) $ and a contribution to the circuit depth due to the CX-gates of $ 2^{n+1} - 2(n+1) $. 
In the following, the derivation of these results is outlined:

A circuit for preparing an arbitrary $ (n + 1) $-qubit state $ \left| {\psi}^{(n+1)} \right\rangle $ from the state
\begin{align*}
 | {\underbrace{ 0 ~ 0 ~ \dots ~ 0 }_{ n+1 {\text{ times}} }} \rangle &= \left| 0 \right\rangle \otimes \left| 0 \right\rangle \otimes \dots \otimes \left| 0 \right\rangle = {\left| 0 \right\rangle}^{\otimes (n+1)}
\end{align*}
is given by inverting a circuit that disentangles an arbitrary $ (n + 1) $-qubit state $ \left| {\psi}^{(n+1)} \right\rangle $ so that the state $ {\left| 0 \right\rangle}^{\otimes (n+1)} $ is produced. 
According to Theorem 9 in \cite{Shende_et_al_Synthesis_of_q_circuits_long}, a circuit for disentangling the qubit representing the least significant bit (LSB) can be given in terms of multiplexors via the form depicted in Fig. {\hyperref[Fig:Shende_et_al_adapted_Disentangling_LSB]{\ref*{Fig:Shende_et_al_adapted_Disentangling_LSB}}}. 
{\emph{Multiplexor}} means that for each bitstring that can be represented by its {\emph{select-qubits}}, indicated by '$ \square $' in the quantum circuit diagrams, a different gate operation is performed for the target qubits in general. 
E. g., an $ R_z $-multiplexor is a gate operation in which in general an $ R_z $-gate operation with a different angle $ \theta $ according to ({\hyperref[eq:Rz_matrix_form]{\ref*{eq:Rz_matrix_form}}}) is applied to the target qubit for each bitstring that is possible w.r.t. the classical configurations of the select-qubits. 
In the diagrams, a line with a slash stands for multiple adjacent horizontal qubit lines and a 1-qubit state $ \left| * \right\rangle $ can be chosen as $ \left| 0 \right\rangle $ or $ \left| 1 \right\rangle $.

\begin{figure*}[t]
\begin{center}
\tikzset{invisible/.style={fill=none,draw=none,line width=0pt,inner xsep=0pt,inner ysep=0pt}}
\begin{quantikz}
 \lstick[5]{$ {{ | {\psi}^{(n+1)} \rangle }} $} & \gategroup[1,steps=1, style={invisible}, label style={label position=above,anchor=mid,yshift=-0.3cm, xshift=-0.2cm}]{MSB} &
  \gategroup[5,steps=3,style={dashed,rounded corners, inner sep=2pt},background,label style={label position=below,anchor=north,yshift=-0.2cm}]{{\text{step } 1}} &[-0.5cm] |[operator]| & |[operator]| & \gategroup[4,steps=3,style={dashed,rounded corners, inner sep=2pt},background,label style={label position=below,anchor=north,yshift=-0.2cm}]{{\text{step } 2}} &[-0.5cm]
 |[operator]| & |[operator]| & & \ \ldots\ & \gategroup[2,steps=3,style={dashed,rounded corners, inner sep=2pt},background,label style={label position=below,anchor=north,yshift=-0.2cm}]{{\text{step } n}} &[-0.5cm]
 |[operator]| & |[operator]| & \gategroup[1,steps=3,style={dashed,rounded corners, inner sep=2pt},background,label style={label position=below,anchor=north,yshift=-0.2cm}]{{\text{step } n + 1 }} &[-0.5cm]
 \gate{{R_z}} & \gate{{R_y}} & \rstick{$ {{\left| {*} \right\rangle}} $}  \\[0.4cm]
 & & & |[operator]| & |[operator]| &  & |[operator]| & |[operator]| & & \ \ldots\ & & \gate{{R_z}}\wire[u][1]{q} & \gate{{R_y}}\wire[u][1]{q}  &  \rstick{$ {{\left| {*} \right\rangle}} $} & \wireoverride{n} & \wireoverride{n} & \wireoverride{n}  \\[0.4cm]
 \wave&&&&&&& & &  & \\
 & & & |[operator]| & |[operator]| & & \gate{{R_z}}\wire[u][3]{q} & \gate{{R_y}}\wire[u][3]{q}  &  \rstick{$ {{\left| {*} \right\rangle}} $} & \wireoverride{n} & \wireoverride{n} & \wireoverride{n} & \wireoverride{n} & \wireoverride{n} & \wireoverride{n} & \wireoverride{n} & \wireoverride{n}  \\[0.4cm]
 & \gategroup[1,steps=1, style={invisible}, label style={label position=below,anchor=mid,yshift=-0.05cm, xshift=-0.2cm}]{LSB} & & \gate{{R_z}}\wire[u][4]{q} & \gate{{R_y}}\wire[u][4]{q} & \rstick{$ {{\left| {*} \right\rangle}} $} & \wireoverride{n} & \wireoverride{n} & \wireoverride{n} & \wireoverride{n} & \wireoverride{n} & \wireoverride{n} & \wireoverride{n} & \wireoverride{n} & \wireoverride{n} & \wireoverride{n} & \wireoverride{n}
\end{quantikz}
\end{center}
\captionsetup{justification=raggedright, singlelinecheck=false}
\caption[]{Quantum circuit schematics of a procedure for disentangling subsequently all qubits of a state $ {{ | {\psi}^{(n+1)} \rangle }} $, corresponding to an inverse initialization procedure. 
The dashed boxes indicate the concatenation steps w.r.t. the procedure in Fig. {\hyperref[Fig:Shende_et_al_adapted_Disentangling_LSB]{\ref*{Fig:Shende_et_al_adapted_Disentangling_LSB}}}.}
\label{Fig:GAEI_inverse_circuit_via_multiplexors}
\end{figure*}
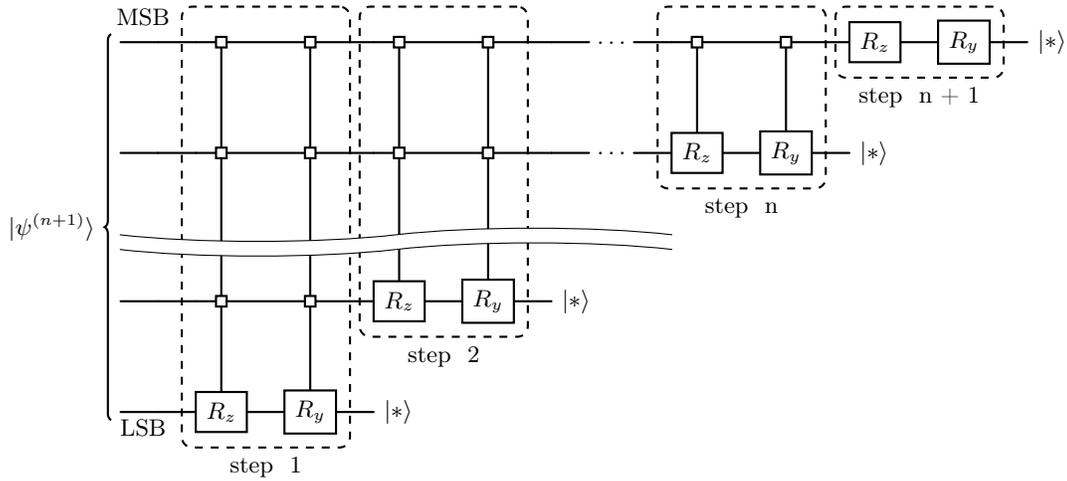

An inverse circuit for the preparation of a general $ (n + 1) $-qubit state from $ {| 0 \rangle}^{\otimes {(n+1)}} $ is then given by applying the gate procedure depicted in Fig. {\hyperref[Fig:Shende_et_al_adapted_Disentangling_LSB]{\ref*{Fig:Shende_et_al_adapted_Disentangling_LSB}}} subsequently $ n + 1 $ times, where the procedure of the concatenation step $ k \in \{ 1, 2, \dots, n + 1 \} $ acts on the $ n+1-(k-1) $ MSBs as shown in Fig. {\hyperref[Fig:GAEI_inverse_circuit_via_multiplexors]{\ref*{Fig:GAEI_inverse_circuit_via_multiplexors}}}. 
The case of one qubit, i.e. $ n = 0 $, for which the multiplexors in the structure of Fig. {\hyperref[Fig:Shende_et_al_adapted_Disentangling_LSB]{\ref*{Fig:Shende_et_al_adapted_Disentangling_LSB}}} have no select-qubits corresponds to Theorem 2 or eq. (9) of \cite{Shende_et_al_Synthesis_of_q_circuits_long}, which states that every 1-qubit state can be obtained from $ | 0 \rangle $ or $ | 1 \rangle $ by the application of an $ R_y $-gate followed by an $ R_z $-gate.

According to Theorem 8 of \cite{Shende_et_al_Synthesis_of_q_circuits_long}, an $ R_j $-gate-multiplexor with $ j \in \{ y, z \} $, where $ R_j $-gate operation acts on the LSB and the other $ n $ qubits are the select-qubits, can be decomposed in the form depicted on the right side of Fig. {\hyperref[Fig:Shende_et_al_adapted_Decomposition_Rj-multiplexor]{\ref*{Fig:Shende_et_al_adapted_Decomposition_Rj-multiplexor}}}, which is also valid for $ n = 1 $, where just $ {R_j} $-multiplexors with $ 0 $ select-qubits result on the right side of the circuit equivalency illustrated in Fig.  {\hyperref[Fig:Shende_et_al_adapted_Decomposition_Rj-multiplexor]{\ref*{Fig:Shende_et_al_adapted_Decomposition_Rj-multiplexor}}}, i.e. just $ {R_j} $-gates, since this case corresponds to Theorem 4 in \cite{Shende_et_al_Synthesis_of_q_circuits_long}.

It is to note that the order of the gates on the right side of the upper row in Fig. {\hyperref[Fig:Shende_et_al_adapted_Decomposition_Rj-multiplexor]{\ref*{Fig:Shende_et_al_adapted_Decomposition_Rj-multiplexor}}} can be reversed, yielding the equivalent structure in the second row. 
Via applying the decomposition of Fig. {\hyperref[Fig:Shende_et_al_adapted_Decomposition_Rj-multiplexor]{\ref*{Fig:Shende_et_al_adapted_Decomposition_Rj-multiplexor}}} recursively, an $ R_j $-multiplexor as it appears in Fig. {\hyperref[Fig:Shende_et_al_adapted_Disentangling_LSB]{\ref*{Fig:Shende_et_al_adapted_Disentangling_LSB}}}, can be broken down to $ R_j $- and CX-gates. 
Following \cite{Shende_et_al_Synthesis_of_q_circuits_long} further, this decomposition can in particular be done with canceling some resulting CX-gates by using the freedom of the gate ordering according to the upper and lower row in Fig. {\hyperref[Fig:Shende_et_al_adapted_Decomposition_Rj-multiplexor]{\ref*{Fig:Shende_et_al_adapted_Decomposition_Rj-multiplexor}}} because it can be exploited for the decomposition of two $ {R_j} $-multiplexors that are separated by a CX-gate between them that via inserting the structure of the upper row in Fig. {\hyperref[Fig:Shende_et_al_adapted_Decomposition_Rj-multiplexor]{\ref*{Fig:Shende_et_al_adapted_Decomposition_Rj-multiplexor}}} for the left $ R_j $-multiplexor and inserting the structure of the lower row in Fig. {\hyperref[Fig:Shende_et_al_adapted_Decomposition_Rj-multiplexor]{\ref*{Fig:Shende_et_al_adapted_Decomposition_Rj-multiplexor}}} for the $ R_j $-multiplexor on the right of the CX-gate, CX-gate structures as illustrated in Fig. {\hyperref[Fig:Shende_et_al_adapted_Decomposition_Rj-multiplexor_Cancelling_CX-gates]{\ref*{Fig:Shende_et_al_adapted_Decomposition_Rj-multiplexor_Cancelling_CX-gates}}} can be generated. 
In such a structure of two CX-gates of equal specification that are separated by a central CX-gate between them that has the same target qubit, the two flanking CX-gates can be canceled, which can be seen by checking that the action on the target qubit for all bitstrings that are possible w.r.t. a classical configuration of the register formed by the control qubits of the three CX-gates is equal for both circuits.

\begin{figure}[t] 
\begin{center}
\tikzset{invisible/.style={fill=none,draw=none,line width=0pt,inner xsep=0pt,inner ysep=0pt}}
\begin{align*}
\begin{quantikz}
 & \gategroup[1,steps=1, style={invisible}, label style={label position=above,anchor=mid,yshift=-0.3cm, xshift=-0.2cm}]{MSB} & |[operator]| &   \\
  & \qwbundle{ } &[-0.2cm]
  |[operator]| &  \\
 & \gategroup[1,steps=1, style={invisible}, label style={label position=below,anchor=mid,yshift=-0.05cm, xshift=-0.2cm}]{LSB} & \gate{{R_j}}\wire[u][2]{q}  &  
\end{quantikz}
 & ~ \widehat{=} ~  
\begin{quantikz}
 & \gategroup[1,steps=1, style={invisible}, label style={label position=above,anchor=mid,yshift=-0.3cm, xshift=-0.2cm}]{MSB} &  & \control{} & & \control{} &   \\
  & \qwbundle{ } &[-0.2cm]
  |[operator]| & & |[operator]| & &    \\
 & \gategroup[1,steps=1, style={invisible}, label style={label position=below,anchor=mid,yshift=-0.05cm, xshift=-0.2cm}]{LSB} & \gate{{R_j}}\wire[u][1]{q}  &  \targ{}\wire[u][2]{q} & \gate{{R_j}}\wire[u][1]{q} & \targ{}\wire[u][2]{q} & 
\end{quantikz} \\[0.3cm]
 & ~ \widehat{=} ~ 
\begin{quantikz}
 & \gategroup[1,steps=1, style={invisible}, label style={label position=above,anchor=mid,yshift=-0.3cm, xshift=-0.2cm}]{MSB} & \control{} & & \control{} & &   \\
  & \qwbundle{ } &[-0.2cm]
   & |[operator]| & & |[operator]| &   \\
 & \gategroup[1,steps=1, style={invisible}, label style={label position=below,anchor=mid,yshift=-0.05cm, xshift=-0.2cm}]{LSB} & \targ{}\wire[u][2]{q} &  \gate{{R_j}}\wire[u][1]{q} & \targ{}\wire[u][2]{q} & \gate{{R_j}}\wire[u][1]{q} &  
\end{quantikz}
\end{align*}
\end{center}
\captionsetup{justification=raggedright, singlelinecheck=false}
\caption[]{Quantum circuit schematics for a decomposition of an $ R_j $-multiplexor into $ R_j $-multiplexors with one select-qubit less. 
The decompositions on the right in the upper and lower row are equivalent. 
Figure adapted from \cite{Shende_et_al_Synthesis_of_q_circuits_long}.}
\label{Fig:Shende_et_al_adapted_Decomposition_Rj-multiplexor}
\end{figure}
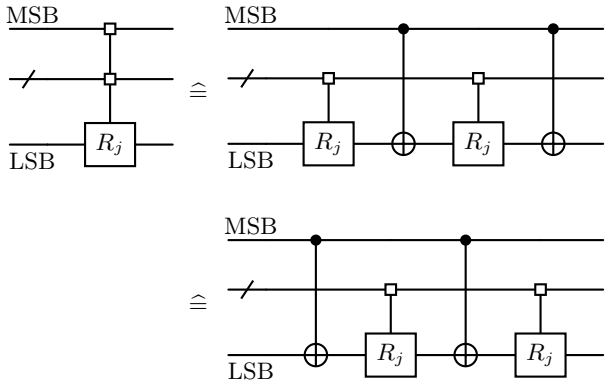

\begin{figure*}[t]
\sbox0{\begin{adjustbox}{width=1.0\textwidth}
\parbox{0.75\textwidth}{
\tikzset{ invisible/.style={fill=none,draw=none,line width=0pt,inner xsep=0pt,inner ysep=0pt}}
\begin{align*}
\begin{quantikz}
 & \gategroup[1,steps=1, style={invisible}, label style={label position=above,anchor=mid,yshift=-0.3cm, xshift=-0.2cm}]{MSB} &[-0.1cm] |[operator]| &[-0.1cm]   \\
\gategroup[1,steps=1, style={invisible}, label style={label position=above,anchor=mid,yshift=0.85cm, xshift=0.35cm}]{k = 0} 
  & & |[operator]| &   \\
  &  & |[operator]| &  \\
 & \gategroup[1,steps=1, style={invisible}, label style={label position=below,anchor=mid,yshift=-0.05cm, xshift=-0.2cm}]{LSB} & \gate{{R_j}}\wire[u][3]{q}  & 
\end{quantikz}
 & ~ \widehat{=} ~ 
\begin{quantikz}
 & \gategroup[1,steps=1, style={invisible}, label style={label position=above,anchor=mid,yshift=-0.3cm, xshift=-0.2cm}]{MSB} &[-0.1cm]  &[-0.1cm] \control{} &[-0.1cm] &[-0.1cm] \control{} &[-0.1cm]   \\
\gategroup[1,steps=1, style={invisible}, label style={label position=above,anchor=mid,yshift=0.85cm, xshift=0.35cm}]{k = 1} 
  &  & |[operator]| & & |[operator]| & &    \\
  &  & |[operator]| & & |[operator]| & &    \\
 & \gategroup[1,steps=1, style={invisible}, label style={label position=below,anchor=mid,yshift=-0.05cm, xshift=-0.2cm}]{LSB} & \gate{{R_j}}\wire[u][2]{q}  &  \targ{}\wire[u][3]{q} & \gate{{R_j}}\wire[u][2]{q} & \targ{}\wire[u][3]{q} & 
\end{quantikz} 
  ~ \widehat{=} ~ 
\begin{quantikz}
 & \gategroup[1,steps=1, style={invisible}, label style={label position=above,anchor=mid,yshift=-0.3cm, xshift=-0.2cm}]{MSB} &[-0.1cm] &[-0.1cm] &[-0.1cm] &[-0.1cm]  &[-0.1cm] \control{} &[-0.1cm] &[-0.1cm] &[-0.1cm] &[-0.1cm] &[-0.1cm] \control{} &[-0.1cm]   \\
 \gategroup[1,steps=1, style={invisible}, label style={label position=above,anchor=mid,yshift=0.85cm, xshift=0.35cm}]{k = 2} 
  &  & & \control{} & & \control{}\gategroup[3,steps=1,style={red, dashed, rounded corners, inner sep=2pt},background,label style={label position=below, anchor=north, yshift=-0.2cm}]{ }    & & \control{}\gategroup[3,steps=1,style={red, dashed, rounded corners, inner sep=2pt},background,label style={label position=below, anchor=north, yshift=-0.2cm}]{ } & &  \control{} & & &   \\
  &  & |[operator]| & & |[operator]| & & & & |[operator]| & & |[operator]| & &    \\
 & \gategroup[1,steps=1, style={invisible}, label style={label position=below,anchor=mid,yshift=-0.05cm, xshift=-0.2cm}]{LSB} & \gate{{R_j}}\wire[u][1]{q}  &  \targ{}\wire[u][2]{q} &  \gate{{R_j}}\wire[u][1]{q} &  \targ{}\wire[u][2]{q} & \targ{}\wire[u][3]{q} &   \targ{}\wire[u][2]{q} & \gate{{R_j}}\wire[u][1]{q} &  \targ{}\wire[u][2]{q} & \gate{{R_j}}\wire[u][1]{q} & \targ{}\wire[u][3]{q}  & 
\end{quantikz}
 \\[0.3cm]
 & ~ \widehat{=} ~ 
\begin{quantikz}
 & \gategroup[1,steps=1, style={invisible}, label style={label position=above,anchor=mid,yshift=-0.3cm, xshift=-0.2cm}]{MSB} &[-0.1cm] &[-0.1cm] &[-0.1cm] &[-0.1cm] &[-0.1cm] &[-0.1cm] &[-0.1cm] &[-0.1cm]  &[-0.1cm] &[-0.1cm] &[-0.1cm] \control{} &[-0.1cm] &[-0.1cm] &[-0.1cm] &[-0.1cm] &[-0.1cm] &[-0.1cm] &[-0.1cm] &[-0.1cm] &[-0.1cm] &[-0.1cm] &[-0.1cm] \control{} &[-0.1cm]  \\
 \gategroup[1,steps=1, style={invisible}, label style={label position=above,anchor=mid,yshift=0.85cm, xshift=0.35cm}]{k = 3} 
  &  & &  & & & \control{} &  & & & & \control{}\gategroup[3,steps=1,style={red, dashed, rounded corners, inner sep=2pt},background,label style={label position=below, anchor=north, yshift=-0.2cm}]{ } & & \control{}\gategroup[3,steps=1,style={red, dashed, rounded corners, inner sep=2pt},background,label style={label position=below, anchor=north, yshift=-0.2cm}]{ } & & & & & \control{} & & & & & &   \\
  &  &  & \control{} &  & \control{}\gategroup[2,steps=1,style={red, dotted, rounded corners, inner sep=2pt},background,label style={label position=below, anchor=north, yshift=-0.2cm}]{ } & &  \control{}\gategroup[2,steps=1,style={red, dotted, rounded corners, inner sep=2pt},background,label style={label position=below, anchor=north, yshift=-0.2cm}]{ } &  & \control{} &  &  & & & & \control{} & & \control{}\gategroup[2,steps=1,style={red, dotted, rounded corners, inner sep=2pt},background,label style={label position=below, anchor=north, yshift=-0.2cm}]{ } & & \control{}\gategroup[2,steps=1,style={red, dotted, rounded corners, inner sep=2pt},background,label style={label position=below, anchor=north, yshift=-0.2cm}]{ } & & \control{} & & &    \\
 & \gategroup[1,steps=1, style={invisible}, label style={label position=below,anchor=mid,yshift=-0.05cm, xshift=-0.2cm}]{LSB} & \gate{{R_j}} & \targ{}\wire[u][1]{q}  & \gate{{R_j}} & \targ{}\wire[u][1]{q} & \targ{}\wire[u][2]{q} & \targ{}\wire[u][1]{q} & \gate{{R_j}} & \targ{}\wire[u][1]{q} & \gate{{R_j}} & \targ{}\wire[u][2]{q} & \targ{}\wire[u][3]{q} &   \targ{}\wire[u][2]{q} & \gate{{R_j}} & \targ{}\wire[u][1]{q} & \gate{{R_j}} & \targ{}\wire[u][1]{q} &  \targ{}\wire[u][2]{q} & \targ{}\wire[u][1]{q} & \gate{{R_j}} & \targ{}\wire[u][1]{q} & \gate{{R_j}}  & \targ{}\wire[u][3]{q}  & 
\end{quantikz}
\end{align*}
}
\end{adjustbox}}\usebox0
\captionsetup{justification=raggedright, singlelinecheck=false}
\caption[]{Illustration of the recursive decomposition of a $ R_j $-multiplexor, defined in Fig. {\hyperref[Fig:Shende_et_al_adapted_Decomposition_Rj-multiplexor]{\ref*{Fig:Shende_et_al_adapted_Decomposition_Rj-multiplexor}}}, for $ n = 3 $ select-qubits, where $ k $ is the recursion step. 
The red surrounded CX-gates in the resulting structures of three CX-gates can be canceled. 
Figure adapted from \cite{Shende_et_al_Synthesis_of_q_circuits_long}.}
\label{Fig:Shende_et_al_adapted_Decomposition_Rj-multiplexor_Cancelling_CX-gates}
\end{figure*}
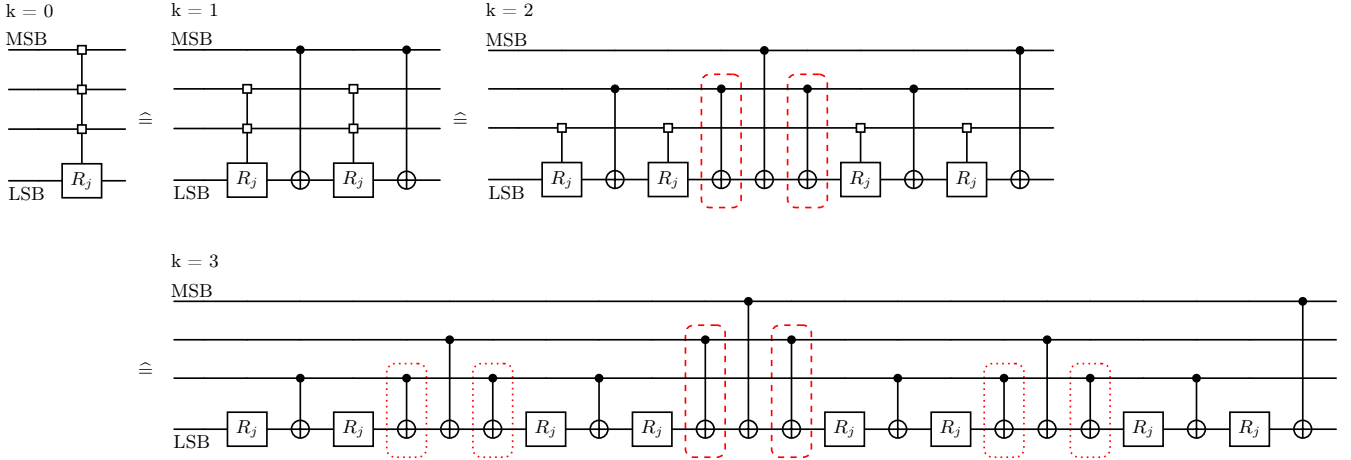
\begin{figure*}[t]
\begin{center}
\tikzset{invisible/.style={fill=none,draw=none,line width=0pt,inner xsep=0pt,inner ysep=0pt}}
\begin{align*}
\begin{quantikz}
 & \gategroup[1,steps=1, style={invisible}, label style={label position=above,anchor=mid,yshift=-0.3cm, xshift=-0.2cm}]{MSB} & |[operator]| & |[operator]| &   \\
  & \qwbundle{ } &[-0.2cm]
  |[operator]| & |[operator]| &  \\
 & \gategroup[1,steps=1, style={invisible}, label style={label position=below,anchor=mid,yshift=-0.05cm, xshift=-0.2cm}]{LSB} & \gate{{R_z}}\wire[u][2]{q}  & \gate{{R_y}}\wire[u][2]{q}  &  
\end{quantikz}
 & ~ \widehat{=} ~  
\begin{quantikz}
 & \gategroup[1,steps=1, style={invisible}, label style={label position=above,anchor=mid,yshift=-0.3cm, xshift=-0.2cm}]{MSB} &  & \control{} & & \control{}\gategroup[3,steps=2,style={red, dashed, rounded corners, inner sep=2pt},background,label style={label position=below, anchor=north, yshift=-0.2cm}]{ } & \control{} & & \control{} & &   \\
  & \qwbundle{ } &[-0.2cm]
  |[operator]| & & |[operator]| & & &  |[operator]| & & |[operator]| &    \\
 & \gategroup[1,steps=1, style={invisible}, label style={label position=below,anchor=mid,yshift=-0.05cm, xshift=-0.2cm}]{LSB} & \gate{{R_z}}\wire[u][1]{q}  &  \targ{}\wire[u][2]{q} & \gate{{R_z}}\wire[u][1]{q} & \targ{}\wire[u][2]{q} & \targ{}\wire[u][2]{q} & \gate{{R_y}}\wire[u][1]{q} &  \targ{}\wire[u][2]{q} & \gate{{R_y}}\wire[u][1]{q}  &
\end{quantikz} 
\end{align*}
\end{center}
\captionsetup{justification=raggedright, singlelinecheck=false}
\caption[]{Illustration of the application of the decomposition in Fig. {\hyperref[Fig:Shende_et_al_adapted_Decomposition_Rj-multiplexor]{\ref*{Fig:Shende_et_al_adapted_Decomposition_Rj-multiplexor}}} for the disentangling structure in Fig. {\hyperref[Fig:Shende_et_al_adapted_Disentangling_LSB]{\ref*{Fig:Shende_et_al_adapted_Disentangling_LSB}}}, where the red surrounded CX-gates in the middle can be canceled.}
\label{Fig:Decomposition_LSB-Disentangling_Cancelling_CX-gates_in_middle}
\end{figure*}
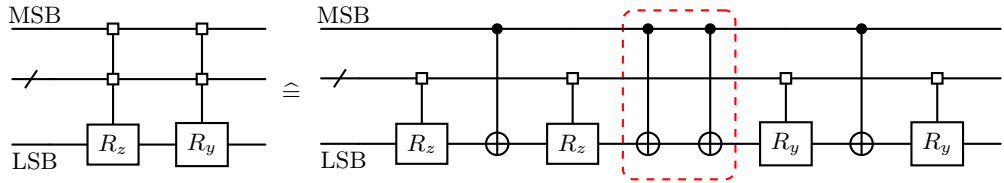

Accordingly, in the decomposition of an $ R_j $-multiplexor w.r.t. $ n + 1 $ qubits, where the $ n $ MSBs are the select-qubits, the decomposition level $ k $, given by performing $ k $ times the step of applying the recursion expression of Fig. {\hyperref[Fig:Shende_et_al_adapted_Decomposition_Rj-multiplexor]{\ref*{Fig:Shende_et_al_adapted_Decomposition_Rj-multiplexor}}} to all present $ R_j $-multiplexors, has $ 2^k $ $ R_j $-multiplexors which involve the $ n+1-k $ LSBs, since the number of multiplexors doubles in each step until after $ n $ steps finally $ 2^n $ $ R_j $-1-qubit gates are obtained. 
In the step $ k $, the same amount of CX-gates as $ R_j $-multiplexors results plus those CX-gates that were added in the steps before, given by $ 0 + 2 + 4 + \dots + {2^{k-1}} $, which can be expressed via the finite geometric series
\begin{align*}
 \sum_{i=0}^{n} {z^{i}} &= \frac{1 - {z^{n + 1}} }{1-z} , \quad z \neq 1
\end{align*}
as
\begin{align*}
 \sum_{i=1}^{k-1} {2^{i}} &= \sum_{i=0}^{k-1} {2^{i}}  - 1 = \frac{ 1 - {2^{k}} }{ 1-2 } - 1 = {2^k} - 2, 
\end{align*}
so that a number of
\begin{align*}
 {2^k} + {2^k} - 2 &= {2^{k+1}} - 2
\end{align*}
CX-gates results, from which the number of CX-gates that cancel if the previously described procedure is applied has to be subtracted.

This amount to be subtracted is given by $ {\sum_{i=2}^{k}} 2 \cdot {2^{i-2}} $ as the sum of all canceled CX-gates until step $ k $ because $ 2 $ CX-gates cancel per occurrence of a structure as indicated in Fig. {\hyperref[Fig:Shende_et_al_adapted_Decomposition_Rj-multiplexor_Cancelling_CX-gates]{\ref*{Fig:Shende_et_al_adapted_Decomposition_Rj-multiplexor_Cancelling_CX-gates}}} and $ 2^{k-2} $ such structures are produced in step $ k $ for $ k > 1 $, whereas $ 0 $ structures result for the recursion steps $ k = 0 $ and $ k = 1 $. 
For this term follows
\begin{align*}
{\sum_{i=2}^{k}} {2^{i-1}} &= {\sum_{i=1}^{k-1}} {2^{i}} =   {\sum_{i=0}^{k-1}} {2^{i}}  -1  =  {\frac{ 1 - 2^{k} }{ 1-2 }} - 1  = 2^{k} - 2 , 
\end{align*}
which is also valid for $ k = 1 $. 
Thus, taking this term into account for the total number of recursion steps $ n $ for $ n > 0 $, yields a total number of
\begin{align*}
 2 \cdot {2^n} - 2 - ( {2^n} - 2 ) &= {2^n}
\end{align*}
CX-gates, whereas $ 0 $ CX-gates are necessary for $ n = 0 $.

After the possible cancellations of CX-gates, the resulting decomposition of an $ R_j $-multiplexor consists of $ 2^n $ $ R_j $-1-qubit gates applied to the LSB, where each is followed by one CX-gate with the LSB as the target qubit according to the pattern indicated in the lower row of Fig. {\hyperref[Fig:Shende_et_al_adapted_Decomposition_Rj-multiplexor_Cancelling_CX-gates]{\ref*{Fig:Shende_et_al_adapted_Decomposition_Rj-multiplexor_Cancelling_CX-gates}}}. 
Inserting the described decomposition for an $ R_j $-multiplexor into the quantum circuit for disentangling the LSB-qubit according to Fig. {\hyperref[Fig:Shende_et_al_adapted_Disentangling_LSB]{\ref*{Fig:Shende_et_al_adapted_Disentangling_LSB}}}, in which there are two such multiplexors, yields then a count of 
\begin{align}
 2 \cdot {2^n}
 \label{eq:GAEI_Disentangling_LSB_required_Rj-gates}
\end{align}
$ R_j $-1-qubit gates.

For the count of CX-gates, it is however to note that it can be exploited again that the order of gates in the decomposition in Fig. {\hyperref[Fig:Shende_et_al_adapted_Decomposition_Rj-multiplexor_Cancelling_CX-gates]{\ref*{Fig:Shende_et_al_adapted_Decomposition_Rj-multiplexor_Cancelling_CX-gates}}} can be reversed, since inserting the decomposition with the gate ordering as depicted in Fig. {\hyperref[Fig:Shende_et_al_adapted_Decomposition_Rj-multiplexor_Cancelling_CX-gates]{\ref*{Fig:Shende_et_al_adapted_Decomposition_Rj-multiplexor_Cancelling_CX-gates}}} for the $ R_z $-multiplexor and the decomposition with the reversed gate ordering for the $ R_y $-multiplexor causes that the CX-gate at the right end in the decomposition according to Fig. {\hyperref[Fig:Shende_et_al_adapted_Decomposition_Rj-multiplexor_Cancelling_CX-gates]{\ref*{Fig:Shende_et_al_adapted_Decomposition_Rj-multiplexor_Cancelling_CX-gates}}} faces another CX-gate of equal specification as illustrated in Fig. {\hyperref[Fig:Decomposition_LSB-Disentangling_Cancelling_CX-gates_in_middle]{\ref*{Fig:Decomposition_LSB-Disentangling_Cancelling_CX-gates_in_middle}}}, so that these two CX-gates resulting in the middle can be canceled, since the CX-gate is its own inverse. 
This aspect results in a count of
\begin{align}
 2 \cdot {2^n} - 2
 \label{eq:GAEI_Disentangling_LSB_required_CX-gates}
\end{align}
CX-gates in the decomposition of the circuit for disentangling the LSB of an $ (n + 1) $-qubit state according to Fig. {\hyperref[Fig:Shende_et_al_adapted_Disentangling_LSB]{\ref*{Fig:Shende_et_al_adapted_Disentangling_LSB}}},  where this statement includes now also the case $ n = 0 $.

\begin{figure*}[t]
\begin{adjustbox}{width=1.0\textwidth}
\tikzset{ invisible/.style={fill=none,draw=none,line width=0pt,inner xsep=0pt,inner ysep=0pt} }
\begin{quantikz}[wire types={q,q,q,n}]
 \lstick{\ket{*}} & \gategroup[1,steps=1, style={invisible}, label style={label position=above,anchor=mid,yshift=-0.3cm, xshift=-0.2cm}]{MSB} &[-0.1cm] \gate{{R_y}} &[-0.1cm] \gate{{R_z}}\gategroup[2,steps=1,style={blue, dashed, rounded corners, inner sep=2pt},background,label style={label position=below, anchor=north, yshift=-0.2cm}]{ } &[-0.1cm] \control{} &[-0.1cm] 
 &[-0.1cm] &[-0.1cm] \control{} &[-0.1cm] &[-0.1cm] &[-0.1cm] &[-0.1cm] \control{} &[-0.1cm] &[-0.1cm] &[-0.1cm] &[-0.1cm] &[-0.1cm] &[-0.1cm] &[-0.1cm] \control{} &[-0.1cm] &[-0.1cm] &[-0.1cm] &[-0.1cm]    \ \ldots\  \\
 \lstick{\ket{*}} & 
   &  & \gate{{R_y}} & \targ{}\wire[u][1]{q} & \gate{{R_y}} &  \gate{{R_z}} & \targ{}\wire[u][1]{q} & \gate{{R_z}}\gategroup[2,steps=1,style={blue, dashed, rounded corners, inner sep=2pt},background,label style={label position=below, anchor=north, yshift=-0.2cm}]{ } & \control{} & & & & \control{} & & & \control{} & & & & \control{} & & \ \ldots\   \\
 \lstick{\ket{*}} & 
  &   &   &  &  &  &  & \gate{{R_y}} & \targ{}\wire[u][1]{q} & \gate{{R_y}} & \targ{}\wire[u][2]{q} & \gate{{R_y}} & \targ{}\wire[u][1]{q} & \gate{{R_y}} & \gate{{R_z}} & \targ{}\wire[u][1]{q} & \gate{{R_z}} & \targ{}\wire[u][2]{q} & \gate{{R_z}} & \targ{}\wire[u][1]{q} & \gate{{R_z}} & \ \ldots\ \\
   \ \vdots\  & & & & & & & & & &  
\end{quantikz}
\end{adjustbox}
\captionsetup{justification=raggedright, singlelinecheck=false}
\caption[]{Beginning of the gate pattern for the general initialization procedure for amplitude encoding. 
The blue dashed boxes encomprise two 1-qubit gates from two different blocks for entangling a qubit that can be executed in parallel. 
Equivalently, the $ R_y $-gates at the beginning of each block for entangling the respectively considered LSB-qubit could be all put in parallel at the beginning of the circuit.}
\label{Fig:GAEI_beginning_gate_pattern}
\end{figure*}
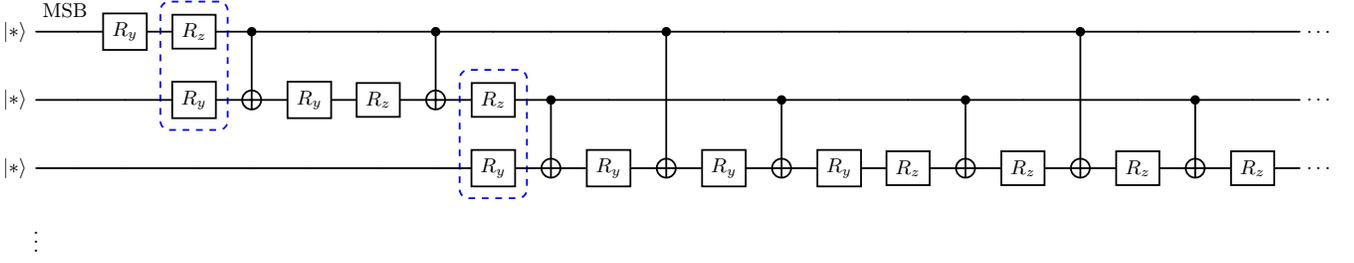

\begin{figure}[t]
\begin{center}
\includegraphics[width=\linewidth]{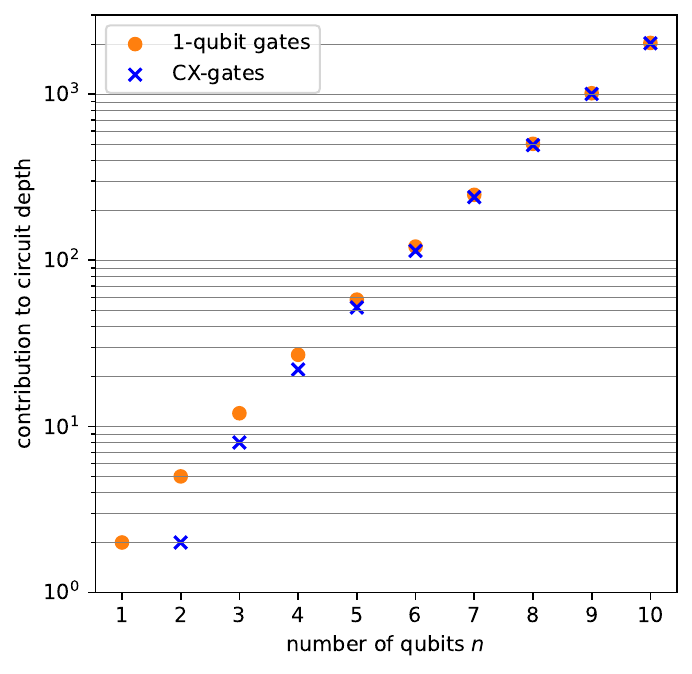}
\end{center}
\captionsetup{justification=raggedright, singlelinecheck=false}
\caption[]{Circuit depth contributions due to the $ R_j $-rotation gates as the native 1-qubit gates and due to the CX-gates for the general initialization procedure for amplitude encoding.}
\label{Fig:Plots_GAE_circuit_depth_contributions}
\end{figure}

Thus, the required numbers of $ {R_j} $- and CX-gates for the initialization procedure according to Fig.  {\hyperref[Fig:GAEI_inverse_circuit_via_multiplexors]{\ref*{Fig:GAEI_inverse_circuit_via_multiplexors}}} are given by the sum of the results for the respective gate counts ({\hyperref[eq:GAEI_Disentangling_LSB_required_Rj-gates]{\ref*{eq:GAEI_Disentangling_LSB_required_Rj-gates}}}) and ({\hyperref[eq:GAEI_Disentangling_LSB_required_CX-gates]{\ref*{eq:GAEI_Disentangling_LSB_required_CX-gates}}}) in the circuit for disentangling the LSB-qubit w.r.t. a summation index from $ 0 $ to $ n $: 
For the number of $ R_j $-gates, it follows
\begin{align}
 2 {\sum_{i=0}^{n}} {2^i} &= 2 {\frac{1-{2^{n+1}}}{ 1-2}} = -2 \cdot {(1-{2^{n+1}})} \nonumber \\
 &= 2^{n+2} - 2
 \label{eq:GAEI_Number_1-qubit_rotation_gates_for_nplus1}
\end{align}
for the initialization of an $ (n+1) $-qubit state according to amplitude encoding, where one half are $ R_y $-gates and the other half are $ R_z $-gates, i.e. a number of $ {2^{n+1}} - 1 $ results for each type, respectively. 
For the number of required CX-gates, it results
\begin{align}
 {\sum_{i=0}^{n}} (2 \cdot {2^{i}} - 2) &= 2 {\sum_{i=0}^{n}} {2^{i}} - 2 \cdot (n + 1) \nonumber \\
 &= 2 \cdot \frac{1 - {2^{n+1}}}{1-2} - 2 \cdot (n + 1) \nonumber \\
  &= {2^{n + 2}} - 2 \cdot (n + 2)
  , 
 \label{eq:GAEI_Number_CX-gates_for_nplus1}
\end{align}
which is also the circuit depth contribution of them, because the CX-gates separate the $ R_j $-gates since their target qubit is the qubit on which the $ R_j $-gates of the respective block for disentangling the correspondingly considered LSB-qubit act. 
With regard to the circuit depth, it is to note for $ n > 0 $ that the $ R_z $-gate and the $ R_y $-gate that mark the end and beginning of the explained decomposition of a block for disentangling the LSB-qubit, respectively, can be put into one gate layer at a transition from one such disentangling block to the next according to the steps indicated in Fig.  {\hyperref[Fig:GAEI_inverse_circuit_via_multiplexors]{\ref*{Fig:GAEI_inverse_circuit_via_multiplexors}}}, since these two $ R_j $-gates act on different qubits. 
This aspect is illustrated in Fig. {\hyperref[Fig:GAEI_beginning_gate_pattern]{\ref*{Fig:GAEI_beginning_gate_pattern}}} and yields a reduction of the circuit depth contribution from the $ R_j $-gates of $ n $ w.r.t. the expression ({\hyperref[eq:GAEI_Number_1-qubit_rotation_gates_for_nplus1]{\ref*{eq:GAEI_Number_1-qubit_rotation_gates_for_nplus1}}}) for the initialization of an $ (n + 1) $-qubit state, i.e. a circuit depth contribution of
\begin{align}
  2^{n+2} - ( n + 2 ) .
 \label{eq:GAEI_Depth_1-qubit_rotation_gates_for_nplus1}
\end{align}

According to this consideration, the circuit depth of the routine for initializing a state according to {{general amplitude encoding}} scales exponentially with the number of qubits. 
The contributions of the $ R_j $-1-qubit rotation gates and CX-gates are shown in Fig. {\hyperref[Fig:Plots_GAE_circuit_depth_contributions]{\ref*{Fig:Plots_GAE_circuit_depth_contributions}}}. 
The specific runtime of the initialization procedure results then via the formula
\begin{align}
 & ({2^{n+1}} - (n + 1)) \nonumber \\
 & ~ \cdot [\text{1-qubit rotation gate execution time}] \nonumber \\ 
 +~ & ({2^{n+1}} - 2(n + 1)) \nonumber \\
 & ~ \cdot [\text{CX-gate execution time}]
 ,
\label{eq:GAEI_formula_time}
\end{align}
which is further discussed in subsection {\hyperref[subsec:Study_Initialization]{\ref*{subsec:Study_Initialization}}}.

In contrast, for the other two presented encoding formats, the numbers of layers of native gates are independent of the number of qubits and no native 2-qubit gates are needed, since the qubits are independent of each other:

\subsubsection{1-Qubit Amplitude Encoding}

For $ n = 1 $ qubit, the expressions ({\hyperref[eq:GAEI_Number_Ry-gates_for_n]{\ref*{eq:GAEI_Number_Ry-gates_for_n}}}) and ({\hyperref[eq:GAEI_Number_Rz-gates_for_n]{\ref*{eq:GAEI_Number_Rz-gates_for_n}}}) for the needed $ R_y $- and $ R_z $-gates yield $ 1 $, respectively and expression ({\hyperref[eq:GAEI_Number_CX-gates_for_n]{\ref*{eq:GAEI_Number_CX-gates_for_n}}}) for the needed CX-gates yields $ 0 $. 
Since all qubits are independent for this encoding scheme, they can be initialized in parallel, which is why the initialization routine for 1-qubit amplitude encoding consists just of one layer of $ R_y $-gates, followed by one layer of $ R_z $-gates.

\subsubsection{Bitstring Encoding}

Bitstring encoding constrains the situation of 1-qubit amplitude encoding further to 1-qubit states that are either $ | 0 \rangle $ or $ | 1 \rangle $, where these two 1-qubit basis states result from each other via the application of an X-gate. 
Since the X-gate corresponds to the specific case of the $ R_x $-gate ({\hyperref[eq:Rx_matrix_form]{\ref*{eq:Rx_matrix_form}}}) for $ {\theta} = {\pi} $, a state in this encoding format can be initialized with only one layer of $ R_x $-gates.

\subsection{Read-out}
\label{subsec:Read-out}

'Read-out' refers to the determining of the numbers represented by a state of a qubit register according to a specific encoding via sufficiently many measurements to achieve a specific accuracy for the determined basis state probabilities. 
Since a superposition state is set by the measurement to the measurement outcome, which is one of the basis states, corresponding to the classical configurations of the computation system, multiple measurements of the system state prepared by the quantum algorithm require in general a corresponding number of executions of the quantum algorithm, also called {\emph{runs}} or {\emph{shots}}. 
If enough qubits with an appropriate connectivity are available, multiple runs of a quantum algorithm could be executed in parallel, but if not, the runtime of the computation problem is also proportional to the necessary number of runs. 
For the efforts for the read-out, this necessary number of runs to infer solution values that are encoded w.r.t. the three previously explained formats is considered in this subsection. 
In it, formulas for upper bounds of the number of runs that is required for a desired accuracy are derived, starting with considering a single qubit that is in a linear combination of the basis states $ \left| 0 \right\rangle $ and $ \left| 1 \right\rangle $ according to (\hyperref[eq:general_state_1_qubit]{\ref*{eq:general_state_1_qubit}}):

\subsubsection{1-Qubit Amplitude Encoding}

For this case, 2 values can be encoded via the amplitudes $ \left\langle i \middle| {\psi}_{\text{qubit}} \right\rangle = { {{{a}}_i} } $
that constitute the probabilities $ Pr{(i)} =  \left\langle {\psi}_{\text{qubit}} \right|  \left( \left| i \right\rangle \left\langle i \right|   \right) \left| {\psi}_{\text{qubit}} \right\rangle =  { \left| \left\langle i \middle| {\psi}_{\text{qubit}} \right\rangle \right|}^2 = {{{|{a_i}|}^2}} $ for finding the classical system state $ \left| {i} \right\rangle $ in a measurement of $ \left| {\psi}_{\text{qubit}} \right\rangle $, respectively. 
So, the amplitudes $ \left\langle i \middle| {\psi}_{\text{qubit}} \right\rangle $ are accessible at least via these probabilities, which can be estimated as the ensemble average
\begin{align}
{\mathcal{P}}(i)
 := \frac{ N_i }{N} 
 \label{eq:inferred_probability}
\end{align}
from the number of times $ N_i $ that $ \left| i \right\rangle $ was detected in $ N $ measurements of the prepared state $ \left| {\psi}_{\text{qubit}} \right\rangle $. 
Since the measurement outcome in this case of one qubit is a binary output according to the occurrence of the state $ \left| 1 \right\rangle $ with success probability $ Pr{(1)} $ and $ Pr{(0)} = 1 - Pr{(1)} $ as the probability of no success, this situation is a Bernoulli random experiment.

It is equivalent to the estimation of the expectation value of the eigenvalue of the observable given by the Pauli-$ Z $-matrix $ {\sigma}_{Z} = \left( \begin{smallmatrix} 1 & 0 \\ 0 & -1 \end{smallmatrix} \right) $ or the projector $ {\frac{1}{2}} ( -{{\sigma}_{Z}} + {\bf{1}} ) =  \left| 1 \right\rangle \left\langle 1 \right| $ (cf. expression for $ Pr{(i)} $ above) with the identity matrix $ {\bf{1}} = \left( \begin{smallmatrix} 1 & 0 \\ 0 & 1 \end{smallmatrix} \right) $ since the basis states $ \left| i \right\rangle $ are the eigenstates of these diagonal operators and uniquely associated to one of the two eigenvalues.

For this situation, the probability that the difference $ | {\mathcal{P}}(1) - Pr{(1)} | $ of the success probability estimated from the measurements and the true success probability is equal or greater than an additive error $ \epsilon \geq 0 $ (e.g. $ \epsilon = 0.1 $ for the probability that the inferred probability $ {\mathcal{P}}(1) $ lies in a symmetrical interval of extension 20 \% around the true probability $ Pr{(1)} $) is given according to \cite{Veltheim_Keski-Vakkuri_Optimizing_measurements} by the Hoeffding inequality (cf. Section S2 from the supplemental material \cite{Veltheim_Keski-Vakkuri_Optimizing_measurements_Sup_Mat} for \cite{Veltheim_Keski-Vakkuri_Optimizing_measurements}):
\begin{align}
Pr( ~ | {\mathcal{P}}(1) - Pr{(1)} | \geq \epsilon ~ ) \leq 2 \cdot e^{ -2N{{\epsilon}^2} }
\label{eq:Hoeffding_ineq_1_qubit}
\end{align}
This probability can be tuned by the number of runs of the random experiment $ N $. 
For the right side of eq. ({\hyperref[eq:Hoeffding_ineq_1_qubit]{\ref*{eq:Hoeffding_ineq_1_qubit}}}), a parameter $ \delta \in [ 0, 1 ] $ can be set as a chosen bound for the probability that a statement $ | {\mathcal{P}}(1) - Pr{(1)} | \geq \epsilon $ is true, which determines the so-called confidence interval (E.g., the choice $ \delta = 0.01 $ means that the probability that $ {\mathcal{P}}(1) $ and $ {Pr{(1)}} $ deviate more than the error bound $ \epsilon $ is not greater than 1 \%.). 
This bound of $ \delta $ can be fulfilled by a sufficient large number of shots $N$, which can be determined by setting $ \delta $ equal to the right side of eq. ({\hyperref[eq:Hoeffding_ineq_1_qubit]{\ref*{eq:Hoeffding_ineq_1_qubit}}}) and re-arranging for $ N $:
\begin{align}
Pr( ~ | {\mathcal{P}}(1) - Pr{(1)} | \geq \epsilon ~ ) & \leq 2 \cdot e^{ -2N{{\epsilon}^2} } 
 \overset{!}{\leq} 
 \delta \nonumber \\
 \Leftrightarrow \quad N & \geq {\frac{1}{2 {{\epsilon}^2}}} \ln{ \left( \frac{2}{\delta}  \right) }
 \label{eq:Number_of_runs_1_qubit}
\end{align}
Since it is known for {{1-qubit amplitude encoding}} that the measurement outcome of one qubit is independent from the measurement outcome of another qubit, ({\hyperref[eq:Number_of_runs_1_qubit]{\ref*{eq:Number_of_runs_1_qubit}}}) gives an upper bound for the necessary number of runs for this encoding.

\subsubsection{General Amplitude Encoding}

Measuring a general superposition state for a register of $ n $ qubits w.r.t. the computational basis yields the basis state $ \left| i \right\rangle $ according to eq.  ({\hyperref[eq:Hoeffding_ineq_1_qubit]{\ref*{eq:Hoeffding_ineq_1_qubit}}}) indicated by the number $ i \in \{ 0, 1, \dots, {2^n} - 1 \} $ with the probability $ Pr(i) = {{| \left\langle i \middle| {\psi}^{(n)} \right\rangle |}^2} = {{|{a_i}|}^2} $, which generalizes the situation according to a binomial distribution for $ n = 1 $ considered before to a random experiment according to a multinomial distribution. 
However, inferring the probabilities according to the expression ({\hyperref[eq:inferred_probability]{\ref*{eq:inferred_probability}}}) from $ N $ runs can be regarded as a Bernoulli experiment for the estimation of each of the probabilities $ Pr(i), i \in \{ 1, \dots, {2^n} - 1 \} $. 
It is to note that this situation can be regarded as only $ {2^n} - 1 $ Bernoulli experiments conducted simultaneously (instead of $ {2^n} $) since the estimate of $ Pr(i=0) $ is fixed thereby as $ {\mathcal{P}}(0) = 1 - \sum_{i=1}^{{2^n} - 1} {\mathcal{P}}(i) $.

However, since these Bernoulli experiments are conducted in parallel, this situation should be a {\emph{multiple comparisions problem}}, which means in general that inferring more properties of the underlying probability distribution is accompanied with a higher probability that at least one of the inferred properties is wrong, e.g. not in a certain confidence interval. 
For the considered situation, in which the estimated $ Pr(i) $ are properties of the underlying multinomial distribution, one can consider e.g. the situation of a register in which all qubits are initialized in the $ \left| 0 \right\rangle $-state and H-gates are applied to all of them, for which $ N = 2 $ measurements, i.e. shots are performed: 
For the prepared state, all probabilities $ Pr(i) $ are equal to $  \frac{1}{ {2}^{n} }  $. 
For one qubit, the results of two measurements could already coincide with the correct probability distribution whereas for more than one qubit, the equal probabilities for all measurement outcomes cannot be realized by two measurements due to insufficient number of shots. 
Specifically, for $ n = 2 $ qubits, there are the four equally probable measurement outcomes '$ \left| 0 \right\rangle $ measured in both runs', '$ \left| 1 \right\rangle $ measured in both runs', '$ \left| 0 \right\rangle $ measured in 1st run and $ \left| 1 \right\rangle $ measured in 2nd run' and '$ \left| 1 \right\rangle $ measured in 1st run and $ \left| 0 \right\rangle $ measured in 2nd run', i.e. a chance of 50 \% that the probability distribution is resolved correctly. 
Accordingly, plugging in $ \epsilon = 0.5 $ as the distance between the true success probability $ Pr(i) = 0.5 $ and the possible incorrect estimations $ {\mathcal{P}}(i) = 0 $ and $ {\mathcal{P}}(i) = 1 $ into eq. ({\hyperref[eq:Hoeffding_ineq_1_qubit]{\ref*{eq:Hoeffding_ineq_1_qubit}}}) yields $ Pr( ~ | {\mathcal{P}}(i) - Pr{(i)} | \geq \epsilon ~ ) \leq  0.736 $ for $ N = 2 $ runs but for $ N = 3 $ already $ Pr( ~ | {\mathcal{P}}(i) - Pr{(i)} | \geq \epsilon ~ ) \leq  0.446 $, which is definitely lower than 50 \%.

Conversely, it is to note that number of runs to resolve a probability distribution with a fixed accuracy should not scale linearly with the number of basis states: 
For the example of equally distributed probabilities w.r.t. the $ 2^n $ basis states, $ 2^n $ measurements can resolve the correct statistics, however, the probability that this case of finding a different basis state in each of the $ 2^n $ measurements occurs should be given by the consideration that in every run, there is one less basis state that would be a valid pick for yielding the considered measurement outcome, so that the probability for it should be (cf. also the corresponding expression via the multinomial coefficient as stated in appendix {\hyperref[App_sec:Probability_Fulfilling_error_bound_for_reference_situation]{\ref*{App_sec:Probability_Fulfilling_error_bound_for_reference_situation}}})
\begin{align*}
 & \quad ~ 1 \cdot \Bigl( 1 - {\frac{1}{2^n}} \Bigr) \cdot \Bigl( 1 - {\frac{2}{2^n}} \Bigr) \cdot \dots \cdot \Bigl( 1 - {\frac{{2^n} - 1}{2^n}} \Bigr) \\
 &=  {\frac{{2^n}}{2^n}} \cdot {\frac{{2^n} - 1}{2^n}} \cdot {\frac{{2^n} - 2}{2^n}} \cdot \dots 
 \cdot {\frac{1}{{2^n}}} \\
  &= {\frac{ ({2^n})! }{ {({2^n})}^{{2^n}} }} ~ \overset{ {2^n} \rightarrow \infty }{ \longrightarrow } ~ 0,
\end{align*}
which shows that the accuracy is in general not maintained for linearly increasing the number of runs.

However, an upper bound for the required number of runs should be made via the {\emph{union bound}}, which states that the probability that event $ A $ or event $ B $ happens, denoted here with the logical 'or' $ \lor $ as $ Pr( {A} \lor {B} ) $, is bounded by the sum of the probabilities for the individual events, i.e. $ Pr( {A} \lor {B} ) \leq Pr( {A} ) + Pr( {B} ) $. 
Hence, the probability that at least one of the inferred probabilities $ {\mathcal{P}}(i) $ is not within a respectively set uncertainty interval can be estimated accordingly. 
Formulated for the choice $ Pr(i) $ with $ i \in \{ 1, 2, \dots, {2^n} - 1 \} $ as the inferred probabilities and equal uncertainty intervals for all, given by $ \epsilon $, it follows with the use of  ({\hyperref[eq:Hoeffding_ineq_1_qubit]{\ref*{eq:Hoeffding_ineq_1_qubit}}}) for the individual $ Pr( | {\mathcal{P}}(i) - Pr{(i)} | \geq \epsilon ) $:
\begin{align}
 Pr \Bigl( ~ \overset{ {2^n} - 1 }{ \underset{i=1}{\lor} }  | {\mathcal{P}}(i) - Pr{(i)} | \geq \epsilon ~ \Bigr) 
 & \leq {\underbrace{ \sum_{ i = 1 }^{ {2^n} - 1 } }_{= {2^n} - 1}} 2 \cdot e^{ -2N{{\epsilon}^2} } 
 \label{eq:Hoeffding_ineq_union bound_multiple_qubits}
\end{align}
Analogously to the step from ({\hyperref[eq:Hoeffding_ineq_1_qubit]{\ref*{eq:Hoeffding_ineq_1_qubit}}}) to ({\hyperref[eq:Number_of_runs_1_qubit]{\ref*{eq:Number_of_runs_1_qubit}}}), it follows further:
\begin{align}
 N & \geq {\frac{1}{2 {{\epsilon}^2}}} \ln{ \left( \frac{2 {({2^n} - 1)} }{\delta}  \right) }
 \label{eq:Number_of_runs_multiple_qubits_absolute_error}
\end{align}

\begin{figure*}[t]
\begin{minipage}[c]{0.485\textwidth}
\centering
\begin{tikzpicture}
\draw (0,0) node[inner sep=0]{\includegraphics[height=0.95\linewidth]{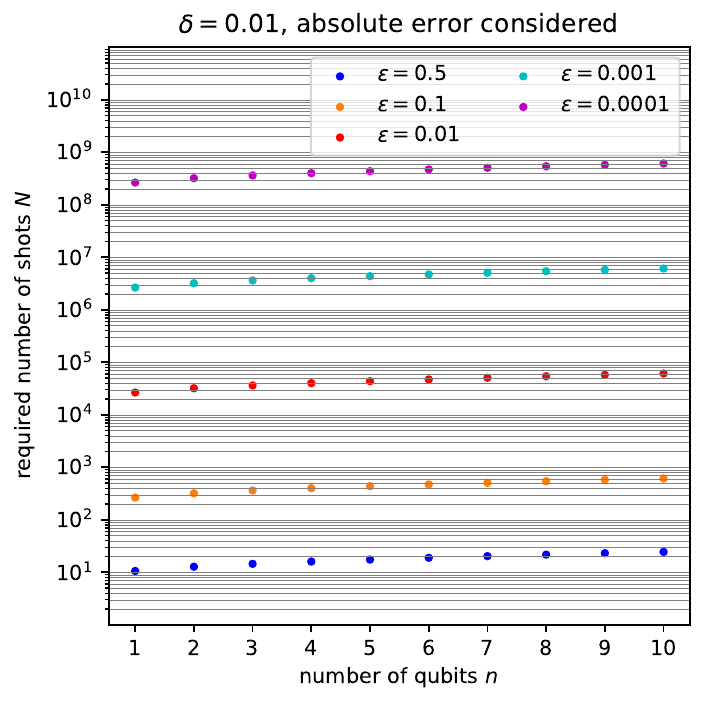}};
\draw[draw=none] (0, 0) -- (0.5\linewidth, 0);
\draw (-0.475\linewidth, 0.45\linewidth) node{(a)};
\end{tikzpicture}
\end{minipage}
\hfill
\begin{minipage}[c]{0.485\textwidth}
\centering
\begin{tikzpicture}
\draw (0,0) node[inner sep=0]{\includegraphics[height=0.95\linewidth]{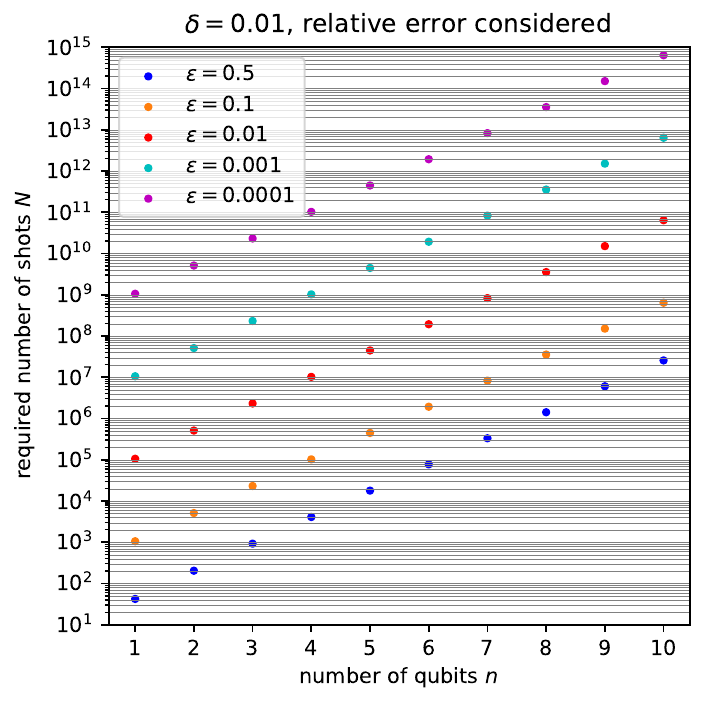}};
\draw[draw=none] (0, 0) -- (0.5\linewidth, 0);
\draw (-0.475\linewidth, 0.45\linewidth) node{(b)};
\end{tikzpicture}
\end{minipage}
\begin{minipage}[c]{0.485\textwidth}
\centering
\begin{tikzpicture}
\draw (0,0) node[inner sep=0]{\includegraphics[height=0.95\linewidth]{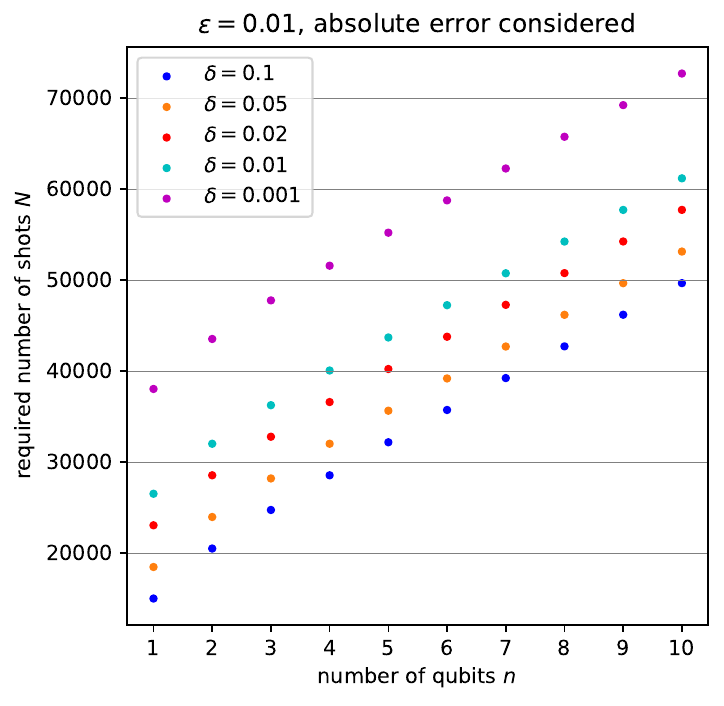}};
\draw[draw=none] (0, 0) -- (0.5\linewidth, 0);
\draw (-0.475\linewidth, 0.45\linewidth) node{(c)};
\end{tikzpicture}
\end{minipage}
\hfill
\begin{minipage}[c]{0.485\textwidth}
\centering
\begin{tikzpicture}
\draw (0,0) node[inner sep=0]{\includegraphics[height=0.95\linewidth]{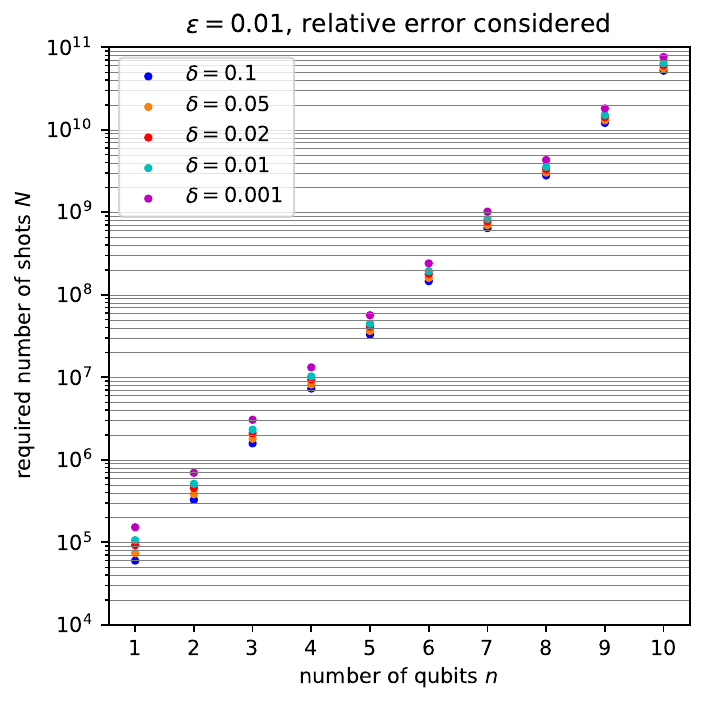}};
\draw[draw=none] (0, 0) -- (0.5\linewidth, 0);
\draw (-0.475\linewidth, 0.45\linewidth) node{(d)};
\end{tikzpicture}
\end{minipage}
\captionsetup{justification=raggedright, singlelinecheck=false}
\caption[]{Required number of shots $ N $ according to equality in formula ({\hyperref[eq:Number_of_runs_multiple_qubits_absolute_error]{\ref*{eq:Number_of_runs_multiple_qubits_absolute_error}}}) for the plots (a) and (c) and according to equality in formula (\hyperref[eq:Number_of_runs_multiple_qubits_relative_error]{\ref*{eq:Number_of_runs_multiple_qubits_relative_error}}) for the plots (b) and (d).}
\label{Fig:Plots_Required_number_of_runs}
\end{figure*}

However, it is to note that the error $ \epsilon $ is an absolute error, which is related to the reference of $ 1 = 100  $ \% like in ({\hyperref[eq:Number_of_runs_1_qubit]{\ref*{eq:Number_of_runs_1_qubit}}}) but if every $ Pr(i) , i \in \{ 0, \dots, {2^n} - 1 \} $ is used to store a number related to a solution value, a relative error should be of interest. 
Since the probability amount of $ 1 $ can be split to $ 2^n $ parts, the relative error according to $ {\frac{\epsilon}{{2^n}}} $ should be considered for the accuracy of the numbers given by the $ Pr(i) $, i.e.
\begin{align}
 N & \geq {\frac{1}{2 {{\epsilon}^2}}} {2^{2n}} \ln{ \left( \frac{2 {({2^n} - 1)} }{\delta}  \right) }
 \label{eq:Number_of_runs_multiple_qubits_relative_error}
\end{align}
or $ {\frac{\epsilon}{{2^{n-1}}}} $ in order to refer the accuracy to the absolute error $ \epsilon $ with reference $ 1 $ used in ({\hyperref[eq:Number_of_runs_1_qubit]{\ref*{eq:Number_of_runs_1_qubit}}}). 
Hence, in terms of the number of encoded values $ 2^n = {\tilde{n}} $, this upper bound for $ N $ scales as $ N \geq {\frac{ {\tilde{n}}^2 }{ 2{{\epsilon}^2} }} {\ln{ \left( \frac{2( {\tilde{n}} - 1)}{\delta} \right) }} $ w.r.t. $ {\epsilon} $ as the relative error according to (\hyperref[eq:Number_of_runs_multiple_qubits_relative_error]{\ref*{eq:Number_of_runs_multiple_qubits_relative_error}}). 
Fig. {\hyperref[Fig:Plots_Required_number_of_runs]{\ref*{Fig:Plots_Required_number_of_runs}}} shows the number of runs $ N $ for the case of equality in formula (\hyperref[eq:Number_of_runs_multiple_qubits_relative_error]{\ref*{eq:Number_of_runs_multiple_qubits_relative_error}}) for different parameter choices and the applicability of formula (\hyperref[eq:Number_of_runs_multiple_qubits_relative_error]{\ref*{eq:Number_of_runs_multiple_qubits_relative_error}}) is further discussed in section {\hyperref[sec:Empirical_Studies_for_GAE]{\ref*{sec:Empirical_Studies_for_GAE}}}.

\subsubsection{Bitstring Encoding}

For {{bitstring encoding}}, it is known that there is either the state $ \left| 0 \right\rangle $ or $ \left| 1 \right\rangle $ for every qubit, so that the state corresponds to a classical configuration of the computation system. 
For a machine with ideal measurement operations, the accuracy of the solution in {{bitstring encoding} should be determined therefore just by the number of qubits used to resolve a number relative to a certain range from which the number is taken but not by the number of runs of the quantum algorithm, i.e. the required number of runs is $ 1 $.

An example of a quantum algorithm that uses this encoding format for the solution and thus needs one run is the Bernstein-Vazirani quantum algorithm \cite{Abhijith_et_al_Q_Algs} for deducing a secret code (see appendix {\hyperref[App_subsec:A_Mapping_of_the_LA_Eq_to_the_BV-Q-Alg]{\ref*{App_subsec:A_Mapping_of_the_LA_Eq_to_the_BV-Q-Alg}}}).

\section{Empirical Studies for General Amplitude Encoding}
\label{sec:Empirical_Studies_for_GAE}

To verify the claims for the initialization time according to formula (\hyperref[eq:GAEI_formula_time]{\ref*{eq:GAEI_formula_time}}) and the upper bound for the required number of runs (\hyperref[eq:Number_of_runs_multiple_qubits_relative_error]{\ref*{eq:Number_of_runs_multiple_qubits_relative_error}}), corresponding simulation experiments were conducted, which are documented in the subsections {\hyperref[subsec:Study_Initialization]{\ref*{subsec:Study_Initialization}}} and {\hyperref[subsec:Study_Read-out]{\ref*{subsec:Study_Read-out}}}, respectively. 
Like all simulations of quantum circuits that are presented in this work, also these simulation studies were done using the means of IBM's software kit {\emph{Qiskit}} \cite{qiskit} in the version 1.2.2 together with the kits {\emph{qiskit\_aer}} in the version 0.15.1 and {\emph{qiskit\_ibm\_runtime}} in the version 0.30.0.

\subsection{Initialization}
\label{subsec:Study_Initialization}

\begin{table*}[t]
\begin{center}
\begin{tabular}[c]{l|c|c|c|c|c|c|c|c|c|c}
 \multicolumn{1}{c|}{  } & \multicolumn{9}{c}{ $ n = $ }  \\[0.15cm]
 \multicolumn{1}{c|}{  } & $ 1 $ & $ 2 $ & $ 3 $ & $ 4 $ & $ 5 $ & $ 6 $ & $ 7 $ & $ 8 $ & $ 9 $ & $ 10 $  \\ \hline
 \multicolumn{1}{l|}{ experiment runs } & $ 20 $ & $ 40 $ & $ 80 $ & $ 160 $ & $ 320 $ & $ 640 $ & $ 1280 $ & $ 2560 $ & $ 5120 $ & $ 10240 $  \\ \hline
 \multicolumn{1}{l|}{ durations for optimization level $0~$ } 
 & $ ~512~ $ & $ ~3680~ $ & $ ~30400~ $ & $ ~95232~ $ & $ ~230368~ $ & $ ~554880~ $ & $ ~1157376~ $ & $ ~2417856~ $ & $ ~4914624~ $ & $ ~9940320~ $ \\ \hline
 \multicolumn{1}{l|}{ durations for optimization level $1~$ } 
 & $ 512 $ & $ 3424 $ & $ 20640 $ & $ 73056 $ & $ 174048 $ & $ 385728 $ & $ 806656 $ & $ 1634432 $ & $ 3345472 $ & $ 6830080 $ \\ \hline
 \multicolumn{1}{l|}{ durations for optimization level $2~$ } 
 & $ 512 $ & $ 3424 $ & $ 21152 $ & $ 56352 $ & $ 139520 $ & $ 284096 $ & $ 612672 $ & $ 1287072 $ & $ 2728544 $ & $ 5448512 $ \\ \hline 
 \multicolumn{1}{l|}{ durations for optimization level $3~$ } 
  & $ 512 $ & $ 3424 $ & $ 20896 $ & $ 56096 $ & $ 123168 $ & $ 278624 $ & $ 610176 $ & $ 1295136 $ & $ 2733088 $ & $ 5475424 $ 
\end{tabular}
\end{center}
\captionsetup{justification=raggedright, singlelinecheck=false}
\caption[]{Found maximal {\texttt{duration}}-values returned by {\emph{Qiskits}}'s function {\texttt{schedule()}} applied to the initialization circuit of a random statevector w.r.t. $ n $ qubits and the different optimization levels of {\emph{Qiskits}}'s circuit compiling procedure. 
For $ n $ qubits, $ {2^n} \cdot 10 $ experiment runs, i.e. sampled random statevectors were considered. 
The durations are referred to the compilation for {\texttt{FakeSherbrooke}} and have to be multiplied by its {\texttt{dt}}-value of $ \approx 2.222 \cdot {10^{-10}} $ s to obtain the estimated execution time for the schedule of the pulses.}
\label{Fig_Table_GAEI_Study_durations}
\end{table*}

Since no way could be found to get to know the actual execution time of quantum circuits submitted to IBM's quantum computers more precisely than up to full seconds (which is the 'quantum time' in the '{\texttt{job.metrics()}}'-information of executed jobs), it can be provided here only a study of the actually expectable runtimes of the initialization procedure for {{general amplitude encoding}} that is based on snapshot data from IBM for their machines and {\emph{Qiskit}}'s own function for estimating runtimes {\texttt{schedule()}}, which also allows to consider circuits with no measurements:

As the {\texttt{backend}}, i.e. the execution device for quantum circuits, the model provided from IBM for their quantum computer {\emph{Sherbrooke}} was used. 
{\emph{Sherbrooke}} is a processor of the type {\emph{Eagle}}, containing 127 qubits, which are formed by superconducting circuits. 
Its model, the so-called {\emph{fake backend}} {\texttt{FakeSherbrooke}} implements the same connectivity of the qubits and provides data of the qubits and the physically implemented operations, which were taken as a snapshot of {\emph{Sherbrooke}}.

To check that formula (\hyperref[eq:GAEI_formula_time]{\ref*{eq:GAEI_formula_time}}) provides a realistic estimation, a random statevector according to {{general amplitude encoding}} for $ n $ qubits was generated via the {\emph{Qiskit}}-function {\texttt{random\_statevector()}}, which was then initialized in a quantum circuit via the function {\texttt{prepare\_state()}} (not via {\texttt{initialize()}}, since this function resorts to the qubit-reset operation, which is not supported by {\emph{Sherbrooke}}).

Via {\texttt{generate\_preset\_pass\_manager(backend, optimization\_level)}}, followed by the application of the function {\texttt{run()}}, this circuit was then compiled to the native gate operations and connectivity of {\emph{Sherbrooke}} via specifying the {\texttt{backend}}-argument as the fake backend. 
The latter was then also set in the subsequent application of the function {\texttt{schedule()}} to it. 
The function {\texttt{schedule()}} provides an estimation for the runtime of a quantum circuit, since it expresses it explicitly as a sequence of electromagnetic pulses of specific durations that realize the operations in the circuit, where the specifications for this are given in the definitions of the fake backend for the considered situation. 
In particular, {\texttt{schedule()}} provides the parameter {\texttt{duration}}, which is a number that represents the execution time of the circuit. 
As it typically applies for all time specifications for a specific backend, it is given in the particular time unit for this backend, which can be addressed as the parameter {\texttt{dt}} of the backend and is given in seconds. 
For {\texttt{FakeSherbrooke}}, {\texttt{dt}} is $ \approx 2.222 \cdot {10^{-10}} $ s.

The described workflow should hence give reasonable estimation of the time that it takes to initialize a random statevector on a quantum computer based on superconducting qubits. 
However, this gives just the time w.r.t. {\bf{one}} random statevector and indeed, it was observed from $ n = 3 $ qubits onwards that the durations vary for different random states, which is probably because {\emph{Qiskit}} finds more optimized compiled circuits for special cases. 
Therefore, multiple runs of this workflow, which is referred here to as a single {\emph{experiment}}, were performed and the maximum of the occurred durations was determined, since this should be the quantity of interest for determining the runtime that has to be expected for the initialization of a general state. 
Specifically for the study presented here, $ {2^n} \cdot 10 $ experiment runs were conducted, so that the number of explored random states grows linearly with the state space.

An aspect that was studied in this framework is the effect of the parameter {\text{{optimization\_level}} in {\texttt{generate\_preset\_pass\_manager(backend, optimization\_level)}}.  
According to the {\emph{Qiskit}}-documentation \cite{qiskit-website_optimization}, no additional optimization of the compiled circuit for the considered hardware is done for {\texttt{optimization\_level = 0}}; whereas for an {\texttt{optimization\_level}} of 1 onwards, a search for optimization possibilities is applied, which is intensified at level 2 and further at level 3, where the latter also makes explicitly use of a decomposition by which any 2-qubit gate can be expressed by at most 3 native 2-qubit gates. 
Table {\hyperref[Fig_Table_GAEI_Study_durations]{\ref*{Fig_Table_GAEI_Study_durations}}} documents the found maximal durations and Fig. {\hyperref[Fig:GAEI_Runtime_Study]{\ref*{Fig:GAEI_Runtime_Study}}} shows the corresponding runtimes compared to values according to the reference formula (\hyperref[eq:GAEI_formula_time]{\ref*{eq:GAEI_formula_time}}).
\begin{figure}[t]
\begin{center}
\includegraphics[width=\linewidth]{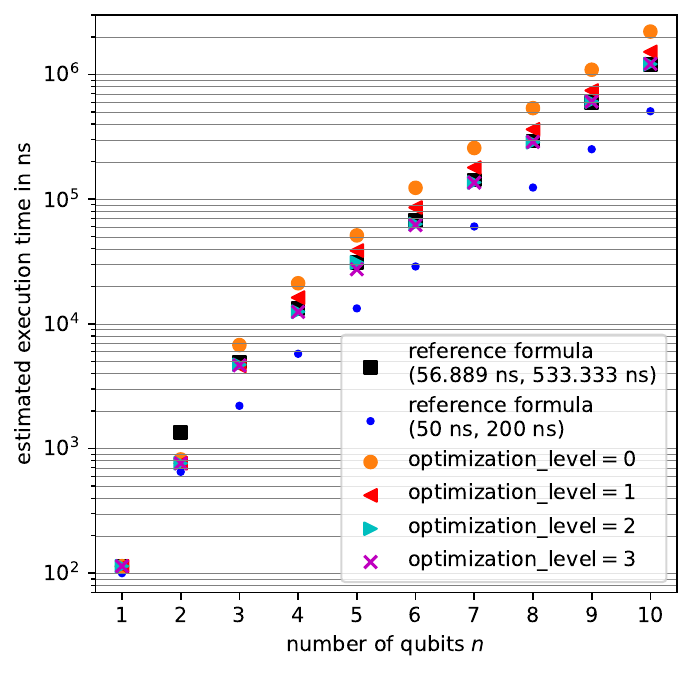}
\end{center}
\captionsetup{justification=raggedright, singlelinecheck=false}
\caption[]{Execution times for the initialization of a general state that were estimated via the reference formula (\hyperref[eq:GAEI_formula_time]{\ref*{eq:GAEI_formula_time}}) and via the {\emph{Qiskit}}'s function {\texttt{schedule()}} according to the found durations in Table {\hyperref[Fig_Table_GAEI_Study_durations]{\ref*{Fig_Table_GAEI_Study_durations}}}. 
For the times from the reference formula, the used 1- and 2-qubit gate execution times are stated in the brackets of the plot labels as the first and second value, respectively.}
\label{Fig:GAEI_Runtime_Study}
\end{figure}

For superconducting hardware, the impression was obtained that a typical execution time scale of native 1-qubit gates should be $ 50 $ ns, whereas a typical scale for the native 2-qubit gates, which are often given just by the CX-gate as it was also assumed in the outlined decomposition procedure of V. V. Shende et al \cite{Shende_et_al_Synthesis_of_q_circuits_long} in subsection {\hyperref[subsec:Initialization]{\ref*{subsec:Initialization}}}, should be $ 200 $ ns \cite{Tennie_et_al_QC_non-lin_DEs, Kandala_et_al_CX}. 
Corresponding to these values for the respective execution times in formula (\hyperref[eq:GAEI_formula_time]{\ref*{eq:GAEI_formula_time}}), the graph {\emph{reference formula (50 ns, 200 ns)}} in Fig. {\hyperref[Fig:GAEI_Runtime_Study]{\ref*{Fig:GAEI_Runtime_Study}}} depicts the associated runtimes for the initialization.

However, {\texttt{configuration().basis\_gates}} called for the fake backend gives the information that the set of native gates of {\emph{Sherbrooke}} is given by the identity-gate, the X-gate, the $ \sqrt{\text{X}} $-gate and the $ {R_z} $-gate as the native 1-qubit gates as well as the ECR-gate as the native 2-qubit gate. 
The latter is the echoed cross-resonance gate, which is equivalent to the CX-gate up to a layer of 1-qubit gates that are set before and after the CX-gate. 
The command {\texttt{target[]}} applied to the fake backend w.r.t. the gate operation of interest returns the gate execution times: 
While it is stated an execution time of $ \approx 56.889 $ ns for the 1-qubit gates except for the $ R_z $-gate, for which $ 0 $ ns is stated, the ECR-gate time is $ \approx 533.333 $ ns. 
Therefore, {\emph{reference formula (56.889 ns, 533.333 ns)}} in Fig. {\hyperref[Fig:GAEI_Runtime_Study]{\ref*{Fig:GAEI_Runtime_Study}}} shows the times according to formula (\hyperref[eq:GAEI_formula_time]{\ref*{eq:GAEI_formula_time}}) for these specific gate times.

In general, the graphs according to the reference formula seem to represent the global behavior of the estimated times from the simulation study, which is not surprising since according to \cite{Griffin_et_al_Review}, the compiling in {\emph{Qiskit}} is based on the decomposition procedure of V. V. Shende et al. \cite{Shende_et_al_Synthesis_of_q_circuits_long}, however the native gate set of {\emph{Sherbrooke}} is different. 
For {\emph{reference formula (56.889 ns, 533.333 ns)}}, it can be seen until $ n = 8 $ that some points of the simulation study lie below this reference. 
This occurs mostly for those for {\texttt{optimization\_level = 2}} and {\texttt{optimization\_level = 3}} but compared to such slight deviations for them, the deviation for $ n = 2 $ seems to be systematical because it is larger and occurs for all optimization levels. 
This deviation for $ n = 2 $ could be possibly explained by the application of the procedure to express a general 2-qubit gate just via at most 3 2-qubit gates, mentioned before for optimization level 3. 
Maybe this procedure is implemented in the compiling procedure of {\emph{Qiskit}} by default also for the other optimization levels, even if no information on this could be found. 
In fact, it was observed for the compiled circuit w.r.t. random states that it involves only one ECR-gate for all optimization levels, whereas the derived reference formula for the circuit depth contribution of the CX-gates yields two. 
The slight deviations for other $ n $ could be attributed to the fact that the empirically obtained maximum times do not represent the actual maximum times but still lie below as a finite number of random states was considered. 
Moreover, this issue could also be due to the different native gates than those considered for the formula and also the aspect that the coupling of the qubits in {\emph{Sherbrooke}} does not exactly correspond to all-to-all-connectivity as assumed in the derivation of formula (\hyperref[eq:GAEI_formula_time]{\ref*{eq:GAEI_formula_time}}), but is a grid. 
Nevertheless, the execution times according to the formula match the order of magnitude of the estimations obtained via the functions of {\emph{Qiskit}} for the considered qubit numbers, which is why the conclusion is drawn that the reference formula should yield reasonable estimations concerning the execution on real hardware of this type.

It is to note concerning the applicability of {{general amplitude encoding}} that the coherence time for superconducting qubits is only of the order of $ 100 $ µs \cite{Tennie_et_al_QC_non-lin_DEs, Kandala_et_al_CX}. 
In fact, from the data on the properties w.r.t. all qubits of the fake backend, a minimum relaxation time {\texttt{t1}} of $ \approx 40.905  $ µs and a minimum dephasing time {\texttt{t2}} of $ \approx 16.932 $ µs were read out.

\subsection{Read-out}
\label{subsec:Study_Read-out}

\begin{figure*}[t]
\begin{minipage}[c]{0.485\textwidth}
\centering
\begin{tikzpicture}
\draw (0,0) node[inner sep=0]{\includegraphics[height=0.95\linewidth]{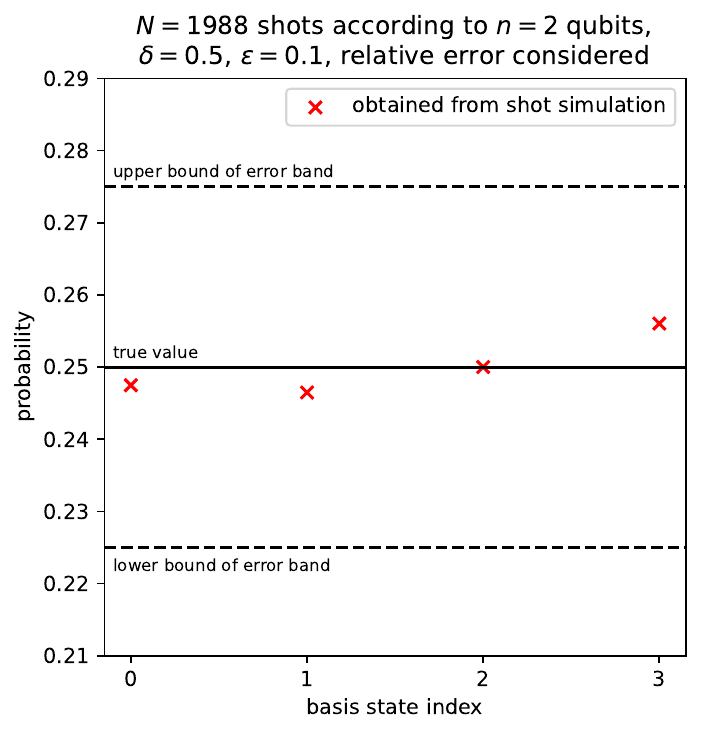}};
\draw[draw=none] (0, 0) -- (0.5\linewidth, 0);
\draw (-0.475\linewidth, 0.45\linewidth) node{(a)};
\end{tikzpicture}
\end{minipage}
\hfill
\begin{minipage}[c]{0.485\textwidth}
\centering
\begin{tikzpicture}
\draw (0,0) node[inner sep=0]{\includegraphics[height=0.95\linewidth]{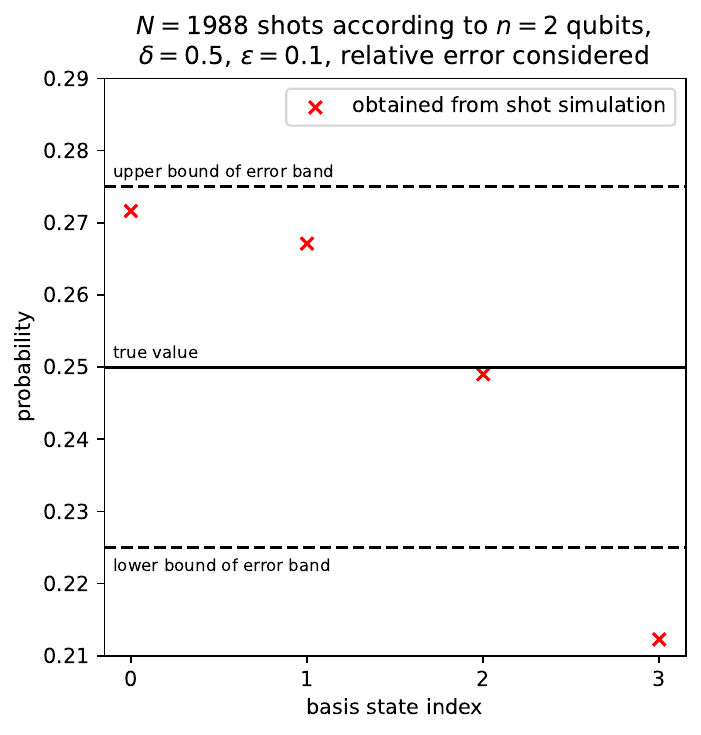}};
\draw[draw=none] (0, 0) -- (0.5\linewidth, 0);
\draw (-0.475\linewidth, 0.45\linewidth) node{(b)};
\end{tikzpicture}
\end{minipage}
\captionsetup{justification=raggedright, singlelinecheck=false}
\caption[]{Exemplary results of a single experiment run for studying the read-out accuracy. 
The number of shots $ N $ was set according to the value for equivalence in formula (\hyperref[eq:Number_of_runs_multiple_qubits_relative_error]{\ref*{eq:Number_of_runs_multiple_qubits_relative_error}}), which was then ceiled to the next integer. 
(a) and (b) are outcomes for the same parameter setting; however, in (a), the desired error bound is fulfilled but in (b), the inferred probability for the basis state index $ 3 $ violates the error band of $ 2 \epsilon $. Specifically, (a) depicts the measurement outcome of $ 492 $, $ 490 $, $ 497 $ and $ 509 $ counts for the indices from $ 0 $ to $ 3 $ and (b) the counts $ 540 $, $ 531 $, $ 495 $ and $ 422 $.}
\label{Fig:Study_Verification_Required_number_of_runs_Examples_for_1_Experiment}
\end{figure*}

To verify formula  (\hyperref[eq:Number_of_runs_multiple_qubits_relative_error]{\ref*{eq:Number_of_runs_multiple_qubits_relative_error}})for the required number of shots to obtain the probabilities of a state vector according to {{general amplitude encoding}} with a given accuracy, shot simulations were conducted: 
Since formula (\hyperref[eq:Number_of_runs_multiple_qubits_relative_error]{\ref*{eq:Number_of_runs_multiple_qubits_relative_error}})was derived w.r.t. a relative error $ \epsilon $ that is referred to the situation that all probabilities are equal, i.e. $ \frac{1}{2^n} $, exactly this situation was considered, which is obtained by the application of a layer of H-gates to $ n $ qubits that were prepared in the $ \left| 0 \right\rangle $-state. 
Moreover, formula (\hyperref[eq:Number_of_runs_multiple_qubits_relative_error]{\ref*{eq:Number_of_runs_multiple_qubits_relative_error}})was derived for an ideal system, i.e. without noise. 
Therefore, it was resorted to simulation also in this empirical study, even if only one layer of H-gates is considered. 
Specifically, the {\texttt{AerSimulator}} was used with the method {\texttt{automatic}}. 
So, all qubits of the set quantum circuit were measured, where a number of shots according to the rounded up value that is given by the case of equality in formula (\hyperref[eq:Number_of_runs_multiple_qubits_relative_error]{\ref*{eq:Number_of_runs_multiple_qubits_relative_error}}) were simulated.

This workflow is referred in this subsection as one {\emph{experiment}}. 
Fig. {\hyperref[Fig:Study_Verification_Required_number_of_runs_Examples_for_1_Experiment]{\ref*{Fig:Study_Verification_Required_number_of_runs_Examples_for_1_Experiment}}} shows exemplary outcomes of such an experiment. 
Specifically, Fig. {\hyperref[Fig:Study_Verification_Required_number_of_runs_Examples_for_1_Experiment]{\ref*{Fig:Study_Verification_Required_number_of_runs_Examples_for_1_Experiment}}} (a) shows a case in which all inferred probabilities lie within the error band of $ 2 \epsilon $ w.r.t. the relative error $ \epsilon $ referred to $ \frac{1}{2^n} $ as desired. 
However, since it is considered $ \delta > 0 $ for the required number of runs according to formula (\hyperref[eq:Number_of_runs_multiple_qubits_relative_error]{\ref*{eq:Number_of_runs_multiple_qubits_relative_error}}) (for $ \delta  \longrightarrow 0 $, of course $ N \longrightarrow \infty $), it can also occur that this error bound is violated as it was the case of the experiment outcome in Fig. {\hyperref[Fig:Study_Verification_Required_number_of_runs_Examples_for_1_Experiment]{\ref*{Fig:Study_Verification_Required_number_of_runs_Examples_for_1_Experiment}}} (b), since the last point is an outlier. 
Thus, in order to account for $ \delta $, multiple runs of such an experiment were conducted. 
The real value for the probability according to $ \delta $, i.e. the rate of the occurrence of a violation of the set error band $ 2 \epsilon $ by at least one inferred basis state probability, should be empirically determinable for a number of experiment runs that goes to $ \infty $ since the inferred value for $ \delta $ should converge then to the real one. 
For this study, $ \frac{1}{\delta} $ experiment runs times an integer factor $ F $ were performed, so that it is to expect according to (\hyperref[eq:Number_of_runs_multiple_qubits_relative_error]{\ref*{eq:Number_of_runs_multiple_qubits_relative_error}}) that not more than $ F $ outcomes of these experiment runs violate the error band by at least one inferred probability (e.g. at most $ 100 $ cases of $ \frac{1}{0.1} \cdot 100 = 1000 $ experiment runs if $ \delta = 0.1 $ and $ F = 100 $ are considered). 
Accordingly, the number of such outlier cases for the many runs of an experiment for a specific setting were recorded and counted. 
The obtained counts are documented in \mbox{Table {\hyperref[Fig_Table_GAE_Required_runs_Study_found_outliers]{\ref*{Fig_Table_GAE_Required_runs_Study_found_outliers}}}.}
\begin{table*}[t]
\begin{minipage}[c]{0.485\textwidth}
\centering
\begin{tikzpicture}
\draw[draw=none] (0, 0) -- (0.5\linewidth, 0);
\draw (-0.475\linewidth, 0.0\linewidth) node{(a)};
\end{tikzpicture}
\end{minipage}
\hfill
\begin{minipage}[c]{0.485\textwidth}
\centering
\begin{tikzpicture}
\draw[draw=none] (0, 0) -- (0.5\linewidth, 0);
\draw (-0.475\linewidth, 0.0\linewidth) node{(b)};
\end{tikzpicture}
\end{minipage}
\hfill
\vspace{-0.5cm}
\begin{minipage}[c]{0.485\textwidth}
\begin{center}
\begin{tabular}[c]{l|c|c|c|c|c} 
  \multicolumn{6}{c}{ $ \delta = 0.5, \epsilon = 0.1, F = 100 $: }  \\[0.15cm]
  & \multicolumn{5}{c}{ $ n = $ }  \\[0.15cm]
  & $ 1 $ & $ 2 $ & $ 3 $ & $ 4 $ & $ 5 $   \\ \cline{1-6}
   $ N $ & $ ~278~ $ & $ ~1988~ $ & $ ~10664~ $ & $ ~52408~ $ & $ ~246799~ $   \\ \cline{1-6}
  \multicolumn{1}{l|}{ outliers$~$ } & $ 25 $ & $ 4 $ & $ 0 $ & $ 0 $ & $ 0 $ 
  \\ 
\end{tabular}
\end{center}
\end{minipage}
\hfill
\begin{minipage}[c]{0.485\textwidth}
\begin{center}
\begin{tabular}[c]{l|c|c|c|c|c}
  \multicolumn{6}{c}{ $ \delta = 0.1, \epsilon = 0.1, F = 100 $: }  \\[0.15cm]
  & \multicolumn{5}{c}{ $ n = $ }  \\[0.15cm]
  & $ 1 $ & $ 2 $ & $ 3 $ & $ 4 $ & $ 5 $   \\ \cline{1-6}
  $ N $ & $ ~600~ $ & $ ~3276~ $ & $ ~15814~ $ & $ ~73009~ $ & $ ~329202~ $   \\ \cline{1-6} 
  \multicolumn{1}{l|}{ outliers$~$ } & $ 16 $ & $ 1 $ & $ 0 $ & $ 0 $ & $ 0 $ 
  \\ 
\end{tabular}
\end{center}
\end{minipage}
\hfill
\vspace{0.5cm}
\begin{minipage}[c]{0.485\textwidth}
\centering
\begin{tikzpicture}
\draw[draw=none] (0, 0) -- (0.5\linewidth, 0);
\draw (-0.475\linewidth, 0.0\linewidth) node{(c)};
\end{tikzpicture}
\end{minipage}
\hfill
\begin{minipage}[c]{0.485\textwidth}
\centering
\begin{tikzpicture}
\draw[draw=none] (0, 0) -- (0.5\linewidth, 0);
\draw (-0.475\linewidth, 0.0\linewidth) node{(d)};
\end{tikzpicture}
\end{minipage}
\hfill
\vspace{-0.5cm}
\begin{minipage}[c]{0.485\textwidth}
\begin{center}
\begin{tabular}[c]{l|c|c|c|c|c}
  \multicolumn{6}{c}{ $ \delta = 0.5, \epsilon = 0.01, F = 100 $: }  \\[0.15cm]
  & \multicolumn{5}{c}{ $ n = $ }  \\[0.15cm]
  & $ 1 $ & $ 2 $ & $ 3 $ & $ 4 $ & $ 5 $   \\ \cline{1-6}
   $ N $ & $ ~27726~ $ & $ ~198793~ $ & $ ~1066306~ $ & $ ~5240762~ $ & $ ~24679842~ $   \\ \cline{1-6}
 \multicolumn{1}{l|}{ outliers$~$ } & $ 15 $ & $ 7 $ & $ 0 $ & $ 0 $ & $ 0 $ 
  \\ 
\end{tabular}
\end{center}
\end{minipage}
\hfill
\begin{minipage}[c]{0.485\textwidth}
\begin{center}
\begin{tabular}[c]{l|c|c|c|c|c}
  \multicolumn{6}{c}{ $ \delta = 0.5, \epsilon = 0.1, F = 1000 $: }  \\[0.15cm]
  & \multicolumn{5}{c}{ $ n = $ }  \\[0.15cm]
  & $ 1 $ & $ 2 $ & $ 3 $ & $ 4 $ & $ 5 $   \\ \cline{1-6}
   $ N $ & $ ~278~ $ & $ ~1988~ $ & $ ~10664~ $ & $ ~52408~ $ & $ ~246799~ $   \\ \cline{1-6}
  \multicolumn{1}{l|}{ outliers$~$ } & $ 214 $ & $ 94 $ & $ 2 $ & $ 0 $ & $ 0 $ 
  \\ 
\end{tabular}
\end{center}
\end{minipage}
\vspace{0.3cm}
\captionsetup{justification=raggedright, singlelinecheck=false}
\caption[]{Numbers of detected outliers in studies of the read-out accuracy for different parameter settings. 
W.r.t. the reference parameter choice (a), $ \delta $ was decreased in the study (b), $ \epsilon $ was decreased in (c) and $ F $ was increased in (d), where the latter should allow to empirically determine the real rate of error bound violations more precisely. 
For the studies (a), (b) and (c), hence at most $ 100 $ outlier cases are expected and for study (d), at most $ 1000 $, since $ {\frac{1}{\delta}} \cdot F  $ experiment runs were done in very study. 
The number of shots $ N $ in a single experiment was set according to the value for equivalence in formula (\hyperref[eq:Number_of_runs_multiple_qubits_relative_error]{\ref*{eq:Number_of_runs_multiple_qubits_relative_error}}), which was then ceiled to the next integer.}
\label{Fig_Table_GAE_Required_runs_Study_found_outliers}
\end{table*}

\begin{table*}[t]
\begin{center}
\begin{tabular}[c]{l|c|c|c|c|c|c|c|c|c|c|c|c}
  \multicolumn{13}{c}{ $ \delta = 0.5, \epsilon = 0.1, F = 100 $: }  \\[0.15cm]
  & \multicolumn{12}{c}{ $ n = $ }  \\[0.15cm]
  & $ 1 $ & $ 2 $ & $ 3 $ & $ 4 $ & $ 5 $ & $ 6 $ & $ 7 $ & $ 8 $ & $ 9 $ & $ 10 $ & $ 11 $ & $ 12 $  \\ \cline{1-13}
   $ N $ & $ ~41~ $ & $ ~462~ $ & $ ~1961~ $ & $ ~5907~ $ & $ ~16001~ $ & $ ~41401~ $ & $ ~99802~ $ & $ ~225206~ $ & $ ~511371~ $ & $ ~1187631~ $ & $ ~2604712~ $ & $ ~5669580~ $  \\ \cline{1-13}
  \multicolumn{1}{l|}{ outliers$~$ } & $ ~100~ $ & $ 92 $ & $ 96 $ & $ 88 $ & $ 93 $ & $ 99 $ & $ 88 $ & $ 98 $ & $ 98 $ & $ 94 $ & $ 88 $ & $ 100 $
\end{tabular}
\end{center}
\captionsetup{justification=raggedright, singlelinecheck=false}
\caption[]{Minimum numbers of required shots $ N $ that were empirically found for reading out a constant probability distribution w.r.t. $ 2^n $ basis states with an accuracy given by the probability $ 1 - {\delta} $ to find all inferred probabilities within an error band of $ 2 {\epsilon} $ w.r.t. a relative error $ {\epsilon} $. 
The same experiment parameters as for case (a) of Table {\hyperref[Fig_Table_GAE_Required_runs_Study_found_outliers]{\ref*{Fig_Table_GAE_Required_runs_Study_found_outliers}}} were used. 
Accordingly, it was set that not more than $ 100 $ outcomes of the $ 200 $ experiment runs that were always conducted for a specific $ N $ and $ n $ shall violate the error bound. 
The specific numbers of detected outliers are also stated.}
\label{Fig_Table_GAE_Read-out_Empirical_Study}
\end{table*}

In every considered situation, the observed number of outliers is significantly below the maximum expected number. 
However, this just confirms the correctness of the given formula (\hyperref[eq:Number_of_runs_multiple_qubits_relative_error]{\ref*{eq:Number_of_runs_multiple_qubits_relative_error}}) since it merely states an upper bound for the required number of shots to achieve a desired accuracy (because of the '$ \leq $'-symbols in the union bound (\hyperref[eq:Hoeffding_ineq_union bound_multiple_qubits]{\ref*{eq:Hoeffding_ineq_union bound_multiple_qubits}}) and the Hoeffding inequality (\hyperref[eq:Hoeffding_ineq_1_qubit]{\ref*{eq:Hoeffding_ineq_1_qubit}}), i.e., the probability for a deviation by the amount $ \epsilon $ can also be less). 
Moreover, the variation of the parameters even indicates that the percentage of outliers gets lower for smaller $ \delta $ and $ \epsilon $ and larger $ n $. 
In regard of this and the many cases for which even no violation of the error bound could be detected, the conclusion is drawn that this formula yields a valid but relatively pessimistic estimation of the required number of runs. 
This is not surprising since it was derived via the application of the union bound property, which holds for probabilities in general. 
However, for low $ n $ and relatively high $ \delta $ and $ \epsilon $, the study indicates that the inferred rates of error bound violation at least approach the same order of magnitude that is given for the rates according to the formula.

However, to provide a suggestion for the actual scaling behavior of the required number of runs $ N $ with the number of qubits $ n $, an empirical study for approximately determining $ N(n) $ was conducted in a corresponding manner for the reference situation according to the parameter settings of Table {\hyperref[Fig_Table_GAE_Required_runs_Study_found_outliers]{\ref*{Fig_Table_GAE_Required_runs_Study_found_outliers}}} (a): 
For the prepared constant probability distribution according to $ \frac{1}{{2^{n}}} $ for all basis state probabilities, where all of them shall be inferred from measurements of the state vector with a precision up to a desired relative error $ \epsilon = 0.1 $, respectively, the number of required shots $ N $ w.r.t. $ n $ for which the error bound is fulfilled with at least the set probability of $ \delta = 0.5 $ was empirically determined. 
Again, $ {\frac{1}{\delta}} \cdot F $ with $ F = 100 $ repetitions of a single extraction of the probability distribution via $ N $ simulated shots for $ n $ qubits were always done, so that it is desired that not more than $ 100 $ of the performed $ 200 $ experiment runs yield an outcome where at least one extracted basis state probability violates the error band. 
W.r.t. a fixed $ n $, it was proceeded such that $ N $ was increased in relatively large steps until $ 100 $ or less of these outlier cases were detected for the $ 200 $ experiment runs. 
Then, the interval w.r.t. $ N $ between the detected $ N $ that meets the conditions set for the error and the step before was rastered. 
If there was a step in this $ N $-interval, for which the error bounds were fulfilled, the interval given by this step and the step before was inspected with finer increments. 
In the case that no step within an interval yielded a successful result w.r.t. the error requirements, the subinterval between the last of these refinement steps and the $ N $-value limiting the considered interval, for which a successful result was detected, was considered. 
So, the overall procedure for approximately determining the required number of shots to reach the set error bound was performed like an interval search. 
Table {\hyperref[Fig_Table_GAE_Read-out_Empirical_Study]{\ref*{Fig_Table_GAE_Read-out_Empirical_Study}}} gives a protocol of the found $ N(n) $ and Fig. {\hyperref[Fig:GAE_Read-out_Empirical_Study]{\ref*{Fig:GAE_Read-out_Empirical_Study}}} shows a plot of them together with the deduced upper bound formula (\hyperref[eq:Number_of_runs_multiple_qubits_relative_error]{\ref*{eq:Number_of_runs_multiple_qubits_relative_error}}) and points according to considered regression functions. 
Regarding the points of the regression fits and in particular the trends for lower $ n $, it seems that at least for the considered reference situation of equal probabilities, an empirical scaling behavior according to $ {\tilde{n}} ~ { \ln{ ( {\tilde{n}} ) } } $ with $ {\tilde{n}} = {2^{n}} $ fits the best, which supports the theoretical reasoning in subsection {\hyperref[subsec:Read-out]{\ref*{subsec:Read-out}}} that the required $ N $ scales worse than linearly with $ {\tilde{n}} $.

\begin{figure}[t]
\begin{center}
\includegraphics[width=\linewidth]{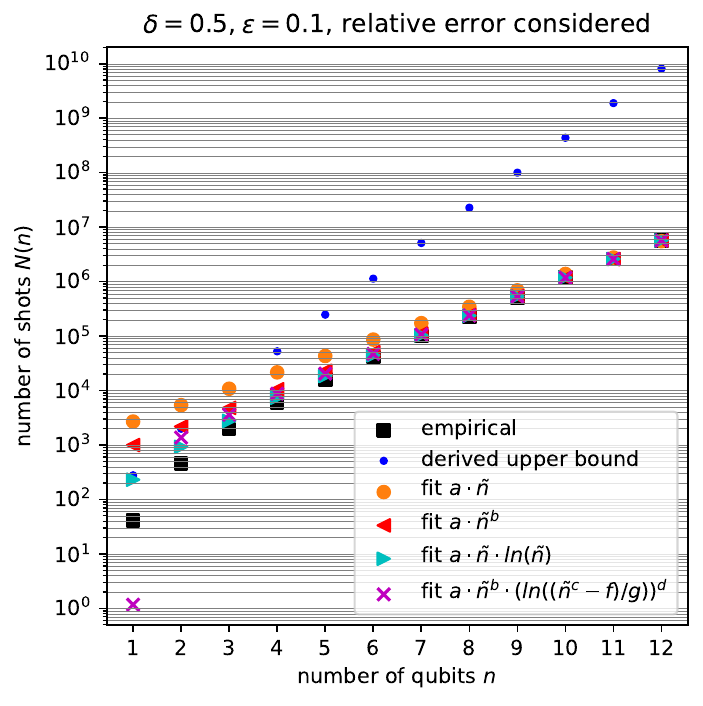}
\end{center}
\captionsetup{justification=raggedright, singlelinecheck=false}
\caption[]{Required number of shots $ N $ to read out the probability distribution in the situation that all $ {\tilde{n}} = {2^{n}} $ probabilities are equal and shall be read out up to a relative error $ \epsilon = 0.1 $, respectively, where a success probability of $ 1 - {\delta} = 0.5 $ was demanded. 
The values denoted by 'empirical' are the found $ N(n) $ in Table {\hyperref[Fig_Table_GAE_Read-out_Empirical_Study]{\ref*{Fig_Table_GAE_Read-out_Empirical_Study}}}, where the same experiment parameters as for case (a) of Table {\hyperref[Fig_Table_GAE_Required_runs_Study_found_outliers]{\ref*{Fig_Table_GAE_Required_runs_Study_found_outliers}}} were used. 
The points of 'derived upper bound' are the values according to equality in formula ({\hyperref[eq:Number_of_runs_multiple_qubits_relative_error]{\ref*{eq:Number_of_runs_multiple_qubits_relative_error}}}), also corresponding to the $ N(n) $ given in Table {\hyperref[Fig_Table_GAE_Required_runs_Study_found_outliers]{\ref*{Fig_Table_GAE_Required_runs_Study_found_outliers}}} (a). 
Furthermore, the points according to the considered fit functions that matched the best to the empirical values are shown, for which the function structure is stated. 
The regression fits were done via the '{\texttt{curve\_fit}}'-function of the {\emph{scipy}}-package for {\emph{python}}. 
Specifically, it was found $ a \approx 1345.964 $ for the function of structure $ a \cdot {\tilde{n}} $; $ (a \approx 453.672 , b \approx 1.134 ) $ for the function of structure $ a \cdot {{\tilde{n}}^{b}} $; $ a \approx 166.452 $ for the function of structure $ a \cdot {\tilde{n}} \cdot {\ln( {\tilde{n}} ) } $ and $ ( a \approx 61.69 , b \approx 1.03 , c \approx 1.273 , d \approx 1.096 , f \approx 2.381 , g \approx 0.035 ) $ for the function of structure $ a \cdot {{\tilde{n}}^{b}} \cdot {{( \ln( ( {{\tilde{n}}^{c}} - f ) / g ) )}^{d}} $.}
\label{Fig:GAE_Read-out_Empirical_Study}
\end{figure}

So, this empirically found scaling behavior for the considered reference situation according to $ {\tilde{n}} ~ {\ln( {\tilde{n}} )} $ is proposed for resource estimations on quantum algorithms that make use of amplitude encoding. 
An empirical study on the dependency w.r.t. the desired $ {\epsilon} $ and $ {\delta} $ is left here for further work.

It is to note that at least for the considered situation of equal probabilities and the consideration of a number of runs $ N $ that is a multiple of the number of basis states $ {\tilde{n}} $, the analytical approach explained in the appendix {\hyperref[App_sec:Probability_Fulfilling_error_bound_for_reference_situation]{\ref*{App_sec:Probability_Fulfilling_error_bound_for_reference_situation}}}could be found, which yields the formula ({\hyperref[eq:GAE_reference_situation_analytical_formula_delta_wrt_N]{\ref*{eq:GAE_reference_situation_analytical_formula_delta_wrt_N}}}) for the success probability that the error bound is fulfilled. 
It is not apparent that this formula could be inverted to obtain a formula for the required number of runs according to $ N({\tilde{n}}, {\epsilon}, {\delta}) $ but the derived relation, corresponding to $ {\delta}({\tilde{n}}, N, {\epsilon}) $, should allow in principle to do an analytical study of the accuracy of the extracted probability distribution, e.g. to additionally verify that the empirically found shot numbers are of the correct order of magnitude. 
However, due to the growing number of contributions in it with $ {\tilde{n}} $ and $ N $, for which many products involving factorials and  $ {{\tilde{n}}}^{N} $ have to be computed, even its application to just moderately larger $ {\tilde{n}} $ or $ N $ apparently demands extended time and computing resources, which is why such a study is left here as a next step for further work as well.

\section{A Novel Encoding Scheme for Treating Generic Problems according to the Lattice Boltzmann Method}
\label{sec:New_Encoding_for_LBM}

Having the perspective on the applicability of QC for CFD, we propose in this section a new design for a lattice Boltzmann quantum algorithm, based on the presented estimations of the efforts implied for different encoding schemes and the considerations in the appendix. 
For an explanation of the LBM, we point to \cite{Succi_Book_long, Succi_Book_short, Krueger_et_al_Book} and for a review on QC approaches to implement the LBM, we refer to the introduction section of \cite{Lacatus_Moeller_LB_col_op_arxiv_v2}.

The concept of our algorithm is derived based on the idea to set a {\emph{bottom level}} of encoding according to bitstring encoding to realize the in general non-linear collision operation like presented in appendix {\hyperref[App_subsec:Representing_Non-lin_Functions]{\ref*{App_subsec:Representing_Non-lin_Functions}}} and a {\emph{top level}} of encoding similar to amplitude encoding for an additional qubit register for the implementation of the linear streaming operation. 
The latter would however not store relevant data in the amplitudes but is just set to generate enough basis states to label the set grid points. 
The bottom level would then represent one realization of a voxel, i.e. the collision process taking place locally at a grid point, and thus involve just the register of the bottom level, whereas the top level would encode multiple realizations of the system of the bottom level, corresponding to the multiple voxels in the system, for which the connectivity of the voxels including boundary conditions for the non-local (but linear) streaming step has to be implemented. 
As a simple example for the illustration of this preliminary idea, the situation of two distribution functions $ f_0 $ and $ f_1 $ that are set for two grid points $ x_0 $ and $ x_1 $, respectively, can be considered: 
The treated state would be of the structure
\begin{align}
 {a_0} {\left| 0 \right\rangle} \otimes { \underbrace{ {\left| {f_0}({x_0}) \right\rangle} \otimes {\left| {f_1}({x_0}) \right\rangle} }_{ \text{bitstrings} } } + {a_1} {\left| 1 \right\rangle} \otimes { \underbrace{ {\left| {f_0}({x_1}) \right\rangle} \otimes {\left| {f_1}({x_1}) \right\rangle} }_{ \text{bitstrings} } } ,
 \label{eq:New_enc_for_LBM_original_idea_structure_state}
\end{align}
where $ {\left| {f_0}({x_i}) \right\rangle} \otimes {\left| {f_1}({x_i}) \right\rangle} $ shall also include ancilla qubits to realize the non-linear collision operation according to the procedure in appendix {\hyperref[App_subsec:Representing_Non-lin_Functions]{\ref*{App_subsec:Representing_Non-lin_Functions}}}. 
The state ({\hyperref[eq:New_enc_for_LBM_original_idea_structure_state]{\ref*{eq:New_enc_for_LBM_original_idea_structure_state}}}) is a pure entangled state except for the case that it is $ {f_0}({x_0}) = {f_0}({x_1}) $ and $ {f_1}({x_0}) = {f_1}({x_1}) $. 
The application of the collision operation at the register that holds the qubits for the parts $ {\left| {f_0}({x_i}) \right\rangle} \otimes {\left| {f_1}({x_i}) \right\rangle} $ would implement the collision operation for all voxels in parallel, yielding the function values after the collision $ {\hat{f}} $. The streaming would then mean a swapping of product state parts $ {\left| {f_j}({x_i}) \right\rangle} $ between the individual parts of the linear combination that forms the entangled state like
\begin{align}
 & {a_0} { | 0 \rangle} \otimes { { { | {{\hat{f}}_0}({x_0}) \rangle} \otimes { | {{\hat{f}}_1}({x_0}) \rangle} } } \nonumber \\
 & ~ + {a_1} { | 1 \rangle} \otimes { { { | {{\hat{f}}_0}({x_1}) \rangle} \otimes { | {{\hat{f}}_1}({x_1}) \rangle} } } \nonumber \\[0.15cm]
 \longrightarrow ~~ & {a_0} { | 0 \rangle} \otimes { { { | {{\hat{f}}_0}({x_0}) \rangle} \otimes { | {{\hat{f}}_1}({x_1}) \rangle} } } \nonumber \\
 & ~ + {a_1} { | 1 \rangle} \otimes { { {| {{\hat{f}}_0}({x_1}) \rangle} \otimes { | {{\hat{f}}_1}({x_0}) \rangle} } }
 .
 \label{eq:New_enc_for_LBM_original_idea_streaming}
\end{align}

However, in the form of the algorithm up to this point, there are two problematic aspects:

The first is a minor one and concerns the implementation of the collision step. 
For the collision step, it is to note that it cannot be implemented using reset-operations, which was mentioned in appendix {\hyperref[App_subsec:Representing_Non-lin_Functions]{\ref*{App_subsec:Representing_Non-lin_Functions}}} as an option if it is operated on a classical state, i.e. not on a superposition state. 
For a superposition of multiple classical states like ({\hyperref[eq:New_enc_for_LBM_original_idea_structure_state]{\ref*{eq:New_enc_for_LBM_original_idea_structure_state}}}), a reset of an ancilla qubit, which should be based on a measurement of it, would in general destroy this state since it would be collapsed to the corresponding part of the superposition. 
If a collision operation shall be applied $ t $ times, it is therefore necessary to augment the bottom level register with $ t $ times the needed number of ancilla qubits for one application according to appendix {\hyperref[App_subsec:Representing_Non-lin_Functions]{\ref*{App_subsec:Representing_Non-lin_Functions}}}, so that an unused set of ancilla qubits can be used for each application. 
So, a part of the linear combination of classical states would have the form
\begin{align}
  {a_i} {\left| i \right\rangle} & \otimes {\left| {f_0}({x_i}) \right\rangle} \otimes {\left| {f_1}({x_i}) \right\rangle} \otimes \dots \nonumber \\
 & \otimes {\underbrace{ {\left| {\tilde{a}} \right\rangle} \otimes {\left| {\tilde{a}} \right\rangle}  \otimes \dots \otimes {\left| {\tilde{a}} \right\rangle} }_{ t \text{ times} }}
 \label{eq:New_enc_for_LBM_ancillas_for_col_step_multiple_times}
\end{align}
where now, in contrast to ({\hyperref[eq:New_enc_for_LBM_original_idea_structure_state]{\ref*{eq:New_enc_for_LBM_original_idea_structure_state}}}), the $ {\left| {f_0}({x_i}) \right\rangle} \otimes {\left| {f_1}({x_i}) \right\rangle} \otimes \dots $ in this expression do not include ancilla qubits since a set of ancilla qubits to implement one collision operation is denoted by $ {\left| {\tilde{a}} \right\rangle} $. 
According to the procedure in appendix {\hyperref[App_subsec:Representing_Non-lin_Functions]{\ref*{App_subsec:Representing_Non-lin_Functions}}}, the maximum number of qubits that is needed in $ {\left| {\tilde{a}} \right\rangle} $ should be equal to the sum of the qubits that are respectively chosen to resolve the values for the individual $ f_j $. 
In this case, a set of ancilla qubits would be effectively used to store the values of the $ f_j $ before the collision step.

However, the second problematic aspect of the sketched algorithm concept is a fundamental problem and concerns the streaming step. 
The streaming step according to an operation in the form like ({\hyperref[eq:New_enc_for_LBM_original_idea_streaming]{\ref*{eq:New_enc_for_LBM_original_idea_streaming}}}) is {\bf{impossible}} to implement in a quantum computer since it is a non-linear operation. 
To show this, the minimal situation of exchanging the states of a qubit according to $ | 0 \rangle $ or $ | 1 \rangle $ in a linear combination of bitstring states is considered. 
It shall be realized
\begin{align}
 {a} {| 0 \rangle} {| {\alpha} \rangle} {| {\beta} \rangle} + {b} {| 1 \rangle} {| {\gamma} \rangle} {| {\delta} \rangle}  ~ & \overset{!}{\longrightarrow} ~ {a} {| 0 \rangle} {| {\alpha} \rangle} {| {\delta} \rangle} + {b} {| 1 \rangle} {| {\gamma} \rangle} {| {\beta} \rangle} 
 \label{eq:New_enc_for_LBM_exchange_qubit_states_ket_notation}
\end{align}
for all $ {\alpha}, {\beta}, {\gamma}, {\delta} \in \{ 0, 1 \} $. 
For simplicity, $ a = b $ is considered and these amplitudes are furthermore neglected, i.e. set to $ a = b = 1 $, in the following consideration since they just contribute as normalization prefactors but do not change the qualitative reasoning. 
The statevector operation in the notation ({\hyperref[eq:New_enc_for_LBM_exchange_qubit_states_ket_notation]{\ref*{eq:New_enc_for_LBM_exchange_qubit_states_ket_notation}}}) corresponds to the explicit vector notation
\begin{align}
 \begin{pmatrix}
 ( 1 - {\alpha} ) ( 1 - {\beta} )  \\
 ( 1 - {\alpha} ) {\beta} \\
 {\alpha} ( 1 - {\beta} ) \\
 {\alpha} {\beta} \\
 ( 1 - {\gamma} ) ( 1 - {\delta} )  \\
 ( 1 - {\gamma} ) {\delta} \\
 {\gamma} ( 1 - {\delta} ) \\
 {\gamma} {\delta}
\end{pmatrix}
  ~ & \overset{!}{\longrightarrow} ~ 
  \begin{pmatrix}
 ( 1 - {\alpha} ) ( 1 - {\delta} )  \\
 ( 1 - {\alpha} ) {\delta} \\
 {\alpha} ( 1 - {\delta} ) \\
 {\alpha} {\delta} \\
 ( 1 - {\gamma} ) ( 1 - {\beta} )  \\
 ( 1 - {\gamma} ) {\beta} \\
 {\gamma} ( 1 - {\beta} ) \\
 {\gamma} {\beta}
\end{pmatrix} .
 \label{eq:New_enc_for_LBM_exchange_qubit_states_vector_notation}
\end{align}
This represents a demanded mapping of input vectors to output vectors, where specifically, just a re-assigning of the $ 16 $ input vectors is performed, which is documented in Table {\hyperref[Table:New_enc_for_LBM_exchange_qubit_states_mapping_vectors]{\ref*{Table:New_enc_for_LBM_exchange_qubit_states_mapping_vectors}}}.

\begin{table}[t]
\begin{center}
\begin{tabular}[c]{r|r|r}
 \multicolumn{1}{ c }{ input bitstring $ {\alpha} {\beta} {\gamma} {\delta} $ } & \multicolumn{1}{ |c }{ input index } & \multicolumn{1}{|c}{ output index } \\ \hline 
  \multicolumn{1}{ c }{ $0000$ } & \multicolumn{1}{ |c }{ $1$ } & \multicolumn{1}{|c}{ $1$ } \\ \hline
  \multicolumn{1}{ c }{ $0001$ } & \multicolumn{1}{ |c }{ $2$ } & \multicolumn{1}{|c}{ $5$ } \\ \hline 
  \multicolumn{1}{ c }{ $0010$ } & \multicolumn{1}{ |c }{ $3$ } & \multicolumn{1}{|c}{ $3$ } \\ \hline
  \multicolumn{1}{ c }{ $0011$ } & \multicolumn{1}{ |c }{ $4$ } & \multicolumn{1}{|c}{ $7$ } \\ \hline
  \multicolumn{1}{ c }{ $0100$ } & \multicolumn{1}{ |c }{ $5$ } & \multicolumn{1}{|c}{ $2$ } \\ \hline
  \multicolumn{1}{ c }{ $0101$ } & \multicolumn{1}{ |c }{ $6$ } & \multicolumn{1}{|c}{ $6$ } \\ \hline 
  \multicolumn{1}{ c }{ $0110$ } & \multicolumn{1}{ |c }{ $7$ } & \multicolumn{1}{|c}{ $4$ } \\ \hline
  \multicolumn{1}{ c }{ $0111$ } & \multicolumn{1}{ |c }{ $8$ } & \multicolumn{1}{|c}{ $8$ } \\ \hline
  \multicolumn{1}{ c }{ $1000$ } & \multicolumn{1}{ |c }{ $9$ } & \multicolumn{1}{|c}{ $9$ } \\ \hline
  \multicolumn{1}{ c }{ $1001$ } & \multicolumn{1}{ |c }{ $10$ } & \multicolumn{1}{|c}{ $13$ } \\ \hline 
  \multicolumn{1}{ c }{ $1010$ } & \multicolumn{1}{ |c }{ $11$ } & \multicolumn{1}{|c}{ $11$ } \\ \hline
  \multicolumn{1}{ c }{ $1011$ } & \multicolumn{1}{ |c }{ $12$ } & \multicolumn{1}{|c}{ $15$ } \\ \hline
  \multicolumn{1}{ c }{ $1100$ } & \multicolumn{1}{ |c }{ $13$ } & \multicolumn{1}{|c}{ $10$ } \\ \hline
  \multicolumn{1}{ c }{ $1101$ } & \multicolumn{1}{ |c }{ $14$ } & \multicolumn{1}{|c}{ $14$ } \\ \hline 
  \multicolumn{1}{ c }{ $1110$ } & \multicolumn{1}{ |c }{ $15$ } & \multicolumn{1}{|c}{ $12$ } \\ \hline
  \multicolumn{1}{ c }{ $1111$ } & \multicolumn{1}{ |c }{ $16$ } & \multicolumn{1}{|c}{ $16$ } \\ 
\end{tabular}
\end{center}
\captionsetup{justification=raggedright, singlelinecheck=false}
\caption[]{Explicit inputs and outputs of the mapping according to ({\hyperref[eq:New_enc_for_LBM_exchange_qubit_states_ket_notation]{\ref*{eq:New_enc_for_LBM_exchange_qubit_states_ket_notation}}}) or ({\hyperref[eq:New_enc_for_LBM_exchange_qubit_states_vector_notation]{\ref*{eq:New_enc_for_LBM_exchange_qubit_states_vector_notation}}}), given in terms of an associated index for the input and the corresponding index of the resulting output.}
\label{Table:New_enc_for_LBM_exchange_qubit_states_mapping_vectors}
\end{table}

In order to derive an implementation of this mapping as a quantum circuit, i.e. as a linear operation, the task is to derive a corresponding matrix that realizes this mapping w.r.t. the computational basis, given by the unit vectors, since the statevector is referred to this basis. 
Here, $ 8 $ linearly independent vectors are needed to give a basis for the vector space formed by the $ {2^3} = 8 $ computational basis states that can be formed by the $ 3 $ considered qubits. 
However, for the set of the $ 16 $ vectors given by ({\hyperref[eq:New_enc_for_LBM_exchange_qubit_states_vector_notation]{\ref*{eq:New_enc_for_LBM_exchange_qubit_states_vector_notation}}}), the found number of linearly independent vectors is only $ 7 $. 
To deduce the matrix to be implemented as a quantum circuit, the further procedure should be now to construct a matrix that mediates ({\hyperref[eq:New_enc_for_LBM_exchange_qubit_states_vector_notation]{\ref*{eq:New_enc_for_LBM_exchange_qubit_states_vector_notation}}}) for the subspace spanned by these $ 16 $ vectors by formulating it for a basis that includes $ 7 $ linearly independent vectors of them and determine then the basis change matrix that transforms between this chosen basis and the basis of the unit vectors. 
Here specifically, the set of the linearly independent vectors according to ({\hyperref[eq:New_enc_for_LBM_exchange_qubit_states_vector_notation]{\ref*{eq:New_enc_for_LBM_exchange_qubit_states_vector_notation}}}) for the indices $ 1, 2, 5, 8, 10, 11 $ and $ 13 $ as in Table {\hyperref[Table:New_enc_for_LBM_exchange_qubit_states_mapping_vectors]{\ref*{Table:New_enc_for_LBM_exchange_qubit_states_mapping_vectors}}} is chosen (in this order), which is then completed to a basis by adding the eighth unit vector $ {(0, 0, 0, 0, 0, 0, 0, 1)}^{T} $ as the eighth basis vector, for which it is furthermore arbitrarily defined that it shall be mapped to itself. 
In this chosen basis, denoted as $ B $ the mapping ({\hyperref[eq:New_enc_for_LBM_exchange_qubit_states_vector_notation]{\ref*{eq:New_enc_for_LBM_exchange_qubit_states_vector_notation}}}) implies the representation matrix
\begin{align}
 {D_{B, B}} &= 
  \begin{pmatrix}
 1 & 0 & 0 & 0 & 0 & 0 & 0 & 0  \\
 0 & 0 & 1 & 0 & 0 & 0 & 0 & 0  \\ 
 0 & 1 & 0 & 0 & 0 & 0 & 0 & 0  \\
 0 & 0 & 0 & 1 & 0 & 0 & 0 & 0  \\
 0 & 0 & 0 & 0 & 0 & 0 & 1 & 0  \\
 0 & 0 & 0 & 0 & 0 & 1 & 0 & 0  \\
 0 & 0 & 0 & 0 & 1 & 0 & 0 & 0  \\
 0 & 0 & 0 & 0 & 0 & 0 & 0 & 1  
\end{pmatrix} .
 \label{eq:New_enc_for_LBM_exchange_qubit_states_mapping_deduced_matrix}
\end{align}
For the vector $ {( 1, 0, 0, 0, 0, 0, 1, 0 )}^{T} $, i.e. the vector according to the index $ 3 $ in Table {\hyperref[Table:New_enc_for_LBM_exchange_qubit_states_mapping_vectors]{\ref*{Table:New_enc_for_LBM_exchange_qubit_states_mapping_vectors}}}, it was found that it is the linear combination according to the representation
\begin{align}
  \begin{pmatrix}
 0 \\
 1 \\
 0 \\
 0 \\
 -1 \\
 1 \\
 0 \\
 0 
 \end{pmatrix} 
 \label{eq:New_enc_for_LBM_exchange_qubit_states_contradiction_example_vector}
\end{align}
in the basis $ B $, for which the application of $ {D_{B, B}} $ yields
\begin{align}
 {D_{B, B}} \cdot 
  \begin{pmatrix}
 0 \\
 1 \\
 0 \\
 0 \\
 -1 \\
 1 \\
 0 \\
 0 
 \end{pmatrix} 
 &= 
   \begin{pmatrix}
 0 \\
 0 \\
 1 \\
 0 \\
 0 \\
 1 \\
 -1 \\
 0 
 \end{pmatrix} .
 \label{eq:New_enc_for_LBM_exchange_qubit_states_contradiction_mapping_matrix}
\end{align}
However, the mapping ({\hyperref[eq:New_enc_for_LBM_exchange_qubit_states_vector_notation]{\ref*{eq:New_enc_for_LBM_exchange_qubit_states_vector_notation}}}) actually demands that the vector with index $ 3 $ according to Table {\hyperref[Table:New_enc_for_LBM_exchange_qubit_states_mapping_vectors]{\ref*{Table:New_enc_for_LBM_exchange_qubit_states_mapping_vectors}}} is mapped to itself, which is a contradiction and therefore means that ({\hyperref[eq:New_enc_for_LBM_exchange_qubit_states_vector_notation]{\ref*{eq:New_enc_for_LBM_exchange_qubit_states_vector_notation}}}) cannot be represented as a matrix and is thus a non-linear operation. 
This is rooted to the fact that the new value that shall be assigned to a qubit in a part of the linear combination depends on the value of the same qubit in the other part of the linear combination.

The importance of the chosen encoding scheme for Boltzmann methods has already been addressed in \cite{Schalkers_Moeller_Encoding_BMs}. 
The consideration presented here is in accordance with the discussion given there and represents an extension of it: 
Specifically, the discussed preliminary encoding scheme ({\hyperref[eq:New_enc_for_LBM_original_idea_structure_state]{\ref*{eq:New_enc_for_LBM_original_idea_structure_state}}}) corresponds to the encoding format considered in \cite{Schalkers_Moeller_Encoding_BMs} in the section 3.2 -- {\emph{Computational basis state encoding}}. 
However, in \cite{Schalkers_Moeller_Encoding_BMs}, it was only shown that the streaming operation cannot be implemented as a unitary operation. 
Here, it is shown that it should in fact be even not linear and thus impossible to implement.

Based on the discussion given in \cite{Schalkers_Moeller_Encoding_BMs}, M. Schalkers and M. Möller also propose there a new algorithm and making a modification to our encoding format  ({\hyperref[eq:New_enc_for_LBM_original_idea_structure_state]{\ref*{eq:New_enc_for_LBM_original_idea_structure_state}}}) via just the same technique that they apply to realize the streaming should also allow to cure the problem of the streaming in our approach: 
The procedure of Schalkers and Möller is to augment the part for the grid point $ i $ in the linear combination with all the information that is needed to compute the quantities at this point after $ t $ time steps, which means that not only the initial quantities of the considered grid point are stored in the respective part of the linear combination but also the information about neighboring points required for the computation w.r.t. grid point $ i $ has to be included in the bitstring state and updated in the computation. 
The latter is done by applying the circuit that implements the collision step also to the sets of qubits in the extended bitstring state that encode the distribution functions $ f_j $ for these other grid points in the vicinity of point $ i $. 
In this modified encoding format, the streaming can then be implemented simply via SWAP-gates (cf. \cite{Schalkers_Moeller_Encoding_BMs}). 
So, the bitstring state holds the information about a stencil, which has to be extended if more time steps shall be computed without subsequent measuring and re-initializing.

E.g., for a 1D simulation with periodic boundary conditions and use of the D1Q3-stencil, ({\hyperref[eq:New_enc_for_LBM_ancillas_for_col_step_multiple_times]{\ref*{eq:New_enc_for_LBM_ancillas_for_col_step_multiple_times}}}) would have to be modified to the form
\begin{align}
  {a_i} {\left| i \right\rangle} & \otimes {\left| {f_0}({x_{i-1}}) \right\rangle} \otimes {\left| {f_1}({x_{i-1}}) \right\rangle} \otimes {\left| {f_2}({x_{i-1}}) \right\rangle} \otimes {\left| {\tilde{a}} \right\rangle}  \nonumber \\
  & \otimes {\left| {f_0}({x_i}) \right\rangle} \otimes {\left| {f_1}({x_i}) \right\rangle} \otimes {\left| {f_2}({x_i}) \right\rangle} \otimes {\left| {\tilde{a}} \right\rangle} \nonumber \\
 & \otimes  {\left| {f_0}({x_{i+1}}) \right\rangle} \otimes {\left| {f_1}({x_{i+1}}) \right\rangle} \otimes {\left| {f_2}({x_{i+1}}) \right\rangle} \otimes {\left| {\tilde{a}} \right\rangle}  
 \label{eq:New_enc_for_LBM_example_D1Q3_1_time_steps}
\end{align}
to be able to implement one time step and to the form
\begin{align}
  {a_i} {\left| i \right\rangle} & \otimes {\left| {f_0}({x_{i-2}}) \right\rangle} \otimes {\left| {f_1}({x_{i-2}}) \right\rangle} \otimes {\left| {f_2}({x_{i-2}}) \right\rangle} \otimes {\left| {\tilde{a}} \right\rangle} \otimes {\left| {\tilde{a}} \right\rangle}  \nonumber \\
  & \otimes  {\left| {f_0}({x_{i-1}}) \right\rangle} \otimes {\left| {f_1}({x_{i-1}}) \right\rangle} \otimes {\left| {f_2}({x_{i-1}}) \right\rangle} \otimes {\left| {\tilde{a}} \right\rangle} \otimes {\left| {\tilde{a}} \right\rangle} \nonumber \\
  & \otimes {\left| {f_0}({x_i}) \right\rangle} \otimes {\left| {f_1}({x_i}) \right\rangle} \otimes {\left| {f_2}({x_i}) \right\rangle} \otimes {\left| {\tilde{a}} \right\rangle} \otimes {\left| {\tilde{a}} \right\rangle} \nonumber \\
 & \otimes  {\left| {f_0}({x_{i+1}}) \right\rangle} \otimes {\left| {f_1}({x_{i+1}}) \right\rangle} \otimes {\left| {f_2}({x_{i+1}}) \right\rangle} \otimes {\left| {\tilde{a}} \right\rangle} \otimes {\left| {\tilde{a}} \right\rangle} \nonumber \\
 & \otimes  {\left| {f_0}({x_{i+2}}) \right\rangle} \otimes {\left| {f_1}({x_{i+2}}) \right\rangle} \otimes {\left| {f_2}({x_{i+2}}) \right\rangle} \otimes {\left| {\tilde{a}} \right\rangle} \otimes {\left| {\tilde{a}} \right\rangle} 
 \label{eq:New_enc_for_LBM_example_D1Q3_2_time_steps}
\end{align}
to be able to implement two time steps (cf. visualization in Fig. {\hyperref[Fig:New_enc_for_LBM_example_D1Q3_time_steps]{\ref*{Fig:New_enc_for_LBM_example_D1Q3_time_steps}}}). 
In general, for a 1D simulation with a D1Q$q$-voxel, the required number of qubits for the subsequent computation of $ t $ time steps would be
\begin{align}
 & { \left\lceil { {\log}_{2}{({N_x})} } \right\rceil } + {\sum_{j}^{q}} {Q_{f_j}} \nonumber \\
 & \quad \cdot { ( 1 + {\underbrace{ \left[ \text{contribution due to the ancilla qubits} \right] }_{ \text{maximally } t }} ) } \nonumber \\
 & \quad \cdot ( 1 + 2t )
\end{align}
where $ {N_x} $ is the number of grid points, $ {Q_{f_j}} $ is the number of qubits used to resolve the values for the distribution function $ f_j $ and $ \left\lceil \dots \right\rceil $ denotes the ceiling function. 
Here, the last factor $ (1 + 2t) $ accounts for the addition of bitstrings for the quantities of the needed points in the vicinity. 
According to the presented consideration, the number of required qubits scales also logarithmically with the number of grid points for higher dimensions.

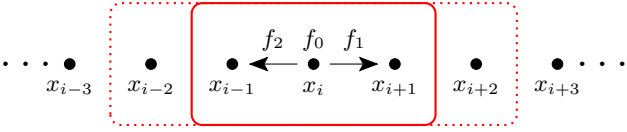
\begin{figure}[t]
\centering
\begin{tikzpicture}
\filldraw[black] ({0.0\linewidth}, {0.0\linewidth}) circle (2pt);
\filldraw[black] ({0.125\linewidth}, {0.0\linewidth}) circle (2pt);
\filldraw[black] ({0.25\linewidth}, {0.0\linewidth}) circle (2pt);
\filldraw[black] ({0.375\linewidth}, {0.0\linewidth}) circle (2pt);
\filldraw[black] ({-0.125\linewidth}, {0.0\linewidth}) circle (2pt);
\filldraw[black] ({-0.25\linewidth}, {0.0\linewidth}) circle (2pt);
\filldraw[black] ({-0.375\linewidth}, {0.0\linewidth}) circle (2pt);
\node at ({1.2*0.375\linewidth}, {0.0\linewidth}) {\LARGE{$ \dots $}};
\node at ({-1.165*0.375\linewidth}, {0.0\linewidth}) {\LARGE{$ \dots $}};
\draw [arrows = {-Stealth[length=7.5pt, inset=1.5pt]}] ({0.2*0.125\linewidth}, {0.0*0.25\linewidth}) -- ({0.8*0.125\linewidth}, {0.0*0.25\linewidth});
\draw [arrows = {-Stealth[length=7.5pt, inset=1.5pt]}] ({-0.2*0.125\linewidth}, {-0.0*0.25\linewidth}) -- ({-0.8*0.125\linewidth}, {-00*0.25\linewidth});
\node at ({0.0*0.25\linewidth}, {-0.15*0.25\linewidth}) {{$ x_{i} $}};
\node at ({0.125\linewidth}, {-0.15*0.25\linewidth}) {{$ x_{i+1} $}};
\node at ({0.25\linewidth}, {-0.15*0.25\linewidth}) {{$ x_{i+2} $}};
\node at ({0.375\linewidth}, {-0.15*0.25\linewidth}) {{$ x_{i+3} $}};
\node at ({-0.125\linewidth}, {-0.15*0.25\linewidth}) {{$ x_{i-1} $}};
\node at ({-0.25\linewidth}, {-0.15*0.25\linewidth}) {{$ x_{i-2} $}};
\node at ({-0.375\linewidth}, {-0.15*0.25\linewidth}) {{$ x_{i-3} $}};
\node at ({0.0*0.125\linewidth}, {0.15*0.25\linewidth}) {{$ {f_0} $}};
\node at ({0.5*0.125\linewidth}, {0.15*0.25\linewidth}) {{$ {f_1} $}};
\node at ({-0.5*0.125\linewidth}, {0.15*0.25\linewidth}) {{$ {f_2} $}};
\draw[rounded corners, dotted, red, thick] ({-2.5*0.125\linewidth}, {-0.75*0.125\linewidth}) rectangle ({2.5*0.125\linewidth}, {0.75*0.125\linewidth});
\draw[rounded corners, red, thick] ({-1.5*0.125\linewidth}, {-0.75*0.125\linewidth}) rectangle ({1.5*0.125\linewidth}, {0.75*0.125\linewidth});
\end{tikzpicture}
\captionsetup{justification=raggedright, singlelinecheck=false}
\caption[]{Illustration of the subproblems in the LBM that are computed in parallel in the proposed algorithm (cf. \cite{Schalkers_Moeller_Encoding_BMs}). The situation is visualized for the example of a 1D simulation using the D1Q3 stencil, which is indicated for the grid point $ i $. 
In the streaming step, the portion $ f_0 $ of the fluid density remains at the point $ i $, while the portions $ f_1 $ and $ f_2 $ are streamed to the right and left neighboring point, respectively. 
The region of points from which information is needed to compute one time step for the central point $ i $ is framed in red and the needed extension of this region for the computation of two time steps is marked by the red dotted frame.}
\label{Fig:New_enc_for_LBM_example_D1Q3_time_steps}
\end{figure}

In this sense, our presented new algorithm could be regarded as an extension of the algorithm proposed in \cite{Schalkers_Moeller_Encoding_BMs}. 
However, we point out that it should be a substantial extension because the algorithm of \cite{Schalkers_Moeller_Encoding_BMs} implements only the lattice gas automaton representation of the LBM, which allowed it also to implement the collision step as a unitary operation. 
However, instead of using just the qubit states $ | 0 \rangle $ and $ | 1 \rangle $ to represent occupation numbers according to $ 0 $ and $ 1 $, we explicitly employ bitstring encoding at the bottom level of encoding to realize the full LBM, i.e. including non-linear collision operators. 
For this, it is to note that this is achieved also in our algorithm without the need of a probabilistic operation like an LCU-procedure but by using that the approximation of an arbitrary function can be implemented for this classical encoding format according to appendix {\hyperref[App_subsec:Representing_Non-lin_Functions]{\ref*{App_subsec:Representing_Non-lin_Functions}}} as a unitary operation at least w.r.t. a register with added ancilla qubits.

The output result of this algorithm would be given in general by one part of the set linear combination of basis states and hence the complete set of the distribution functions $ f_j $ treated in the bottom level for the considered grid point, for which the following 3 features are to be noted:
\begin{itemize}
\item[(i)] 
The probability to obtain the result for a specific grid point would be given by the amplitudes $ a_i $ of the top level register and could accordingly be chosen since they store no information about the flow quantities.
\item[(ii)] 
The region that has to be considered around a grid point for the subproblem that one part of the overall linear combination computes can be further extended in a variable manner, so that a part of the linear combination yields the resulting quantities for more than just a single grid point.
\item[(iii)] 
The algorithm can be run in two modes, depending on whether the correlation analysis described in appendix {\hyperref[App_subsec:Representing_Non-lin_Functions]{\ref*{App_subsec:Representing_Non-lin_Functions}}} is applied for the collision circuit. If so, as shown in appendix {\hyperref[App_subsec:Representing_Non-lin_Functions]{\ref*{App_subsec:Representing_Non-lin_Functions}}}, in general less ancilla qubits than the number of qubits set to encode a set of $ f_j $-values are required and furthermore, only the information of one specific $ f_j $ is needed from the outermost points of the considered region around a grid point, so that this mode represents a resource saving mode. 
Furthermore, it is to note that the circuits for streaming and collision have to be applied in a specific time step only to those subsets of qubits in the bottom level register that represent grid points for which the correct information is present for the computation. 
For the outermost points of the chosen extended region around the central grid point, the streaming cannot be performed correctly in the first time step and likewise also for the grid points next to them in the next time step and so forth, so that an application of streaming and collision in the computation of time step $ t $ is meaningless for points that are $ t - 1 $ steps away from the outermost points of the considered regions around the central grid point. 
So, a cascade around the central point (or points in the case that the region was chosen such that the quantities of multiple points are correctly recorded at the final time step according to (ii)) w.r.t. the correct information results for going back in time starting from the final time. 
For this mode, just the $ f_j $ at the final time $ t $ of the central point (or the multiple points for which it is set that the information is correctly propagated in the $ t $ time steps) are given explicitly in the resulting output bitstring. 
In contrast, if the correlation analysis is not performed and the same number of ancilla qubits as the number of qubits to represent the $ f_j $ is used for the collision circuit, such a set of ancilla qubits can be used to directly store copies of the $ f_j $ before a collision step. 
Thus, in this mode, not only the information about the central grid points at the final time $ t $ is given out in explicit form but also the information before the collision operations for all time steps plus such history information about the other included points according to the described information cascade can be read out with one run of the algorithm, as it could be desired e.g. for CAA.
\end{itemize}

Concerning the expected runtime scaling, it is to note that the collision step as well as the streaming step of the LBM are performed in parallel for all grid points, respectively, where the depths of the circuits for collision and streaming do not depend on the number of grid points. 
However, an efficient procedure for initializing a superposition of specific basis states according to the structure ({\hyperref[eq:New_enc_for_LBM_original_idea_structure_state]{\ref*{eq:New_enc_for_LBM_original_idea_structure_state}}})has to be elaborated. 
Even if the algorithm should allow the subsequent computation of multiple time steps, this setup is considered appropriate in particular for simulations in which a turbulent flow field shall be read out at each time step and taken as input for the next step, like e.g. in CAA, as it was deduced in this work that for amplitude encoding, the circuit depth for initializing an arbitrary state scales linearly with the number of input values $ \tilde{n} $ and reading an arbitrary state out should scale at least like $ {{\tilde{n}}} ~ {\ln({\tilde{n}})} $, whereas here, there is the freedom to set the probability amplitudes for obtaining the results for the individual grid points. 
An elaborated analysis of this algorithm is subject of ongoing work.

\section{Conclusion}
\label{sec:Conclusion}

\begin{table*}[t]
\begin{center}
\begin{tabular}[c]{l|c|c|c}
 \multicolumn{1}{ c| }{ } & \multicolumn{1}{ c|}{ general amplitude encoding } & \multicolumn{1}{c|}{ 1-qubit amplitude encoding } & \multicolumn{1}{c}{ bitstring encoding } \\[0.1cm] \hline
 \multicolumn{1}{ l| }{  } & \multicolumn{1}{ l| }{ general superposition state, } & \multicolumn{1}{ l| }{ no general superposition state } & \multicolumn{1}{ l }{ no superposition state } \\ 
 \multicolumn{1}{ l| }{ description } & \multicolumn{1}{ l| }{ i.e., entanglement can be } & \multicolumn{1}{ l| }{ but product state, i.e. no } & \multicolumn{1}{  }{  } \\ 
  &  \multicolumn{1}{ l| }{ present }     &  \multicolumn{1}{ l| }{ entanglement } &   \\[0.1cm] \hline
 \multicolumn{1}{l|}{ circuit depth contribution of }    
 &  &  &    \\ 
 \multicolumn{1}{l|}{ 1-qubit rotation gates in the }  & $ 2^{n+1} - (n+1) $  & $ 2 $ & $ 1 $   \\
 \multicolumn{1}{l|}{ procedure for initializing a }  &  &  &    \\
 \multicolumn{1}{l|}{ general state in the }  &  &  &    \\
 \multicolumn{1}{l|}{ respective encoding }  &  &  &    \\[0.1cm] \hline
 \multicolumn{1}{l|}{ circuit depth contribution of }  &  &  &  \\  
 \multicolumn{1}{l|}{ CX-gates in the procedure }  
 & $ 2^{n+1} - 2(n+1)  $ & $ 0 $ & $ 0 $  \\
   \multicolumn{1}{l|}{ for initializing a general state }  &  &  &    \\
     \multicolumn{1}{l|}{ in the respective encoding }  &  &  &    \\[0.1cm] \hline
  \multicolumn{1}{l|}{ upper bound for the required }  &  &   \\
  \multicolumn{1}{l|}{ number of runs $ N $ of a }  & $ N \geq {\frac{1}{2 {{\epsilon}^2}}}  \ln{ \left( \frac{2 {({2^n} - 1)} }{\delta}  \right) } $  & $ N \geq {\frac{1}{2 {{\epsilon}^2}}}  \ln{ \left( \frac{2  }{\delta}  \right) } $ & $ 1 $  \\ 
  \multicolumn{1}{l|}{ quantum algorithm that }  & \multicolumn{1}{l|}{ with $ \epsilon $ as an absolute error }   & \multicolumn{1}{l|}{ with $ \epsilon $ as an absolute error } &  \\ 
 \multicolumn{1}{l|}{ produces a final state in the }  & \multicolumn{1}{l|}{ w.r.t. $1$, i.e. $ \epsilon \rightarrow \frac{ \epsilon }{ {2^{n}} } $ for a } & \multicolumn{1}{l|}{ w.r.t. $1$, i.e. $ \epsilon \rightarrow \frac{ \epsilon }{ {2} } $ for a }  &    \\
 \multicolumn{1}{l|}{ respective encoding }  & \multicolumn{1}{l|}{ relative error w.r.t. $ \frac{1}{2^n} $ } & \multicolumn{1}{l|}{ relative error w.r.t. $ \frac{1}{2} $ }  &   \\ 
\end{tabular}
\end{center}
\captionsetup{justification=raggedright, singlelinecheck=false}
\caption[]{Summary of the main points from section {\hyperref[sec:Theoretical_Considerations]{\ref*{sec:Theoretical_Considerations}}}. 
$ n $ denotes the number of qubits and $ \delta $ defines a bound for the probability that a basis state probability that was inferred by measurements differs from the real value by more than the error $ \epsilon $.}
\label{Fig_Table_Summary}
\end{table*}

This work provides a framework to quantify the required resources that are implied by different encoding schemes in QC, for which the main aspects are summarized in Table {\hyperref[Fig_Table_Summary]{\ref*{Fig_Table_Summary}}}. 
In particular, we are not aware that the upper bound for the required number of runs for general amplitude encoding was already stated somewhere in the literature before.

For the statements on general amplitude encoding, simulation studies with {\emph{Qiskit}} were performed, which are assessed as a verification of these statements since the graph of the execution times estimated from {\emph{Qiskit}} for IBM's 'fake backend' of their quantum computer {\emph{Sherbrooke}} matches the graph of the reference formula, where deviations like times for high optimization levels of the circuit compilation that lie slightly below the result of the derived reference formula can be explained, and the number of times for which the desired accuracy $ \epsilon $ of the probabilities read out from the derived number of runs $ N $ is not fulfilled was always relatively much below the expected value $ \delta $, where a decreasing of this number is indicated by the conducted simulation experiments w.r.t. a decrease of the desired error $ \epsilon $ or set probability bound $ \delta $ and increase of the number of qubits $ n $. 
The latter observation confirms the expectation that this derived sufficient number of runs is a quite conservative estimation. 
Therefore, it should be investigated whether more sophisticated approaches of probability theory can be found that allow to give a lower upper bound. 
In order to provide nevertheless a scaling that can be used for more accurate resource estimations, a simulation study for the reference situation of equal probabilities was performed, indicating a scaling similar to $ {\tilde{n}} ~ { \ln( {\tilde{n}} ) } $ w.r.t. the $ {\tilde{n}} = 2^{n} $ amplitudes available for storing values. 
Concerning further work on this issue, this empirical study could be extended to an examination of the dependency of $ {\delta} $ and $ {\epsilon} $ and also an investigation based on the analytical result of appendix {\hyperref[App_sec:Probability_Fulfilling_error_bound_for_reference_situation]{\ref*{App_sec:Probability_Fulfilling_error_bound_for_reference_situation}}} could be considered.

Subject of ongoing work is also the proposed algorithm for the LBM. 
To enable the treatment of non-linear problems, it resorts to the simultaneous processing of multiple states that correspond to the classical representation of data as bitstrings, respectively. 
However, an open task is to devise an efficient routine for initializing a state of this required structure. 
Regarding this aspect, the complexity of this quantum algorithm has to be compared then also to a corresponding classical realization of the used concept for parallelization, where it is operated just on one large bitstring.

\acknowledgements

This project was made possible by the DLR Quantum Computing Initiative and the Federal Ministry for Research, Technology and Space; \url{qci.dlr.de/projects/toquaflics}.

H. A. Kösel acknowledges helpful exchange with Aaron Nagel.

Furthermore, we acknowledge the provision of the {\emph{quantikz}}-package \cite{quantikz_arxiv_v1} for \LaTeX , which was used to create the quantum circuit diagrams in this work, and the software development kit {\emph{Qiskit}} from IBM \cite{qiskit}, which was used for this work in the version 1.2.2 together with the kits {\emph{qiskit\_aer}} in the version 0.15.1 and {\emph{qiskit\_ibm\_runtime}} in the version 0.30.0 for simulation studies of quantum circuits.

\appendix

\section{Remarks on Bitstring Encoding}
\label{App_sec:Remarks_on_Bitstring_Encoding}

In this appendix, two considerations on the use of bitstring encoding are stated, which are illustrated via a minimal example, respectively.

The first, given in subsection {\hyperref[App_subsec:Representing_Non-lin_Functions]{\ref*{App_subsec:Representing_Non-lin_Functions}}}, concerns its use to approximately represent non-linear functions. 
Secondly, subsection {\hyperref[App_subsec:A_Mapping_of_the_LA_Eq_to_the_BV-Q-Alg]{\ref*{App_subsec:A_Mapping_of_the_LA_Eq_to_the_BV-Q-Alg}}}, presents that the upwind solving procedure for the linear advection equation can be mapped to the Bernstein-Vazirani quantum algorithm, which is a quantum algorithm that uses bitstring encoding. 
For the latter, however, it shall be directly stated at this point that no advantage of the described implementation was found but it was nevertheless included here as a didactical example that illustrates that it depends crucially on the given type of problem whether an implementation in the framework of QC is more efficient than a classical one.

\subsection{Representing Non-linear Functions}
\label{App_subsec:Representing_Non-lin_Functions}

On the one hand, the evolution of a quantum system with time is mediated by the time-evolution operator, which is set as a unitary and thus linear operator. 
On the other hand, measurement processes correspond to Hermitian projection operators, which set the system state to an eigenstate of the operator for the measured observable that is associated to the eigenvalue that corresponds to the measurement result, and are thus also linear. 
Therefore, it should not be possible to define an operation that realizes a non-linear operation for an arbitrary quantum state, as e.g. taking the square of values that are encoded as the amplitudes of a qubit state. 
There are approaches to implement quantum circuits that realize non-linear operations for an arbitrary state according to {{general amplitude encoding}} via setting up copies of the state, for which then an interaction operation between these copies is applied  \cite{Jaksch_et_al_VQAs_for_CFD, Tennie_et_al_QC_non-lin_DEs, Lloyd_et_al_Q_alg_for_non-lin_DEs_arXiv_v2}. 
Regarding this, it is however to note that the several copies of the state have to be realized by setting up multiples of the quantum circuit that prepares the considered state since the no-cloning theorem forbids the existence of a general operation that copies an arbitrary quantum state to another register \cite{Nielsen_Chuang_Book}.

However, since bitstring encoding corresponds to the classical storing of data, the representation of non-linear functions of it can be implemented in a corresponding manner. 
Of course, it is to note that such a procedure is therefore not a quantum algorithm unless its implementation can be expressed as a circuit of lower depth by using H-gates, however, due to its relatively simple implementation, it should be considered as a work-around to easily handle non-linear operations in quantum algorithms.
\begin{table*}[t]
\begin{center}
\begin{tabular}[c]{r|r|r|r|r}
 \multicolumn{1}{ c }{ input bitstring } & \multicolumn{1}{ |c }{ index for input value $ x $ in units $ 2/7 $ } & \multicolumn{1}{|c}{ $ {x^2} $ in units $ 2/7 $ } & \multicolumn{1}{  |c }{ index that is the closest to $ {x^2} \cdot 7/2 $ } & \multicolumn{1}{ |c }{ output bitstring } \\ \hline 
  \multicolumn{1}{ r }{ $000$ } & \multicolumn{1}{ |r }{ $0$ } & \multicolumn{1}{|r}{ $0$ } & \multicolumn{1}{  |r }{ $0$ } & \multicolumn{1}{ |r }{ $000$ } \\ \hline
  \multicolumn{1}{ r }{ $001$ } & \multicolumn{1}{ |r }{ $1$ } & \multicolumn{1}{|r}{ $ \approx 0.286 $ } & \multicolumn{1}{  |r }{ $0$ } & \multicolumn{1}{ |r }{ $000$ } \\ \hline
  \multicolumn{1}{ r }{ $010$ } & \multicolumn{1}{ |r }{ $2$ } & \multicolumn{1}{|r}{ $ \approx 1.143 $ } & \multicolumn{1}{  |r }{ $1$ } & \multicolumn{1}{ |r }{ $001$ } \\ \hline
  \multicolumn{1}{ r }{ $011$ } & \multicolumn{1}{ |r }{ $3$ } & \multicolumn{1}{|r}{ $ \approx 2.571 $ } & \multicolumn{1}{  |r }{ $3$ } & \multicolumn{1}{ |r }{ $011$ } \\ \hline
  \multicolumn{1}{ r }{ $100$ } & \multicolumn{1}{ |r }{ $4$ } & \multicolumn{1}{|r}{ $ \approx 4.571 $ } & \multicolumn{1}{  |r }{ $5$ } & \multicolumn{1}{ |r }{ $101$ } \\ \hline
  \multicolumn{1}{ r }{ $101$ } & \multicolumn{1}{ |r }{ $5$ } & \multicolumn{1}{|r}{ $ \approx 7.143 $ } & \multicolumn{1}{  |r }{ $7$ } & \multicolumn{1}{ |r }{ $111$ } \\ \hline
  \multicolumn{1}{ r }{ $110$ } & \multicolumn{1}{ |r }{ $6$ } & \multicolumn{1}{|r}{ $ \approx 10.286 $ } & \multicolumn{1}{  |r }{ $7$ } & \multicolumn{1}{ |r }{ $111$ } \\ \hline
  \multicolumn{1}{ r }{ $111$ } & \multicolumn{1}{ |r }{ $7$ } & \multicolumn{1}{|r}{ $ 14 $ } & \multicolumn{1}{  |r }{ $7$ } & \multicolumn{1}{ |r }{ $111$ } \\
\end{tabular}
\end{center}
\captionsetup{justification=raggedright, singlelinecheck=false}
\caption[]{Mapping of input bitstrings to output bitstrings for the implementation of the function $ x^2 $ for values of the interval $ [0, 2] $.}
\label{Table:Minimal_example_BS_for_non-lin_functions_mapping}
\end{table*}
\begin{figure*}[t]
\begin{quantikz}
 \lstick{\ket{3\text{rd digit}}} &[-0.1cm] 
 &[-0.1cm] \ocontrol{} &[-0.1cm] \control{}
 &[-0.1cm] 
  \control{} &[-0.1cm] \control{} &[-0.1cm] \ocontrol{} &[-0.1cm] \control{} &[-0.1cm] \ocontrol{} &[-0.1cm] \control{} &[-0.1cm] \ocontrol{} &[-0.1cm] \control{} &[-0.1cm] \swap{3} 
 &[-0.1cm] &[-0.1cm]  &[-0.1cm] &[-0.1cm]   \\
 \lstick{\ket{2\text{nd digit}}} & 
 & \ocontrol{} & \ocontrol{}
 &  \control{} & \ocontrol{} & \control{} & \control{} & \ocontrol{} & \ocontrol{} & \control{} & \control{} & & \swap{3}  & & &  \\
 \lstick{\ket{1\text{st digit}}} & 
 & \ocontrol{}  & \ocontrol{}  &  \ocontrol{} & \control{} & \control{} & \control{} & \control{} & \control{} & \control{} & \control{} & & & \swap{3} & &   \\ 
 \lstick{\ket{0}} & \gate{X} & \targ{}\wire[u][3]{q} & \targ{}\wire[u][3]{q} &  & & & & & & & & \targX{} & & & \gate{ \text{reset to } \ket{0} } &  \\
 \lstick{\ket{0}} & & & & \targ{}\wire[u][4]{q} & \targ{}\wire[u][4]{q} & \targ{}\wire[u][4]{q} & \targ{}\wire[u][4]{q} & & & & & & \targX{} 
 & & \gate{ \text{reset to } \ket{0} } &  \\
 \lstick{\ket{0}} &  &  & &  & & & & \targ{}\wire[u][5]{q} & \targ{}\wire[u][5]{q} & \targ{}\wire[u][5]{q} & \targ{}\wire[u][5]{q} & & & \targX{} & \gate{ \text{reset to } \ket{0} } &  
\end{quantikz}
\captionsetup{justification=raggedright, singlelinecheck=false}
\caption[]{Example for the general circuit pattern for the implementation of a mapping from input bitstrings to output bitstrings. 
Specifically, the mapping of Table {\hyperref[Table:Minimal_example_BS_for_non-lin_functions_mapping]{\ref*{Table:Minimal_example_BS_for_non-lin_functions_mapping}}} is realized.}
\label{Fig:Minimal_example_BS_for_non-lin_functions_mapping}
\end{figure*}
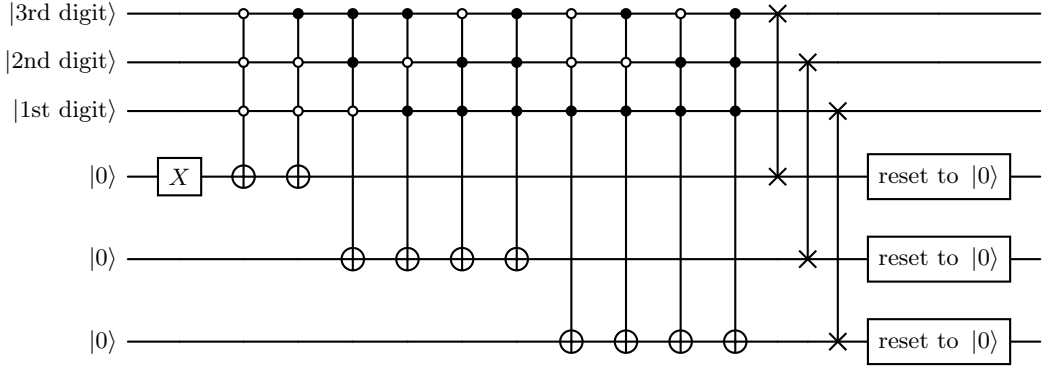

In the following, a general implementation procedure is illustrated for the minimal example to represent the 1D function $ f(x) = x^2 $ for real arguments: 
The input and output values are referred to the same value interval, which is discretized via $ 2^n $ values that are labeled by the bitstrings formed by $ n $ qubits. 
Specifically, the interval $ [0, {\phi}] $ with $ {\phi} = 2 $ and a discretization via equidistant points according to $ i \cdot {\frac{\phi}{L}}, ~ i \in \{ 0, 1, ... , L \} , ~ L = {2^n} - 1 $ are chosen here. 
W.r.t. the bitstring encoding according to $ \{ 0, 1, ... , L \} $ of input and output values, $ {\frac{\phi}{L}} $ is then the unit, i.e. the multiplication factor by which the encoded values have to be multiplied to obtain back the scale of the encoded problem. 
Here, $ n = 3 $ qubits are taken, so that there are $  8 $ possible input values and also $ 8 $ associated output values that approximate the result of the non-linear function w.r.t. the considered interval $ [0, {\phi}] $. 
For all input states, the resulting output states have to be calculated once in order to set up a circuit that mediates a corresponding mapping. 
E.g., the input $ x = 0 \cdot {\frac{2}{7}} $ is mapped under $ x^2 $ to $ 0 \cdot {\frac{2}{7}} $, which is represented by the bitstring for $ i = 0 $ and the input $ x = 1 \cdot {\frac{2}{7}} $ is mapped to $ {\frac{2}{7}} \cdot {\frac{2}{7}} \approx {0.286} \cdot {\frac{2}{7}} $, which is also represented by the bitstring for $ i = 0 $ since $ {0.285} $ is closer to $ i = 0 $ than to $ i = 1 $. 
The complete mapping is given in Table {\hyperref[Table:Minimal_example_BS_for_non-lin_functions_mapping]{\ref*{Table:Minimal_example_BS_for_non-lin_functions_mapping}}} and corresponds to the matrix
\begin{align*}
\begin{pmatrix}
 1 & 1 & 0 & 0 & 0 & 0 & 0 & 0  \\
 0 & 0 & 1 & 0 & 0 & 0 & 0 & 0  \\
 0 & 0 & 0 & 0 & 0 & 0 & 0 & 0  \\
 0 & 0 & 0 & 1 & 0 & 0 & 0 & 0  \\
 0 & 0 & 0 & 0 & 0 & 0 & 0 & 0  \\
 0 & 0 & 0 & 0 & 1 & 0 & 0 & 0  \\
 0 & 0 & 0 & 0 & 0 & 0 & 0 & 0  \\
 0 & 0 & 0 & 0 & 0 & 1 & 1 & 1  \\
\end{pmatrix} , 
\end{align*}
from which the approximation of $ x^2 $ is directly visible according to a horizontally mirrored parabola given by the entries $1$ until the upper bound of the value interval induces a cut-off. 
This non-unitary matrix has to be implemented now in a quantum circuit.
\begin{figure}[t]
\begin{quantikz}
 \lstick{\ket{3\text{rd digit}}}
 &[-0.1cm]  &[-0.1cm] 
 &[-0.1cm] \control{} &[-0.1cm] \ocontrol{}
 &[-0.1cm]  &[-0.1cm] 
 \swap{4}  &[-0.1cm]  &[-0.1cm]  \\
 \lstick{\ket{2\text{nd digit}}} & 
 & \ocontrol{} & 
 \control{} & \ocontrol{} & \swap{2} 
 &  & &  \\
 \lstick{\ket{1\text{st digit}}}
 & \control{}  & \ocontrol{}  & \ocontrol{} & \control{} &  &  & &  \\ 
 \lstick{\ket{0}}  & \targ{}\wire[u][1]{q} &  & \targ{}\wire[u][3]{q} & \targ{}\wire[u][3]{q} &  
 \targX{} &  & \gate{ \text{reset to } \ket{0} } &   \\
 \lstick{\ket{0}} & \gate{X} & \targ{}\wire[u][3]{q} & & & &  \targX{} 
 & \gate{ \text{reset to } \ket{0} } &  
\end{quantikz}
\captionsetup{justification=raggedright, singlelinecheck=false}
\caption[]{Specific circuit implementation of the mapping of Table {\hyperref[Table:Minimal_example_BS_for_non-lin_functions_mapping]{\ref*{Table:Minimal_example_BS_for_non-lin_functions_mapping}}} based on exploiting correlations between the digits of the inputs and outputs.}
\label{Fig:Minimal_example_BS_for_non-lin_functions_mapping_better}
\end{figure}
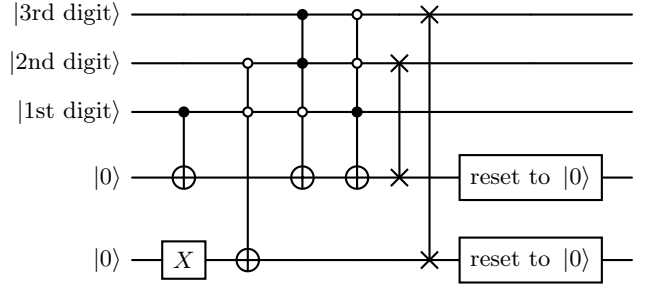

A general procedure would be to introduce as many ancilla qubits as are needed to represent a bitstring and prepare the individual ancilla qubits correspondingly in the state $ | 0 \rangle $ or $ | 1 \rangle $ in dependence of the specific input state via multi-controlled X-gates as can be seen in Fig. {\hyperref[Fig:Minimal_example_BS_for_non-lin_functions_mapping]{\ref*{Fig:Minimal_example_BS_for_non-lin_functions_mapping}}}. 
Fig. {\hyperref[Fig:Minimal_example_BS_for_non-lin_functions_mapping]{\ref*{Fig:Minimal_example_BS_for_non-lin_functions_mapping}}} depicts as well that if it is needed that the input ports are the output ports, the qubit states can be exchanged afterwards via SWAP-gates and the ancilla qubits can then be reset back to the $ | 0 \rangle $-states from which it was started. 
For the circuit in Fig. {\hyperref[Fig:Minimal_example_BS_for_non-lin_functions_mapping]{\ref*{Fig:Minimal_example_BS_for_non-lin_functions_mapping}}}, it was furthermore counted for each binary digit whether there are more cases for the outputs of all inputs in which the considered digit is $1$ than cases in which it is $0$. 
In such a case like for the third digit in the considered example, the ancilla qubit for this output digit should be initialized directly in the $ | 1 \rangle $-state via the application of an $X$-gate in order to minimize the number of the following required multi-controlled X-operations, as it was done for the top ancilla qubit in Fig. {\hyperref[Fig:Minimal_example_BS_for_non-lin_functions_mapping]{\ref*{Fig:Minimal_example_BS_for_non-lin_functions_mapping}}}, which delivers the third output digit at the end.

However, a more efficient implementation can be obtained by looking for correlations between the binary digits of inputs and outputs for the specific mapping like done for obtaining the circuit in Fig. {\hyperref[Fig:Minimal_example_BS_for_non-lin_functions_mapping_better]{\ref*{Fig:Minimal_example_BS_for_non-lin_functions_mapping_better}}}. 
E.g., for the considered example, it can be noticed from Table {\hyperref[Table:Minimal_example_BS_for_non-lin_functions_mapping]{\ref*{Table:Minimal_example_BS_for_non-lin_functions_mapping}}} that the first digit, i.e. the most left digit, does not change, i.e., no gate operation has to be applied to this qubit and thus only two ancilla qubits to account for the other binary digits are needed. 
To lower the amount of operations, it is also again reasonable to prepare the ancilla qubit for the third digit of the output via the application of an X-gate in the $ \left| 1 \right\rangle $-state since in six of the eight results for the output, the third digit is $ 1 $.

Moreover, w.r.t. a subspace that is given by a specific digit for the first digit of the input, the digit in the middle of the 3-bit string for the output has in three of the four cases the same value as the first input digit. 
Therefore, it is also purposeful to set the ancilla qubit for the second digit to the same value as the qubit for the third digit of the 3-qubit input state via a corresponding CX-gate. 
Based on this configuration, the remaining cases w.r.t. the input that are not yet mapped to the correct output state can be treated via the application of multi-controlled X-gates. 
For the considered example, there are still the following cases: 
For the subspace given by the value $0$ for the digits of the first and second digit of the input, the third digit of the output still has to be changed back from the value $ 1 $ to $ 0 $. 
Further, for the input cases according to $ 011 $ and $ 100 $, the value of the second digit of the output, which was given so far by the value of the first digit of the input, has still to be changed. 
Thus, by such an analysis on correlations, the number and complexity of the required gates and the number of ancilla qubits can be reduced.
\begin{figure}[t]
\begin{center}
\includegraphics[width=\linewidth]{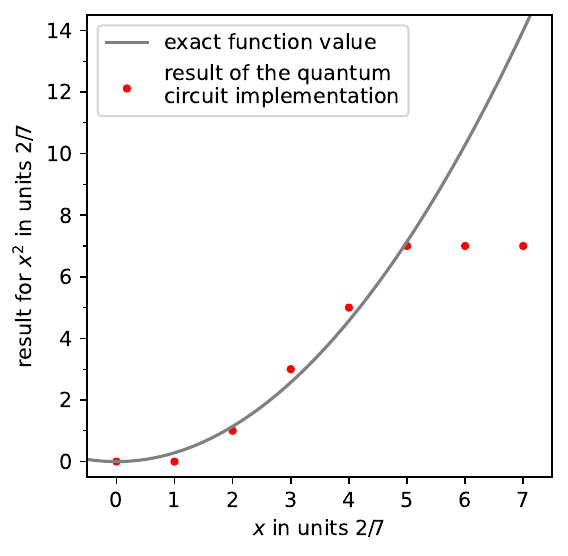}
\end{center}
\captionsetup{justification=raggedright, singlelinecheck=false}
\caption[]{Results obtained from the statevector simulation for the circuits of the Figs. {\hyperref[Fig:Minimal_example_BS_for_non-lin_functions_mapping]{\ref*{Fig:Minimal_example_BS_for_non-lin_functions_mapping}}} and {\hyperref[Fig:Minimal_example_BS_for_non-lin_functions_mapping_better]{\ref*{Fig:Minimal_example_BS_for_non-lin_functions_mapping_better}}}, illustrating the approximate realization of the function $ x^2 $ according to Table {\hyperref[Table:Minimal_example_BS_for_non-lin_functions_mapping]{\ref*{Table:Minimal_example_BS_for_non-lin_functions_mapping}}}.}
\label{Fig:Results_minimal_example_BS_for_non-lin_functions}
\end{figure}

From the implementation of the circuits according to the Figs. {\hyperref[Fig:Minimal_example_BS_for_non-lin_functions_mapping]{\ref*{Fig:Minimal_example_BS_for_non-lin_functions_mapping}}} and {\hyperref[Fig:Minimal_example_BS_for_non-lin_functions_mapping_better]{\ref*{Fig:Minimal_example_BS_for_non-lin_functions_mapping_better}}} and the calculation of the resulting statevector via Qiskit, Fig. {\hyperref[Fig:Results_minimal_example_BS_for_non-lin_functions]{\ref*{Fig:Results_minimal_example_BS_for_non-lin_functions}}} was obtained, which visualizes the approximation principle 
for the considered non-linear function.

\subsection{A Mapping of the Linear Advection Equation to the Bernstein-Vazirani Quantum Algorithm}
\label{App_subsec:A_Mapping_of_the_LA_Eq_to_the_BV-Q-Alg}

For the potential of QC w.r.t. bitstring encoding, often the conceptual possibility is mentioned that QC allows to feed in all classically possible inputs in parallel and in principle also operate on them simultaneously. 
A situation that illustrates this relatively well is the Bernstein-Vazirani (BV) problem \cite{Abhijith_et_al_Q_Algs}. 
The BV problem is the task to decipher a 'secret' bitstring, which is encoded in the so-called {\emph{BV oracle}}. 
This oracle can be implemented as shown in Fig. {\hyperref[Fig:Scheme_BV_q_alg]{\ref*{Fig:Scheme_BV_q_alg}}} in the way that if a digit of the secret code is $1$, a CX-gate that has the corresponding bit as the control bit is applied to the additional bit, indicated here by the label $ t $, as the target qubit. 
Based on the knowledge that a bitstring in which the potential control bits, indicated in Fig. {\hyperref[Fig:Scheme_BV_q_alg]{\ref*{Fig:Scheme_BV_q_alg}}} with $ b $, are all $ 0 $ will undergo no changes by the BV oracle, a piece of information about the secret code can be obtained from observing the result of the $t$-bit if a bitstring is fed in. 
An execution of the oracle operation for a chosen input bitstring is also called {\emph{query}} or {\emph{call}} of the oracle.

The QC version of this setup is obtained from putting H-gates, which are specific to QC since they generate a superposition state, before and after the oracle block. 
Classically, $n$ different bitstrings have to be fed in to the oracle to deduce the secret $n$-bit code. 
For the quantum algorithm, it can be shown that by taking the qubit-configuration $ | {b_1} {b_2} \dots {b_n} \rangle \otimes | t \rangle = | {0} {0} \dots {0} \rangle \otimes | 1 \rangle $ as the input, the secret code is directly given as the output of the $b$-qubits. 
Thus, there is a speed-up from $n$ required runs for the classical version to just $1$ run for the quantum algorithm.

As this article has a perspective on QC for CFD problems, it shall be noted here that a mapping of the simulation of the linear advection eq. to the implementation of the BV quantum algorithm can be drawn: 
For the problem of the 1D linear advection eq.
\begin{align}
{\frac{ \partial }{ {\partial}t }} {\varphi}(x, t) &= -c \cdot {\frac{ \partial }{ {\partial}x }} {\varphi}(x, t) 
 \label{eq:LA_eq}
\end{align}
with a given initial field $ {\varphi}(x, t = {t_0}) $, it can be derived, e.g. via Fourier transformation of the equation that the analytical solution is just given by a spatial shift of the initial field with the time according to
\begin{align}
 {\varphi}(x, t) &= {\varphi}(x - ct, {t_0}) .
 \label{eq:LA_eq_solution}
\end{align}
A simple numerical treatment of eq. ({\hyperref[eq:LA_eq]{\ref*{eq:LA_eq}}}) is the upwind discretization w.r.t. the space and the explicit Euler scheme w.r.t. the time, which reads:
\begin{align}
 & {\varphi}({x_i}, {t_{n}} + {\Delta}t ) \\
 &= {\varphi}({x_i}, {t_{n}} ) - 
 {\frac{  {\Delta}t \cdot c }{{\Delta}x}} 
 \cdot \left\{ 
 \begin{array}{ll}
 { {\varphi}({x_i}, {t_{n}}) - {\varphi}({x_{i-1}}, {t_{n}})  } , & c > 0 \\
 { {\varphi}({x_{i+1}}, {t_{n}}) - {\varphi}({x_{i}}, {t_{n}})  } , & c < 0
 \end{array}
 \right.
 \label{eq:LA_eq_considered_discretization} 
\end{align}
Here, it shall be $ {\Delta}x > 0 $ and $ {\Delta}t > 0 $ for the spatial and temporal time step. 
The numerically stable case for $ |c| = \frac{{\Delta}x}{{\Delta}t} $, i.e. a CFL number of $ 1 $, reproduces the exact result ({\hyperref[eq:LA_eq_solution]{\ref*{eq:LA_eq_solution}}}) and for this situation, it is apparent that it can be implemented by the BV quantum circuit:

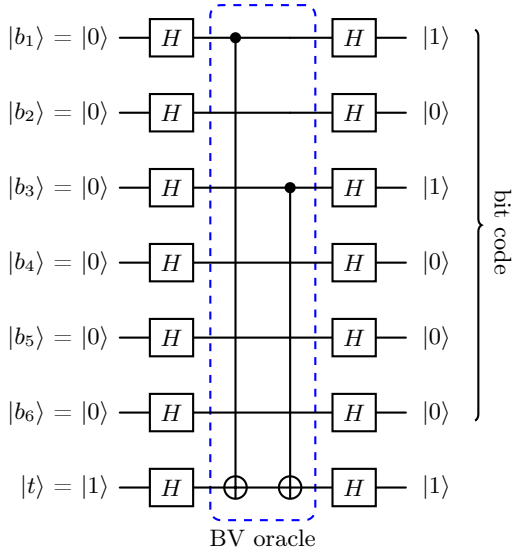
\begin{figure}[t]
\begin{quantikz}
 \lstick{\ket{b_1} = \ket{0} }
  &[-0.1cm] \gate{H}  &[-0.1cm] \control{} \gategroup[7,steps=2,style={blue, dashed, rounded corners, inner sep=2pt},background,label style={label position=below, anchor=north, yshift=-0.2cm}]{ BV oracle }
  &[-0.1cm]  &[-0.1cm] 
 \gate{H} &[-0.1cm] \midstick{\ket{1}}  \rstick[6, label style={xshift=1.1cm, yshift=0.6cm, rotate=-90}]{bit code} \\
 \lstick{\ket{b_2} = \ket{0} } & \gate{H}
  &  & 
  & \gate{H} & \midstick{\ket{0}}  \\
   \lstick{\ket{b_3} = \ket{0} } & \gate{H}
  &   & \control{}  &  \gate{H} &  \midstick{\ket{1}}  \\ 
   \lstick{\ket{b_4} = \ket{0} } & \gate{H} & & & \gate{H} & \midstick{\ket{0}} \\ 
   \lstick{\ket{b_5} = \ket{0} } & \gate{H} & & & \gate{H} & \midstick{\ket{0}} \\
 \lstick{\ket{b_6} = \ket{0} }  & \gate{H} &  &  & \gate{H}  & \midstick{\ket{0}}   \\
 \lstick{\ket{t} = \ket{1}} & \gate{H} & \targ{}\wire[u][6]{q} &  \targ{}\wire[u][4]{q} & \gate{H}  &  \midstick{\ket{1}} 
\end{quantikz}
\captionsetup{justification=raggedright, singlelinecheck=false}
\caption[]{Circuit schematics of the BV quantum algorithm for the example of the secret 6-bit code '101000'. The dashed blue box encomprises an implementation of the BV oracle.}
\label{Fig:Scheme_BV_q_alg}
\end{figure}

For the party that sets up the BV oracle, the secret code has to be already known and for the considered situation of the simulation of eq. ({\hyperref[eq:LA_eq]{\ref*{eq:LA_eq}}}), the known quantity is the initial field $ {\varphi}(x, t = {t_0}) $. 
Accordingly, for this quantum algorithm, the initial field is not encoded as an initial quantum state but a bitstring representation of it is encoded as the BV oracle gate operation itself.

Based on this encoding of the initial field, the BV oracle block can be augmented to account subsequently for the time marching, i.e., the time-marching can be implemented before the second layer of H-gates in the BV quantum algorithm, as shown exemplarily in Fig. {\hyperref[Fig:implementation_example_BV_for_LA-eq]{\ref*{Fig:implementation_example_BV_for_LA-eq}}} for a shift by $ 4 $ time steps of the example bitstring considered in Fig. {\hyperref[Fig:Scheme_BV_q_alg]{\ref*{Fig:Scheme_BV_q_alg}}}. 
If a value at a certain grid point is represented according to the bitstring encoding format by $ d $ qubits, $ k $ time steps can be implemented by swapping all qubits in the $b$-register subsequently by $ d \cdot k $ positions or just by swapping the qubit pairs for which the two qubits are in different states, where the sequence of SWAP-operations has to start with the qubit pair that involves the highest position for the case $ c = {\frac{{\Delta}x}{{\Delta}t}} $ and the lowest position for the case  $ c = -{\frac{{\Delta}x}{{\Delta}t}} $.

Based on this scheme for the time marching, periodic boundary conditions or an outlet for a domain boundary can be implemented relatively easy. 
If a SWAP-gate of the considered extension to realize the given number of time steps $ k $ would reach over a boundary of the $ b $-register, the counting of the $ b $-qubits for determining the partner qubit of the SWAP-operation has to be continued from the other side of the $ b $-register according to a counting with modulo-operation w.r.t. the number of qubits in the $ b $-register in order to implement periodic boundary conditions and such a mismatching SWAP-operation has to be replaced by a CX-operation with the considered $ b $-qubit as the control qubit and the $ t $-qubit as the target qubit in order to implement domain boundaries as outlets. 
Minimal implementation examples for these two boundary condition cases are depicted in Fig. {\hyperref[Fig:implementation_example_BV_for_LA-eq]{\ref*{Fig:implementation_example_BV_for_LA-eq}}} (a) and (b) in the red dotted circuit region, respectively and results w.r.t. these two procedures for an extended example, which were obtained from the statevector of the simulated circuit, are shown in the Figs. {\hyperref[Fig:Results_minimal_example_BV_for_LA-eq]{\ref*{Fig:Results_minimal_example_BV_for_LA-eq}}} (a) and (b), respectively.

\begin{figure*}[t]
\begin{minipage}[c]{0.485\textwidth}
\centering
\begin{tikzpicture}
\draw[draw=none] (0, 0) -- (0.5\linewidth, 0);
\draw (-0.475\linewidth, 0) node{(a)};
\end{tikzpicture}
\begin{quantikz}
 \lstick{\ket{b_1} = \ket{0} }
  &[-0.1cm] \gate{H}  &[-0.1cm] \control{} \gategroup[7,steps=2,style={blue, dashed, rounded corners, inner sep=2pt},background,label style={label position=below, anchor=north, yshift=-0.2cm}]{  }
  &[-0.1cm]  &[0.1cm]  
  \swap{4}\gategroup[7,steps=2,style={red, dotted, rounded corners, inner ysep=2pt, inner xsep=6pt},background,label style={label position=below, anchor=north, yshift=-0.2cm}]{  }
  &  \swap{2} & 
 \gate{H} &[-0.1cm] \midstick{\ket{1}}  \rstick[6, label style={xshift=1.1cm, yshift=0.5cm, rotate=-90}]{result} \\
 \lstick{\ket{b_2} = \ket{0} } & \gate{H}
  & & & & & \gate{H} & \midstick{\ket{0}}  \\
   \lstick{\ket{b_3} = \ket{0} } & \gate{H}
  &   & \control{}  &   & \targX{} & \gate{H} & \midstick{\ket{0}}  \\ 
   \lstick{\ket{b_4} = \ket{0} } & \gate{H} & & & & &  \gate{H} & \midstick{\ket{0}} \\ 
   \lstick{\ket{b_5} = \ket{0} } & \gate{H} & & & \targX{} & &  \gate{H} & \midstick{\ket{1}} \\
 \lstick{\ket{b_6} = \ket{0} }  & \gate{H} &  &  & & &  \gate{H}  & \midstick{\ket{0}}   \\
 \lstick{\ket{t} = \ket{1}} & \gate{H} & \targ{}\wire[u][6]{q} &  \targ{}\wire[u][4]{q} & & &  \gate{H}  &  \midstick{\ket{1}} 
\end{quantikz}
\end{minipage}
\hfill
\begin{minipage}[c]{0.485\textwidth}
\centering
\begin{tikzpicture}
\draw[draw=none] (0, 0) -- (0.5\linewidth, 0);
\draw (-0.475\linewidth, 0) node{(b)};
\end{tikzpicture}
\begin{quantikz}
 \lstick{\ket{b_1} = \ket{0} }
  &[-0.1cm] \gate{H}  &[-0.1cm] \control{} \gategroup[7,steps=2,style={blue, dashed, rounded corners, inner sep=2pt},background,label style={label position=below, anchor=north, yshift=-0.2cm}]{  }
 &[-0.1cm] &[0.1cm]  
  \gategroup[7,steps=2,style={red, dotted, rounded corners, inner ysep=2pt, inner xsep=6pt},background,label style={label position=below, anchor=north, yshift=-0.2cm}]{  }
   &[-0.1cm] \swap{4} & 
 \gate{H} &[-0.1cm] \midstick{\ket{0}}  \rstick[6, label style={xshift=1.1cm, yshift=0.5cm, rotate=-90}]{result} \\
 \lstick{\ket{b_2} = \ket{0} } & \gate{H}
  & & & & & \gate{H} & \midstick{\ket{0}}  \\
   \lstick{\ket{b_3} = \ket{0} } & \gate{H}
  &   & \control{}  & \control{}  &  & \gate{H} & \midstick{\ket{0}}  \\ 
   \lstick{\ket{b_4} = \ket{0} } & \gate{H} & & & & &  \gate{H} & \midstick{\ket{0}} \\ 
   \lstick{\ket{b_5} = \ket{0} } & \gate{H} & & &  & \targX{} &  \gate{H} & \midstick{\ket{1}} \\
 \lstick{\ket{b_6} = \ket{0} }  & \gate{H} &  &  & & &  \gate{H}  & \midstick{\ket{0}}   \\
 \lstick{\ket{t} = \ket{1}} & \gate{H} & \targ{}\wire[u][6]{q} &  \targ{}\wire[u][4]{q} & \targ{}\wire[u][4]{q} & &  \gate{H}  &  \midstick{\ket{1}} 
\end{quantikz}
\end{minipage}
\captionsetup{justification=raggedright, singlelinecheck=false}
\caption[]{Minimal example for the implementation of the solving procedure ({\hyperref[eq:LA_eq_considered_discretization]{\ref{eq:LA_eq_considered_discretization}}}) for the linear advection eq. ({\hyperref[eq:LA_eq]{\ref*{eq:LA_eq}}}) with (a) a periodic boundary condition and (b) an outlet boundary condition for the domain boundary in convection direction via the BV quantum algorithm. 
The dotted red surrounded gate block extends the BV oracle of the example of Fig. {\hyperref[Fig:Scheme_BV_q_alg]{\ref*{Fig:Scheme_BV_q_alg}}}, which encodes an initial bitstring, by a shift of the digits by 4 positions to the right or downwards w.r.t. the $ | b \rangle $-qubits in the circuit, where the respective boundary conditions are taken into account. Hence, the resulting bitstring code is '100010' for (a) and '000010' for (b). 
Specifically for (a), the qubit $ | {b_3} \rangle $, which would hold the digit $ 1 $ after the circuit in Fig. {\hyperref[Fig:Scheme_BV_q_alg]{\ref*{Fig:Scheme_BV_q_alg}}}, is at a further position and thus also the position for which the qubit state has to be swapped is in principle further ahead but due to the rule for periodic boundary conditions, the associated qubit for the SWAP is the qubit $ | {b_1} \rangle $, i.e. the other qubit for which the value $ 1 $ would result after the circuit of Fig. {\hyperref[Fig:Scheme_BV_q_alg]{\ref*{Fig:Scheme_BV_q_alg}}}. 
For this first qubit, the associated SWAP-partner qubit is then further ahead than the qubit $ | {b_3} \rangle $ and the index of its determined exchange partner qubit is still below $ 7 $, so that the SWAP-operation of the first qubit for which the digit of the initial bitstring is $ 1 $ has to be applied first and then the SWAP-operation involving the qubit $ | {b_3} \rangle $ as the second qubit with a digit of $ 1 $ in the initial bitstring while the inverse order of these SWAP-gates would not yield the correct result. Specifically for (b), the annihilation of a digit $ 1 $ can be implemented by another CX-gate as the CX-gate is its own inverse (cf. Fig. {\hyperref[Fig:Decomposition_LSB-Disentangling_Cancelling_CX-gates_in_middle]{\ref*{Fig:Decomposition_LSB-Disentangling_Cancelling_CX-gates_in_middle}}}).}
\label{Fig:implementation_example_BV_for_LA-eq}
\end{figure*}
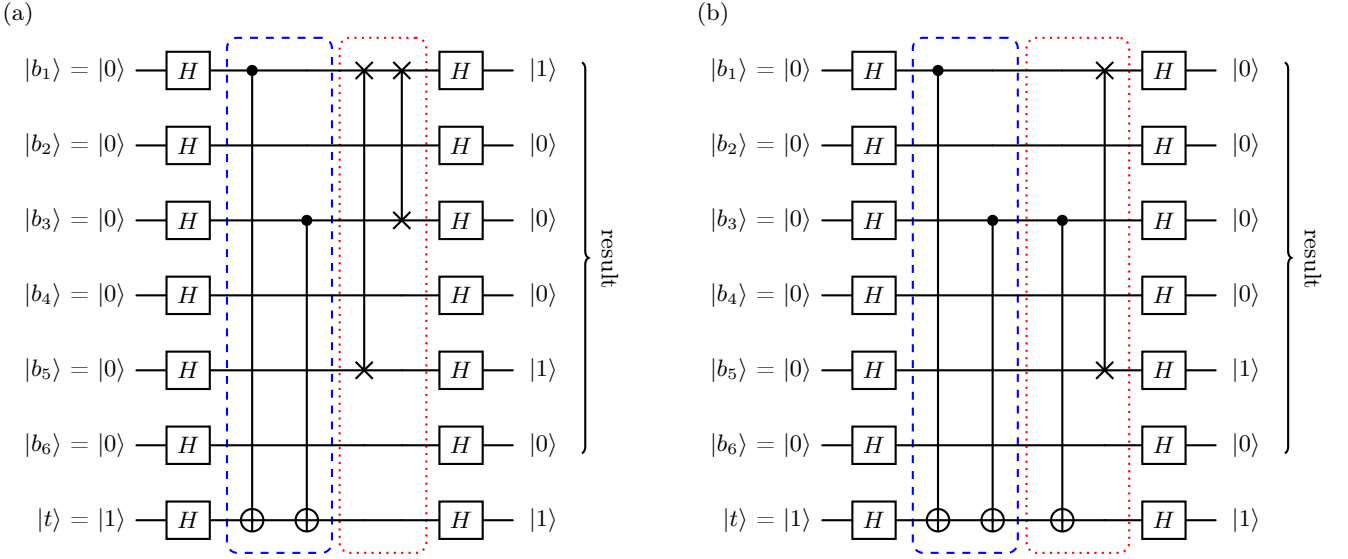

\begin{figure*}[t]
\begin{minipage}[c]{0.485\textwidth}
\centering
\begin{tikzpicture}
\draw (0,0) node[inner sep=0]{\includegraphics[height=0.95\linewidth]{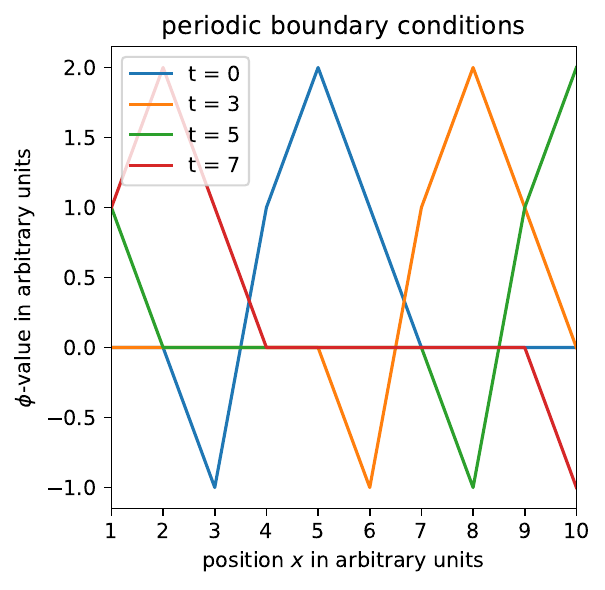}};
\draw[draw=none] (0, 0) -- (0.5\linewidth, 0);
\draw (-0.475\linewidth, 0.45\linewidth) node{(a)};
\end{tikzpicture}
\end{minipage}
\hfill
\begin{minipage}[c]{0.485\textwidth}
\centering
\begin{tikzpicture}
\draw (0,0) node[inner sep=0]{\includegraphics[height=0.95\linewidth]{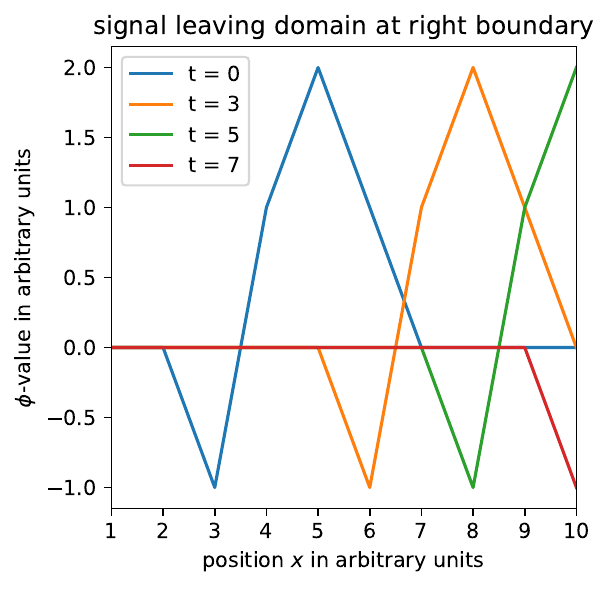}};
\draw[draw=none] (0, 0) -- (0.5\linewidth, 0);
\draw (-0.475\linewidth, 0.45\linewidth) node{(b)};
\end{tikzpicture}
\end{minipage}
\captionsetup{justification=raggedright, singlelinecheck=false}
\caption[]{Results obtained from the statevector simulation of an extended minimal example according to the schemes indicated in Fig. {\hyperref[Fig:implementation_example_BV_for_LA-eq]{\ref*{Fig:implementation_example_BV_for_LA-eq}}} (a) and (b), respectively. Here, two qubits are used to represent a signal value of the set $ \{ -1, 0, 1, 2 \} $ at a specific spatial point according to bitstring encoding.}
\label{Fig:Results_minimal_example_BV_for_LA-eq}
\end{figure*}

However, it is apparent from this quantum circuit implementation that it is practically just a complication of a simple classical implementation, since it corresponds to classical bitstring encoding of the initial field, which is then shifted, just that the more complex H- and CX-gates are needed in the QC version, whereas only X-gates and no $t$-bit would be required in a classical implementation. 
The presented quantum circuit will thus not be more efficient than the direct classical implementation. 
The solving scheme can be represented in the framework of the implementation of the BV quantum algorithm but the actually considered problems are different. 
In the BV problem, a quantity is extracted from the BV oracle, for which it is set by definition that this can be done classically only by the application of multiple inputs, from which this quantity is then computed, whereas the simulation of the eq. ({\hyperref[eq:LA_eq]{\ref*{eq:LA_eq}}}) according to ({\hyperref[eq:LA_eq_considered_discretization]{\ref*{eq:LA_eq_considered_discretization}}}) corresponds to one matrix-vector multiplication to compute a transformed quantity from a known initial one, for which just an encoding as operations for the representation as a BV oracle was used here.

\section{Probability of Fulfilling the Error Bound for the Situation of Equal Probabilities}
\label{App_sec:Probability_Fulfilling_error_bound_for_reference_situation}

For the situation of equal probabilities of $ \frac{1}{{\tilde{n}}} $ w.r.t. $ {\tilde{n}} $ possible outcomes of a single random experiment that is considered as a reference for the read-out accuracy in the subsections {\hyperref[subsec:Read-out]{\ref*{subsec:Read-out}}} and {\hyperref[subsec:Study_Read-out]{\ref*{subsec:Study_Read-out}}}, an analytical consideration is given here:

Since there are again $ {\tilde{n}} $ possible outcomes after one run of the random experiment, there are $ {{{\tilde{n}}}^{N}} $ possible outcomes for conducting $ N $ times such a random experiment with $ {\tilde{n}} $ outcomes for one run. 
The number of combinations w.r.t. the $ N $ runs in which the outcome $ i \in \{ 1, \dots, {\tilde{n}} \} $ of one random experiment occurs $ {k_i} \in \{ 0, \dots , N \} $ times is given by the multinomial coefficient
\begin{align}
 \begin{pmatrix}
 N \\
 {k_{1}}, \dots , {k_{\tilde{n}}}
 \end{pmatrix} &= \frac{ N! }{ {{k_{1}}!} \cdot \dots \cdot {{k_{\tilde{n}}}!} } , 
\label{eq:multinomial_coefficient}
\end{align}
where 
\begin{align}
 {\sum_{i=1}^{ {\tilde{n}} }} {k_i} &= N
 \label{eq:multinomial_coefficient_normalization}
\end{align} 
and correspondingly, $ {\sum_{i=1}^{ {\tilde{n}} }} {{\mathcal{P}}_i} = 1 $ is fulfilled for the inferred probabilities $ {{\mathcal{P}}_i} = \frac{{k_i}}{N} $. 
The probability of an outcome according to a combination
\begin{align}
 ({k_1}, \dots , {k_{\tilde{n}}})
 \label{eq:k_i-combination}
\end{align}
is then
\begin{align}
 {{Pr(1)}^{k_1}} \cdot \dots \cdot {{Pr({\tilde{n}})}^{k_{\tilde{n}}}} 
 \label{eq:general_probability_of_one_k_i-combination}
\end{align}
and the summation over all possible outcomes that are represented by the combination of counts ({\hyperref[eq:k_i-combination]{\ref*{eq:k_i-combination}}}) is given by ({\hyperref[eq:multinomial_coefficient]{\ref*{eq:multinomial_coefficient}}}), so that multiplying ({\hyperref[eq:multinomial_coefficient]{\ref*{eq:multinomial_coefficient}}}) and ({\hyperref[eq:general_probability_of_one_k_i-combination]{\ref*{eq:general_probability_of_one_k_i-combination}}}) gives the success probability of the event that an outcome represented by the counts ({\hyperref[eq:k_i-combination]{\ref*{eq:k_i-combination}}}) is found in the $ N $ runs. 
For the consideration of error intervals, the task is then to identify all combinations ({\hyperref[eq:k_i-combination]{\ref*{eq:k_i-combination}}}) that represent a probability distribution within the error band.

In general, to be able to reach an accuracy of the inferred $ {{\mathcal{P}}_i} $ given by an error interval $ [ {Pr(i)} - {{\epsilon}_i}, {Pr(i)} + {{\epsilon}_i} ] $ around the real probability $ Pr(i) $, $ N $ has to be that large that
\begin{align}
 \left| {\frac{{k_i}}{N}} - {Pr(i)}  \right| & \leq {{\epsilon}_i}
 \label{eq:general_necessary_condition_for_error_bound}
\end{align}
is possible in principle for all $ i $. 
E.g., for the simple example of $ {\tilde{n}} = 2 $ and $ Pr(i) = \frac{1}{{\tilde{n}}} $ for both $ i $, $ N = 2 $ runs are sufficient to be able to resolve the $ Pr(i) $ for any error $ \epsilon $, where the success probability $ 1 - {\delta} $ is given by $ 0.5 $. 
However, at this point, it is already apparent that the reachable accuracy depends on the specific situation, e.g. the considered error $ \epsilon $, and can even decrease for an increase of $ N $ (cf. also Fig. {\hyperref[Fig:GAE_reference_situation_analytical_formula_delta_wrt_N_example_plots]{\ref*{Fig:GAE_reference_situation_analytical_formula_delta_wrt_N_example_plots}}} ): 
E.g., for setting $ N $ to $ 3 $, it is still possible to resolve the $ Pr(i) $ if e.g. $ \epsilon = 0.25 $ is considered since $ {{\mathcal{P}}_i} = \frac{{1}}{3} $ and $ {{\mathcal{P}}_{2 - i + 1}} = \frac{{2}}{3} $ can occur but it is not possible anymore if e.g. $ \epsilon = 0.1 $ is considered, i.e. $ 1 - {\delta} = 0 $ in the latter example.

For a general given distribution $ Pr(i) $, a procedure for identifying the combinations ({\hyperref[eq:k_i-combination]{\ref*{eq:k_i-combination}}}) that fulfil (\hyperref[eq:multinomial_coefficient_normalization]{\ref*{eq:multinomial_coefficient_normalization}}) and ({\hyperref[eq:general_necessary_condition_for_error_bound]{\ref*{eq:general_necessary_condition_for_error_bound}}}) for a given $ N $ would be to check all configurations w.r.t. the possible values for the $ k_i $, e.g. by considering all values of $ \{ 0, \dots , N \} $ for a chosen $ {k_i} $ and inspecting then subsequently the combinations that are formable by distributing the rest of the amount $ N $ to the other $ {k_i} $. 
However, this analysis would yield all individual matching $ ( {k_1}, \dots , {k_{\tilde{n}}} ) $, where it would be looked at some of them multiple times in particular.

For the special case that it is $ Pr(i) = \frac{1}{{\tilde{n}}} $ and a relative error according to $ \frac{ \epsilon }{ {\tilde{n}} } $ with $ \epsilon < 1 $ is considered for $ {{\epsilon}_i} $ for all $ i $, the information about the individual matching combinations $ ( {k_1}, \dots , {k_{\tilde{n}}} ) $ is not needed and the procedure of finding relevant contributions to the success probability $ 1 - {\delta} $ can be reduced if the consideration is further restricted to numbers of experiment runs $ N $ that are a multiple of $ {\tilde{n}} $ according to $ N = z \cdot {\tilde{n}}, z \in \mathbb{N} $: 
For $ z = 1 $, there is only one valid combination $ ( {k_1}, \dots , {k_{\tilde{n}}} ) $, which is $ ( {1}, \dots , {1} ) $, because otherwise one $ i $ would be detected $ 0 $ times, so that $ {\mathcal{P}}(i) = 0 $, which would be out of the error interval $ [ {\frac{1}{ {\tilde{n}} }} - {\frac{\epsilon}{ {\tilde{n}} }}, {\frac{1}{ {\tilde{n}} }} + {\frac{\epsilon}{ {\tilde{n}} }} ] $, since it was set $ \epsilon < 1 $. 
For $ z = 2 $, it depends whether it is already $ {\frac{1}{N}} = {\frac{1}{2{\tilde{n}}}} \geq {\frac{1}{{\tilde{n}}}} - {\frac{{\epsilon}}{{\tilde{n}}}} ~ \Leftrightarrow ~ {\frac{1}{2}} \leq {\epsilon} $. 
If not, there is still only the combination $ ( {k_1}, \dots , {k_{\tilde{n}}} ) = ( {z}, \dots , {z} ) $ fulfilling the error bound. 
But if it is the case, also $ {k_i} $-values of $ {2} \pm {1} $ represent outcomes for the $ {\mathcal{P}}(i) $ that are within the error interval, i.e., 'excitations' of $ {\frac{1}{N}} $ referred to the level $ {\frac{2}{N}} = {\frac{1}{{\tilde{n}}}} $ are allowed w.r.t. the $ {\mathcal{P}}(i) $, where ({\hyperref[eq:multinomial_coefficient_normalization]{\ref*{eq:multinomial_coefficient_normalization}}}) still has to be fulfilled of course. 
This situation that deviations of $ {\frac{1}{N}} $ around $ {\frac{1}{{\tilde{n}}}} $ are allowed is called here the first 'extension level', where the situation that only the configuration $ ( {k_1}, \dots , {k_{\tilde{n}}} ) = ( {z}, \dots , {z} ) $ resides within the error band is referred to as the extension level $ 0 $. 
Generalized, at the $ j $th extension level, it holds:
\begin{align}
 \frac{j}{N} \leq \frac{{\epsilon}}{{\tilde{n}}}  \quad \Leftrightarrow \quad \frac{j}{z} \leq \epsilon
 \label{eq:condition_for_extension_level_j}
\end{align}
The basic idea for identifying all valid combinations $ ( {k_1}, \dots , {k_{\tilde{n}}} ) $ is now to reduce the efforts for inspecting such combinations by a consideration of valid 'excitations'.

\begin{figure*}[t]
\begin{minipage}[c]{0.485\textwidth}
\centering
\begin{tikzpicture}
\draw (0,0) node[inner sep=0]{\includegraphics[height=0.95\linewidth]{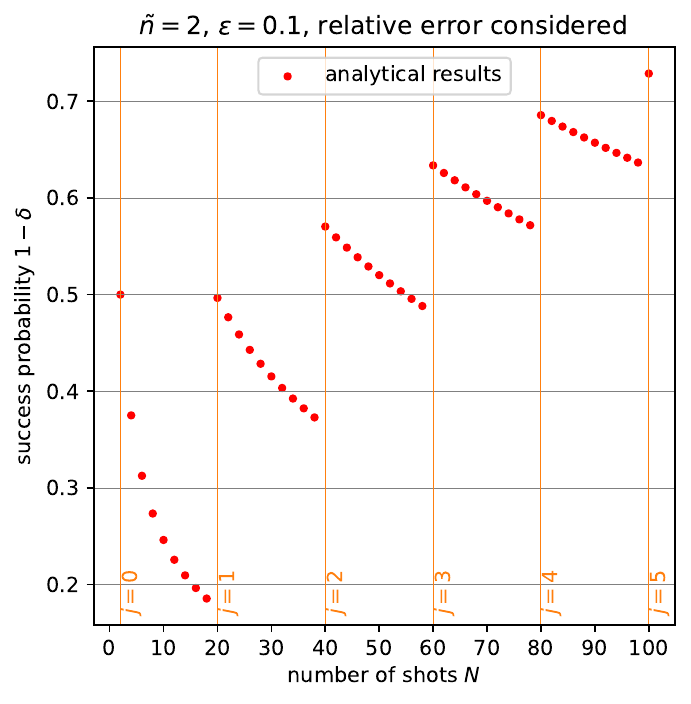}};
\draw[draw=none] (0, 0) -- (0.5\linewidth, 0);
\draw (-0.475\linewidth, 0.45\linewidth) node{(a)};
\end{tikzpicture}
\end{minipage}
\hfill
\begin{minipage}[c]{0.485\textwidth}
\centering
\begin{tikzpicture}
\draw (0,0) node[inner sep=0]{\includegraphics[height=0.95\linewidth]{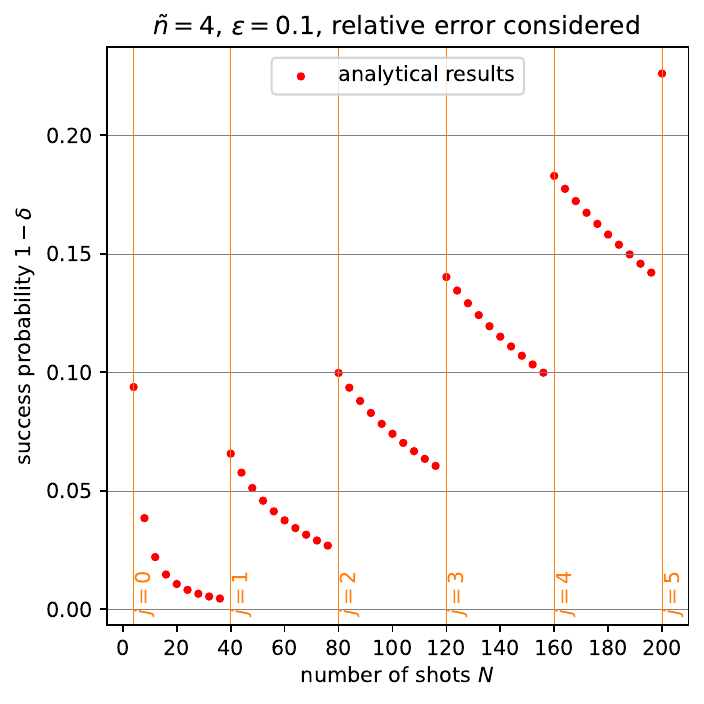}};
\draw[draw=none] (0, 0) -- (0.5\linewidth, 0);
\draw (-0.475\linewidth, 0.45\linewidth) node{(b)};
\end{tikzpicture}
\end{minipage}
\captionsetup{justification=raggedright, singlelinecheck=false}
\caption[]{Analytical results according to the formula ({\hyperref[eq:GAE_reference_situation_analytical_formula_delta_wrt_N]{\ref*{eq:GAE_reference_situation_analytical_formula_delta_wrt_N}}}) for the success probability $ 1 - {\delta} $ to extract a constant probability distribution w.r.t. $ {\tilde{n}} $ possible outcomes of one run of the random experiment via $ N $ runs such that no inferred probability deviates more than $ \frac{\epsilon}{{\tilde{n}}} $ from the real value $ \frac{1}{{\tilde{n}}} $. 
Specifically, exemplary results for $ {\epsilon} = 0.1 $ with (a) $ {\tilde{n}} = 2 $ and (b) $ {\tilde{n}} = 4 $ are shown. 
The reached maximum extension level $ j $ is also indicated.}
\label{Fig:GAE_reference_situation_analytical_formula_delta_wrt_N_example_plots}
\end{figure*}

Since ({\hyperref[eq:multinomial_coefficient_normalization]{\ref*{eq:multinomial_coefficient_normalization}}}) has to hold, an excitation of some $ {k_i} $ in a positive sense w.r.t. the level $ j = 0 $ has to be compensated by a negative excitation of other $ {k_i} $. 
E.g. for $ j = 2, {\tilde{n}} = 4 $, two of the four $ {k_i} $ can be excited negatively by one according to the value $ { \frac{N}{{\tilde{n}}} } - {1} $, but if the third $ {k_i} $ is $ { \frac{N}{{\tilde{n}}} } $, the value of the fourth $ {k_i} $ has to be $ { \frac{N}{{\tilde{n}}} } + {2} $ to account for the amount of excitation. 
Thus, the sum of all excitations has to be zero. 
This example also illustrates that there is in general a degeneracy for a specific combination of excitations w.r.t. the $ {k_i} $, among which the excitations are distributed.

So, all possible outcomes of the $ N $ runs of the random experiment with $ {\tilde{n}} $ outcomes per single run are given via the sum
\begin{align}
 {\sum_{r}} ~ \begin{pmatrix}
 N \\ {K_r}
 \end{pmatrix} 
 \cdot 
 { \left[ {\text{degeneracy factor for }} {K_r} \right] } .
\label{eq:GAE_reference_situation_analytical_formula_delta_wrt_N_schematical_sum}
\end{align}
Here, $ r $ is an index that numbers the excitation configurations that yield inferred probability distributions $ ({\frac{{k_{1}}}{N}}, \dots , {\frac{{k_{\tilde{n}}}}{N}}) $ within the error band $ 2{\epsilon} $, where for the extension level $ j $, every variation around $ \frac{N}{{\tilde{n}}} $ up to an extend $ \pm j $ is possible for the $ {k_i} $. 
Correspondingly, $ {K_r} $ denotes one exemplary configuration $ ( {k_1}, \dots , {k_{\tilde{n}}} ) $ for the considered excitation configuration $ r $. 
Since the excitations are just distributed to the $ {k_i} $, where the number of configurations for this is given by the degeneracy factor, the appearing values in all $ ( {k_1}, \dots , {k_{\tilde{n}}} ) $ for a specific excitation are the same, so that the multinomial coefficient $ \bigl( \begin{smallmatrix} N \\ {k_1}, \dots , {k_{\tilde{n}}} \end{smallmatrix} \bigr) $ is the same for all of them.

The number of such combinations for the distribution of excitations, corresponding to the degeneracy factor, is given again via a multinomial coefficient. 
If the index $ s \in \mathbb{Z} $ labels the 'excitation level' and $ {v_s} $ is its occupation number, i.e. the number of the states $ i $ for which the respective number $ {\frac{N}{\tilde{n}}} + s $ was measured as the associated $ {k_i} $ in the $ N $ runs, then the degeneracy factor for a specific excitation configuration $ r $ is given via
\begin{align}
 \begin{pmatrix}
 {\tilde{n}} \\ {v_{0}}, {v_{1}}, {v_{-1}}, {v_{2}}, {v_{-2}}, \dots \text{until } {v_{\infty}}, {v_{-{\infty}}}
 \end{pmatrix} .
\end{align}
Although this expression goes symbolically until $  {v_{\infty}}, {v_{-{\infty}}} $, it is well-defined since the $ {\tilde{n}} $ states indicated by $ i $ are associated to a specific excitation, respectively, so that it has to hold
\begin{align}
 {\sum_{s = -j}^{j}} ~ {v_s} & \overset{!}{=} {\tilde{n}} , 
 \label{eq:v_s_normalization}
\end{align}
i.e., at most $ {\tilde{n}} $ of the $ {v_i} $ can be unequal to $ 0 $, whereas it results just a factor of $ 0! = 1 $ for the other $ {v_s} $. 
In ({\hyperref[eq:v_s_normalization]{\ref*{eq:v_s_normalization}}}), the summation can be restricted from $ {\sum_{s = -{\infty}}^{\infty}} $ to $ {\sum_{s = -j}^{j}} $ because for the extension level $ j $, only the $ v_s $ with $ |s| \leq j $ can be unequal to $ 0 $. 
In general, it is $ {v_s} \in \{ 0, \dots , {\tilde{n}} \} $ possible w.r.t. the values, where ({\hyperref[eq:v_s_normalization]{\ref*{eq:v_s_normalization}}}) imposes a constraint. 
But moreover, as already stated, to ensure ({\hyperref[eq:multinomial_coefficient_normalization]{\ref*{eq:multinomial_coefficient_normalization}}}) for the associated $ ( {k_1}, \dots , {k_{\tilde{n}}} ) $, it is also required that the total amount of excitation is $ 0 $ according to
\begin{align}
  {\sum_{s = -j}^{j}} ~ s \cdot {v_s} & \overset{!}{=} {0} .
 \label{eq:v_s_1st_moment}
\end{align}
So, the eqs. ({\hyperref[eq:v_s_normalization]{\ref*{eq:v_s_normalization}}}) and ({\hyperref[eq:v_s_1st_moment]{\ref*{eq:v_s_1st_moment}}}) set the requirements for the identification of the valid configurations $ ({v_0}, {v_{1}}, {v_{-1}}, \dots ) $. 
These are labeled here as the $ V_r $ and are unique in contrast to the $ K_r $, which just denote one exemplary $ ({k_1}, \dots , {k_{\tilde{n}}}) $ for a specific $ V_r $. 
An exemplary $ ({k_1}, \dots , {k_{\tilde{n}}}) $ for such a found $ ({v_0}, {v_{1}}, {v_{-1}}, \dots ) $, indicated by $ r $, is then given by setting $ {v_s} $ of the $ {k_i} $-entries to the value $ {z + {s} } $ for all occurring $ s $ for which the excitation occupation number $ v_s $ is not $ 0 $ according to
\begin{align}
 & {K_r} = \nonumber \\
 & ~ {\bigl( {\underbrace{ {z + {{\hat{s}}_{1}} }  , {z + {{\hat{s}}_{1}} }  , ~ \dots  }_{ {v_{{\hat{s}}_{1}}} \text{ times} }} , {\underbrace{ {z + {{\hat{s}}_{2}} }  , {z + {{\hat{s}}_{2}} }  , ~ \dots }_{ {v_{{\hat{s}}_{2}}} \text{ times} }} , ~ \dots \bigr)} , 
 \label{eq:Example_combination_K_r}
\end{align}
where the $ {\hat{s}} $ stand for the $ s $ for which $ {v_s} > 0 $.

If all outcomes in one run of the random experiment are equally probable according to $ \frac{1}{ {\tilde{n}} } $, the probability ({\hyperref[eq:general_probability_of_one_k_i-combination]{\ref*{eq:general_probability_of_one_k_i-combination}}}) for each individual outcome of the random experiment conducted $ N $ times is $ \frac{1}{ {{{\tilde{n}}}^{N}} } $. 
Thus, the success probability $ 1 - {\delta} $ to obtain in $ N = {z} \cdot {\tilde{n}} $ runs an outcome that yields an inferred probability distribution within the error band is
\begin{align} 
 1 - {{\delta}} 
 &= { \frac{1}{ {\tilde{n}}^{N} } } ~ {\sum_{r}} ~ \begin{pmatrix}
 N \\ {K_r}
 \end{pmatrix} 
 \begin{pmatrix}
 {\tilde{n}} \\ 
 {V_r}
 \end{pmatrix} .
 \label{eq:GAE_reference_situation_analytical_formula_delta_wrt_N}
\end{align}

Figs. {\hyperref[Fig:GAE_reference_situation_analytical_formula_delta_wrt_N_example_plots]{\ref*{Fig:GAE_reference_situation_analytical_formula_delta_wrt_N_example_plots}}} (a) and (b) show exemplarily the $ 1 - {{\delta}(N, {\tilde{n}}, {\epsilon})} $ resulting from this combinatorical consideration for $ {\tilde{n}} = 2 $ and $ {\tilde{n}} = 4 $, respectively. 
Formula ({\hyperref[eq:GAE_reference_situation_analytical_formula_delta_wrt_N]{\ref*{eq:GAE_reference_situation_analytical_formula_delta_wrt_N}}}) was implemented such that for a given $ N = {z} \cdot {\tilde{n}} $, the corresponding maximum $ j $ was determined via the condition ({\hyperref[eq:condition_for_extension_level_j]{\ref*{eq:condition_for_extension_level_j}}}) and using the restriction implied thereby for the $ s $ w.r.t. the $ v_s $ that can be unequal to $ 0 $, all configurations $ ({v_0}, {v_{1}}, {v_{-1}}, \dots ) $ that are possible w.r.t. the values $ \{ 0, \dots , {\tilde{n}} \} $ were checked concerning the fulfilling of ({\hyperref[eq:v_s_normalization]{\ref*{eq:v_s_normalization}}}) and ({\hyperref[eq:v_s_1st_moment]{\ref*{eq:v_s_1st_moment}}}) in order to find the $ V_r $.

\newpage
\renewcommand{\refname}{Bibliography}

\end{document}